\documentclass[fleqn,usenatbib]{mnras}

\usepackage{mathptmx}
\usepackage{txfonts}

\usepackage[T1]{fontenc}

\DeclareRobustCommand{\VAN}[3]{#2}
\let\VANthebibliography\thebibliography
\def\thebibliography{\DeclareRobustCommand{\VAN}[3]{##3}\VANthebibliography}

\usepackage{graphicx}	
\usepackage{amssymb}	
\usepackage{textgreek} 	
\usepackage{siunitx}		

\usepackage{booktabs} 	

\usepackage{xargs}    
\usepackage{xcolor, soul}
\sethlcolor{orange}

\title[I. Constraining Quasar Halo Masses]{Resolving the physics of Quasar Ly\textalpha \hspace{1pt} Nebulae (RePhyNe): I. Constraining Quasar host halo masses through Circumgalactic Medium kinematics}

\author[S. de Beer et al.]{
S. de Beer,$^{1}$\thanks{E-mail: s.debeer@campus.unimib.it}
S. Cantalupo,$^{1}$
A. Travascio,$^{1}$
G. Pezzulli,$^{2}$
M. Galbiati,$^{1}$
M. Fossati,$^{1}$
M. Fumagalli,$^{1,3}$
\newauthor
T. Lazeyras,$^{1}$ 
A. Pensabene,$^{1}$
T. Theuns,$^{4}$
W. Wang$^{1}$
\\
$^{1}$Dipartimento di Fisica G. Occhialini, Universita` degli Studi di Milano-Bicocca, Piazza della Scienza 3, 20126 Milano, Italy\\
$^{2}$Kapteyn Astronomical Institute, University of Groningen, Landleven 12, 9747 AD Groningen, The Netherlands \\
$^{3}$INAF - Osservatorio Astronomico di Trieste, via G. B. Tiepolo 11, 34143 Trieste, Italy \\
$^{4}$Institute for Computational Cosmology, Department of Physics, Durham University, South Road, Durham, DH1 3LE, UK
}

\date{Accepted XXX. Received YYY; in original form ZZZ}

\pubyear{2022}

\begin{document}
\label{firstpage}
\pagerange{\pageref{firstpage}--\pageref{lastpage}}
\maketitle

\begin{abstract}
Ly$\alpha$ nebulae ubiquitously found around $z>2$ quasars can supply unique constraints on the properties of the Circumgalactic Medium, such as its density distribution, provided the quasar halo mass is known. We present a new method to constrain quasar halo masses based on the line-of-sight velocity dispersion maps of Ly$\alpha$ nebulae. 
By using MUSE-like mock observations obtained from cosmological hydrodynamic simulations under the assumption of maximal quasar fluorescence,
we show that the velocity dispersion radial profiles of Ly$\alpha$-emitting gas are strongly determined by gravity and that they are thus self-similar with respect to halo mass when rescaled by the virial radius.
Through simple analytical arguments and by exploiting the kinematics of HeII1640$\mathrm{\AA}$ emission for a set of observed nebulae, we show that Ly$\alpha$ radiative transfer effects plausibly do not change the shape of the velocity dispersion profiles but only their normalisation without breaking their self-similarity.
Taking advantage of these results, we define the variable $\eta^{140-200}_{40-100}$ as the ratio of the median velocity dispersion in two specifically selected annuli and derive an analytical relation between $\eta^{140-200}_{40-100}$ and the halo mass which can be directly applied to observations. 
We apply our method to 37 observed quasar Ly$\alpha$ nebulae at $3<z<4.7$ and find that their associated quasars are typically hosted by  $\sim10^{12.16\pm0.14}\mathrm{M}_{\odot}$ haloes independent of redshift within the explored range. This measurement, which is completely independent of clustering methods, is consistent with the lowest mass estimates based on quasar auto-correlation clustering at z$\sim3$ and with quasar-galaxies cross-correlation results. 
\end{abstract}




\section{Introduction}
Within the standard paradigm of structure formation we expect that the dark matter and gaseous structures in our universe form due to gravitational collapse, where the matter contracts into sheets and filaments which constitute the Cosmic Web \citep{Bond1996}.
Quasi-spherical haloes are expected to form in the nodes of this web, mostly where filaments intersect. These haloes are continuously fed with primordial and recycled gas through the filaments of the Cosmic Web, with galaxies forming in the central regions of the haloes  \citep{Rees1977, Silk1977, White1991}.
Some fraction of this gas must reach and feed the galaxies in order to power star formation. However, the detailed physical processes which shape gas accretion onto galaxies and the physical properties of the Circumgalactic Medium (CGM) are still uncertain, especially at $z>2$, during the peak of galaxy formation. 
This is partly due to the diffuse nature of the gas in the CGM, which has hampered its direct study in emission and thus the possibility to probe its morphology and detailed physical properties.  

Several numerical studies from the last decades have suggested that gaseous filaments penetrating high-redshift haloes can remain relatively cold ($< 10^5K$) and dense compared to the surrounding halo gas which is shock-heated to the virial temperature \citep{Keres2005, Agertz2009, Dekel2009, Keres2009}. More recent studies have however argued that the lack of spatial resolution in these models could have affected these results \citep{Joung2012, Mandelker2016, Mandelker2018, Mandelker2019, Peeples2019, Hummels2019, Vossberg2019, Corlies2020, Gronke2022, Li2020, Fielding2020}, artificially prolonging the filaments' life and their ability to directly feed galaxies with cold gas.   

Until recently, with a few exceptions, the only way to probe the high-redshift CGM has been via absorption features in the spectra of background quasars and galaxies \citep{Hennawi2006, Rubin2010, Steidel2010, Rudie2012, Fumagalli2013, Turner2014, Bielby2017, Dutta2020, Dutta2021, Lofthouse2023}.
These studies have confirmed the need for a multi-phase CGM but have not been able, given
the sparseness of background sources, to directly probe the detailed morphology and physical properties (such as the density and ``clumpiness'') of the CGM. 
This limitation can partly be mitigated by using lensed background quasars which afford multiple sight lines through one halo's CGM \citep{Smette1992,Monier1998,Rauch2001,Ellision2004,Zahedy2016,Rubin2018} or gravitational arc tomography \citep{Lopez2018,Mortensen2021,Tejos2021,Bordoloi2022,Fernandez2022}.
The situation has changed in recent years due to new, highly sensitive instrumentation, such as the Multi Unit Spectroscopic Explorer (MUSE) on the ESO Very Large Telescope (VLT) \citep{Bacon2010} and the Keck Cosmic Web Imager (KCWI) \citep{Martin2010, Morrissey2018} which can reach sensitivity levels 
that are at least one order of magnitude deeper than previously possible. For instance, deep MUSE observations (20 hours of integration time or longer) have revealed the CGM in Ly$\alpha$ emission around individual galaxies \citep{Wisotzki2016}, the so called Ly$\alpha$ haloes, at surface brightness (SB) levels of about $10^{-19}$ erg s$^{-1}$ cm$^{-2}$ arcsec$^{-2}$. The faintness of this emission prevented previous detection with the exception of large statistical stacks of Narrow-Band imaging observations \citep{Steidel2011} and deep, long slit spectroscopic observations of multiple foreground and background quasar pairs \citep{Hennawi2013}.
Because such Ly$\alpha$ emission could be due to three different emission mechanisms (recombination radiation, collisional excitation and ``continuum-pumping'', see e.g., \citet{Cantalupo2017} for a review), and because of the resonant nature of Ly$\alpha$ radiation, directly translating these observational constraints into a measurement, e.g. of gas density, is extremely challenging. Indeed, it would require detailed, high-resolution radiative transfer models combined with high-resolution cosmological simulations which are not easily achievable with current computational facilities. 

Luckily, there is a phenomenon that both simplifies the interpretation of Ly$\alpha$ emission and increases its brightness by orders of magnitudes compared to the CGM of typical galaxies: quasar fluorescence \citep{Haiman2001, Bunker2003, Cantalupo2005}. The intense ionising radiation of a bright quasar is able to almost fully ionise its CGM, at least within the quasar ionisation cones, and Cosmic Web filaments on scales of several hundreds of kpc \citep{Cantalupo2014, umehata, Bacon2021}. The resulting recombination emission from the cold ($< 10^5K$) gas is easily detectable with Narrow-band imaging and shallow MUSE surveys ($<1$ hour of integration time) since it can reach SB levels up to $10^{-17}$ erg s$^{-1}$ cm$^{-2}$ arcsec$^{-2}$ at $z>2$ \citep{Cantalupo2012, Cantalupo2014, Farina2017, borisova, ArrigoniB2019, Cantalupo2019, Drake2019, umehata, Cai2019, Farina2019, Fossati2021}. In addition, other CGM emission lines also become easier to detect, including the non-resonant He-H$\alpha$ emission \citep{Cantalupo2019} and H-H$\alpha$ \citep[e.g.][]{Leibler2018, Langen2022}, and metal emission lines such as the CIV doublet (1548.2 $\si{\angstrom}$, 1550.8  $\si{\angstrom}$) \citep{Travascio2020, Guo2020, Fossati2021}. The non-resonant emission lines can then be used to constrain the kinematics, test the recombination-radiation nature of Ly$\alpha$ emission, constrain densities and the ``clumpiness'' of the medium, even below the spatial resolution scale \citep[e.g.][]{Cantalupo2019}.  
In the last few years observations of quasar fields with integral-field-spectroscopy have revealed the ubiquity of CGM Ly$\alpha$ emission around quasars at $z>2$, including quasars with absolute magnitudes within the range of $-27.2 < M_i < -23.7$, some of the faintest known SDSS quasars \citep[e.g.][]{Mackenzie2021}, at all explored redshifts up to the red-wavelength cut-off range of MUSE at $z\sim6$ \citep[e.g.][]{Farina2019}. The availability of quasars and the ease of detection has produced an impressively large statistical sample of more than a few hundred quasar Ly$\alpha$ nebulae in less than a decade, which can be used to directly probe the CGM's physical properties. The extended and diffuse morphology of these nebulae suggest the presence of a pervasive and diffuse cold component of the CGM, as the recombination process becomes inefficient at higher temperatures. 

The Ly$\alpha$ SB due to recombination radiation depends on the integral of the cold gas density squared along the line of sight. 
 As such, the emission is very sensitive to the ``clumpiness'' of the medium, or, similarly, to the ``broadness'' of the density probability distribution function along the line of sight and within the spatial resolution element \citep[e.g.][]{Cantalupo2019}. This information is encoded in the Ly$\alpha$ SB radial profile for example.
 However, it is difficult to compare the implied cold gas density profiles with expectations from current galaxy formation models without knowing the associated dark matter halo mass \citep[e.g.][]{PezzulliCantalupo2019} and references therein).
 This is because more massive haloes have a higher average density and a larger size along the line of sight at a fixed radial distance with respect to smaller haloes (see Section \ref{subsubsec:degSBCl} for more details). In addition, different halo masses could result in different temperatures of the CGM's hot ($> 10^5K$) component, halo baryon fraction and/or cold gas fraction \citep{Crain2007, Kulier2019}, all elements which could have an important effect on the expected Ly$\alpha$ SB profile. 
 The knowledge of the quasar host halo mass is thus fundamental to making precise inferences about other properties of the CGM, such as its density distribution, from the Ly$\alpha$ SB.
 Current methodologies of estimating quasar host halo masses rely on measuring quasar auto-correlation functions (or quasar-galaxy cross correlation functions) which in principle can provide precise estimates of halo masses. However, there are significant discrepancies between different works at similar redshifts that are not yet fully understood. In particular, quasar auto correlation studies suggest halo masses between $10^{12} \mathrm{M}_{\odot}$ and $10^{13} \mathrm{M}_{\odot}$ \citep{Shen2007, Eftekharzadeh2015, Timlin2018} between $z\sim3$ and $z\sim3.5$. In contrast, quasar galaxy cross-correlation studies consistently measure typical quasar halo mass values below $10^{12.5} \mathrm{M}_{\odot}$ at redshifts from $z\sim3$ to $z\sim4$ \citep{Trainor2012,FontRibera2013,garcia2017,He2017} and do not indicate a significant evolution with redshift. At $z > 3.5$, the constraints provided by these studies start diverging from quasar auto-correlation measurements. These discrepancies between quasar auto-correlation and quasar galaxy cross-correlation results at higher redshifts are not yet well understood but have significant implications on the inferred physical properties of the CGM, such as its density distribution.
 
 On the other hand, recent high resolution simulations focusing on the accreting gas in the CGM have demonstrated that the existence of turbulent gas with broad density distributions  is theoretically possible \citep{Hummels2019, Vossberg2019, Corlies2020, Augustin2021}. However, more work is needed to properly confirm the existence and formation mechanism of such gas.
Thus our understanding of the CGM physical properties using its emission will greatly benefit from an alternative and independent methodology to measure quasar host halo masses.

The goal of the RePhyNe project (``Resolving the physics of Quasar Ly$\alpha$ Nebulae''), presented here, is to develop and test, with the help of cosmological simulations, a new methodology which uses the kinematics derived from the CGM emission itself in order to provide alternative and complementary methods to clustering studies to constrain the quasar host halo masses (Paper I, this study). With this information in hand, we can thus provide new constraints on the density distribution of cold gas within the CGM of quasars in the second part of the RePhyNe paper series (hereafter Paper II).  

This paper is structured as follows: In Section \ref{sec:Methods} 
we detail the methods used to derive mock MUSE-like observations which are then used to obtain the line-of-sight velocity dispersion profiles of emitting gas. In Section \ref{sec:results}, we present our main results, including the analytical relation based on the self-similarity with respect to mass of the velocity dispersion profiles rescaled by the virial radius. In Section \ref{sec:Application}, we apply our method to obtain new constraints on the halo mass associated with observed quasars in the MAGG and MQN sample \citep{Lofthouse2020, borisova}.
In Section \ref{sec:Discussion}, we discuss the advantages and limitations of our mass estimation method and we compare our results to the literature.
Finally, in Section \ref{sec:SumConcl} we summarise our work. 
For the sake of consistency with the cosmological simulations used to derive our analytical relation (EAGLE and ENGINE, see Section \ref{subsec:CosmSims}) we assume the same flat $\Lambda$CDM cosmology and use the parameters from the 2013 Planck results \citep{Planck2014}.
In particular, we use $H_0 = 67.7$ km s$^{-1}$ Mpc$^{-1}$, $\Omega_b = 0.04852$, and $\Omega_m = 0.307$.
Furthermore, we define the virial radius $r_{vir}$ of a halo as $r_{200}$, the radius at which the average density of the spherical halo reaches 200 times the critical density of the universe at the given redshift.

\section{Developing a new mass estimation method} \label{sec:Methods}

In this section, we develop and test a new quasar halo mass estimation method which uses
the CGM emission kinematics. As such, it is independent of and complementary to previous quasar clustering studies at $z>2$. 
Because current observations only probe the cold ($10^4 < T < 10^5$ K) part of the CGM, the proposed halo mass estimation method will focus on the kinematics of this component and, in particular, on its associated Ly$\alpha$ radiation, which is the most commonly detected emission.
An analytical expression quantifying the degeneracy between halo mass and the CGM's physical properties, further motivating the need to fix the host halo's mass, is derived in Section \ref{subsubsec:degSBCl}.
In order to take the possibly complex morphology and kinematics of gas accretion within dark matter haloes into account, we calibrate our mass estimate method using mock observations of Ly$\alpha$ nebulae in hydrodynamic cosmological simulations of cosmic volumes. The procedure for generating the mock observations is detailed in Section \ref{subsec:GeneratingMockObs} and the subsequent kinematical analysis is presented in Section \ref{sec:results}. Finally, in Section \ref{subsec:velDispAnnuli} we introduce the new quasar halo mass estimation method.

\subsection{Degeneracy between halo mass and gas clumping factor}\label{subsubsec:degSBCl}
Because the observed SB is the integral of the emissivity over the line of sight, we expect that to the first order it will depend both on halo mass (which determines both the gas density at a given projected distance from the ionising source and the ``integration length") and the cold gas density distribution along the line of sight. The latter can be parameterised through the so called ``clumping factor" (see, e.g. \citet{Cantalupo2017} for a review):
\begin{equation}
    C_l\equiv <n^2>/<n>^2,
\end{equation}
where $n$ is the gas density and $l$ is the spatial scale (or volume) over which the integral is performed. $C_l$ is by definition equal to one if the density on scales $l$ is constant and greater than one otherwise. Because our main goal is to constrain the CGM gas density distribution from the observed CGM emission SB, it is important to understand its possible degeneracy with other variables. We can derive a simple expectation concerning this degeneracy through analytical considerations as developed in \citet{PezzulliCantalupo2019} (to which we refer the reader for more details).  
In particular, by rewriting Equation 12 in \citet{PezzulliCantalupo2019} to include the clumping factor\begin{footnote}{in order to include the clumping factor, we substitute $f_v$ in Equation 12 in \citet{PezzulliCantalupo2019} with $f_v/C_l$ following the explanation in Section 2.1.2 of that paper. We note that $C_l$ here refers to the ``internal clumping factor" of individual clumps as discussed in detail in Section \ref{subsec:corrBroadeningLya}.}\end{footnote}, one can obtain the following relation between clumping factor, halo mass, CGM cold gas fraction $f_{CGM, cold}$, and the volume filling factor of cold gas ($f_v$) for a fixed SB profile proportional to $R^{-\beta}$, with $\beta = 1.5$:
\begin{equation}\label{eq:fCold}
    C_l \propto f_v \: f_{CGM,cold}^{-2} \: M_h^{-5/6}.
\end{equation}
The CGM cold gas fraction used in this relation refers to the total mass of cold gas in the CGM normalised by the total baryonic mass associated to the halo
\begin{equation}
    f_{CGM,cold} = \frac{M_{cold}}{(\Omega_b/\Omega_m)M_h}
\end{equation}
and the exponent of the halo mass in Equation \ref{eq:fCold} can be derived as $(1 + \beta)/3$. 
This relation implies that for a fixed SB profile there is a degeneracy between clumping factor, halo mass, CGM cold gas fraction, and the volume filling factor of the cold gas. Thus information concerning these attributes is required to derive the CGM gas density distribution from observed SB profiles. 
However, the exact behaviour of both quantities is currently unknown in observations and deriving an expectation from simulations is also non-trivial as both quantities may be sensitive to the simulation's feedback recipes and numerical resolution. 
In Paper II we will study these two quantities in more detail using cosmological simulations of differing resolutions. 
Under the plausible assumption that these quantities vary slowly within the halo range relevant for this study, Equation \ref{eq:fCold} implies that 
current uncertainties in the quasar host halo masses as derived by clustering measurements (see Introduction for more details) result in relatively large uncertainties in our ability to constrain the physical properties of the CGM. This highlights the importance of finding complementary methods to constrain quasar host halo masses at $z>3$ as discussed in this work.

\subsection{Generating Mock Observations} \label{subsec:GeneratingMockObs}
\subsubsection{Cosmological simulations} \label{subsec:CosmSims}

Quasar clustering estimates suggest that quasar host halo masses are in the range of $10^{12} \mathrm{M}_{\odot} - 10^{13} \mathrm{M}_{\odot}$ at $z>3$ \citep{Shen2007, Eftekharzadeh2015, Timlin2018}. To follow the formation and evolution of these large haloes cosmological simulations with a volume of at least 50 comoving Mpc$^3$ (cMpc) at $z>3$ are needed. At the same time, these simulations require a high enough resolution to resolve the kinematic components of the CGM. For this reason, we use the EAGLE \citep{Schaye2015, Crain_2015, McAlpine_2016, team2017eagle} and ENGINE SPH simulation suites which contain haloes with a mass of up to  $10^{13.25}$ M\textsubscript{\(\odot\)}. 
Although the CGM is likely not fully resolved in the EAGLE simulations, \citet{Rahmati2015} have shown that observed global column density distribution function of HI and the observed radial covering fraction profiles of strong HI absorbers around bright quasars are well reproduced. This suggests that the simulations are at least able to capture the large scale distribution of the gas in the CGM. 
In particular, the ENGINE simulation uses the EAGLE baryonic physics implementation applied to a 50 cMpc$^3$ volume with the same number of particles as the EAGLE fiducial 100 cMpc$^3$ simulation, resulting in a higher mass resolution. 
The specific EAGLE simulation used is called RefL0100N1504 (Ref): a box with a side length of 100 cMpc containing $1504^{3}$ particles with the standard EAGLE stellar and AGN feedback implementation where both stellar and AGN-feedback are modelled with a stochastic injection of thermal energy \citep{Schaye2015}. The two ENGINE simulations used are RECALL0050N1504 (RECAL) and NoAGNL0050N1504 (NoAGN). Both are boxes with a side length of 50 cMpc containing $1504^3$ particles with the recalibrated EAGLE stellar feedback implementation. The difference between the two simulations is that RECAL also has AGN-feedback implemented, while in the NoAGN simulation the AGN-feedback is turned off. The reason for including the NoAGN simulations is that it allows us to quantify the effect of the EAGLE AGN-feedback implementation on the obtained mass estimates as discussed in Section \ref{subsec:effAGNfeedback}.
Although the RECAL and NoAGN simulations have the same initial conditions, their simulated haloes and gaseous structures are not identical. This coupled with the projection to two dimensions (see Section \ref{subsec:mockCubes}) ensures that we do not analyse two sets of identical Ly$\alpha$ nebulae.
A basic overview of the simulations properties is given in Table \ref{tab:simProp}.  
\begin{table}
    \centering
    \begin{tabular}{p{0.07\textwidth}||p{0.065\textwidth}|p{0.05\textwidth}|p{0.08\textwidth}|p{0.08\textwidth}}
        \toprule
        Prefix & L & N & $m_g$ & $m_{dm}$\\
         & [cMpc] &  & [$M_{\odot}$] & [$M_{\odot}$]\\
        \midrule
        Ref & 100 & $1504^3$ & $1.81 \times 10^6$ & $9.70 \times 10^6$\\
        RECAL & 50 & $1504^3$ & $2.26 \times 10^5$ & $1.21 \times 10^6$\\
        NoAGN & 50 & $1504^3$ & $2.26 \times 10^5$ & $1.21 \times 10^6$\\
        \bottomrule
    \end{tabular}
    \caption{Resolutions and box sizes of the EAGLE (Ref) and ENGINE (RECAL \& NoAGN) simulations used in this work. From left to right the columns show: simulation name prefix, the comoving box size, the number of dark matter particles and the initial equal number of baryonic particles, the initial baryonic particle mass and the dark matter particle mass. The mass resolution of the RECAL and NoAGN simulations is 8 times higher than that of the Ref simulation.}
    \label{tab:simProp}
\end{table}

In order to compare our results to current observations, we analyse two snapshots from each of the three simulations corresponding to redshifts $z = 3.528$ and $z = 3.017$. These two snapshots are chosen to be compatible with the redshift of previous observations of Ly$\alpha$ nebulae \citep{ArrigoniB2019, borisova, Marino2019, Fossati2021} and maximise the halo mass coverage, as the range of halo masses contained in the simulation box increases with decreasing redshift. 

As is common in current cosmological simulations, the multi-phase inter stellar medium (ISM) is not resolved, so the properties of the star-forming gas is defined by an effective equation of state: $P_{\mathrm{eos}} \propto \rho_g^{4/3}$, where $P_{\mathrm{eos}}$ is the gas pressure and $\rho_g$ is the gas volume density. This implies that the temperature of these gas particles is artificially set by the effective pressure imposed on the unresolved, multi-phase ISM \citep{Schaye2015} and not by the hydrodynamical interaction with the ambient gas. 
In the EAGLE and ENGINE simulations gas is defined to be star-forming, and thus placed on the effective equation of state, if its density lies above the following metallicity (Z) dependant threshold:

\begin{equation}\label{eq:SFthr}
\centering
    n_{*}(Z) = \bigg(\frac{0.002}{Z} \bigg)^{0.64} 10^{-1} \mathrm{cm}^{-3} \, .
\end{equation}

As this threshold separates the CGM from the ISM in the cosmological simulation, its actual value is of relevance for the predicted SB of the CGM emission and it will be further explored in Paper II. However, this threshold has a negligible effect on the quasar host halo mass estimate as is demonstrated in Appendix \ref{app:effectSFden}.

The simulations allow us to separately explore the kinematics of the different components of quasar haloes, including dark matter, cold ($T<10^5$K) and hot ($T>10^5$K) gas, of which solely the cold component is currently traceable by CGM emission observations at $z>2$. 
The kinematics of the dark matter is directly linked to the gravitational potential of the halo and thus to its mass. In later sections, we explore the relation between the cold and the dark matter kinematics in order to test if the former can be used as a proxy for the latter.  

\subsubsection{Modelling the CGM emission of quasars} \label{subsec:LyaEm}

As discussed in the introduction, one of the advantages of studying the Ly$\alpha$ CGM emission around quasars is their intense ionising radiation. This leads to the majority of the hydrogen in their CGM being highly ionised, simplifying Ly$\alpha$ modelling with respect to the CGM of star-forming galaxies for example. 
In the ``highly-ionised'' case the the contributions due to collisional excitation \citep{Haiman2000, Fardal2001, Dijkstra2006, Cantalupo2008, Rosdahl2012} can be neglected at temperatures above $T = 10^4$K based on the derivation in \citet{PezzulliCantalupo2019}. There is not enough neutral hydrogen for collisional excitation to make a significant contribution to the emission, even though the collisional excitation coefficient dominates the recombination coefficient at those temperatures  \citep{Cantalupo2008}.
We also neglect scattering (or ``photon-pumping'') of Ly$\alpha$ and continuum photons of galaxies and quasars \citep{Cantalupo2014}, which is difficult to model properly with current numerical models. Sophisticated radiative transfer modelling would be required and such modelling depends on the optical depth and precise kinematics of the emitting gas clumps on sub-kiloparsec scales which are currently not resolved by cosmological simulations \citep{Hummels2019, Corlies2020, Zahedy2021}. Observations of non-resonant lines such as He-H$\alpha$ and H-H$\alpha$ confirm that recombination is the main emission source at $z\sim2.3$ \citep{Leibler2018, Langen2022}. It is, however, not clear if the same holds at $z>3$, but upcoming JWST observations will help clarify the issue. 

By not modelling any radiative transfer effects we also do not model the broadening of the Ly$\alpha$ line caused by its resonant nature \citep{Cantalupo2005}. 
However, our mass estimation method is designed to be independent of the line-broadening as is explained in Sections \ref{subsec:corrBroadeningLya} \& \ref{subsec:velDispAnnuli}. Moreover, as discussed below, the kinematical analysis is predominantly independent of the actual value of the SB, as long as the SB value is high enough to be detectable. For the reasons stated above and for simplicity sake, we thus only include Ly$\alpha$ emission from recombination radiation and we leave further discussion concerning emission mechanisms to Paper II. 

We calculate the emissivity of the gas due to Ly$\alpha$ recombination radiation by assuming an ionising source, such as a bright quasar, resides in the centre of each halo and assuming \textit{maximal fluorescence}, i.e. the central ionising source is bright enough to ionise the entirety of the surrounding medium within an opening angle of 100\%. The Ly$\alpha$ emissivity $\epsilon_{\mathrm{Ly}\alpha}$ is calculated using the following relation 
\begin{equation}
    \epsilon_{\mathrm{Ly}\alpha} = \frac{1 - Y/2}{1-Y}\frac{h\nu_{\mathrm{Ly}\alpha}}{4\pi} n_{\mathrm{H}}^2\alpha_{\mathrm{eff}}(T) \, \, .
    \label{eq:LyaEm}
\end{equation}
Where $Y$ is the number fraction of primordial helium, $(1-Y/2)(1-Y)$ is a correction term due to the presence of primordial helium, $h$ is Planck's constant, $\nu_{\mathrm{Ly}\alpha}$ is the Ly$\alpha$ rest-frame frequency (1215.67$\AA$), $n_{\mathrm{H}}$ is the number density of hydrogen and $\alpha_{\mathrm{eff}}(T)$ is the case A effective recombination coefficient. We note that using the case B effective recombination coefficient would produce very similar results \citep[e.g.][]{PezzulliCantalupo2019}.  
The case A recombination coefficient is taken from \citet{HuiGnedin1997} Appendix A and the (weakly temperature dependent) fraction of recombination events that result in the emission of a Ly$\alpha$ photons is taken from \citet{Cantalupo2005}
\begin{equation}
    \alpha_{\mathrm{eff}}(T) = 0.35 \times \alpha(T) \, ,
\end{equation}
\begin{equation} \label{eq:alhpaRec}
    \alpha(T) = 1.269 \times 10^{-13} \mathrm{cm^3 s^{-1}} \frac{\tau^{1.503}}{[\, 1.0 + (\tau/0.522)^{0.470}]\,^{1.923}} \, ,
\end{equation}
\begin{equation}
    \tau = 2 \times \frac{157807 \, \mathrm{K}}{T} \, .
\end{equation} 

A central, ionising source, such as we are assuming, would also heat the gas through photo-electric heating. 
We refer to this temperature as the photo-heating temperature and its exact value is largely determined by the shape of the ionising spectrum \citep{Osterbrock2006}.
\citet{PezzulliCantalupo2019} have calculated this photo-heating floor for the ionised CGM as a function of density, metallicity and the ionising QSO spectrum for a distance of 50 kpc away from the ionising source. Their results can be found in the aforementioned work. We adopt a value of $T = 5 \times 10^4 K$ for the photo-heating floor which corresponds to a metallicity of $0.1 Z_{\odot}$ and cool phase number density of 1 cm$^{-3}$ assuming a ``standard'' QSO spectrum \citep{Lusso2015}. While an over- or underestimation of the photo-heating floor does effect the emissivity of the gas, it does not influence our ultimate halo mass estimation. We discuss in Paper II how varying the imposed photo-heating floor changes implications concerning the density of the CGM based on our halo mass estimates and calculated Ly$\alpha$ emissivity.

\subsubsection{SPH particle to grid conversion} \label{subsec:P2C}
The large majority of Ly$\alpha$ nebulae known to date have been discovered using integral-field-spectroscopy, e.g. with the MUSE or the KCWI instruments \citep{Wisotzki2016, borisova, Leclercq2017, ArrigoniB2019, Fossati2021}. Hence the observations are essentially 3D spatial-spatial-velocity grids. In order to compare the simulation to the data, the SPH-particle based simulations are also converted to grids. We use the code \texttt{P2C}\footnote{\url{https://gitlab.com/sdebeer/P2C}} (Particles to Chombo; originally developed by S. Cantalupo) to this aim. A brief description of the code is given below.

\texttt{P2C} converts particle fields, such as SPH outputs generated by the EAGLE, Sherwood \citep{Bolton2017} or AREPO \citep{Springel2010} codes, into adaptively-refined-meshes in the standard ``Chombo'' format \citep{Chombo}, which can be used as an input for state-of-the-art visualisation softwares such as VisIt \citep{VisIt}. In particular, after a regular base grid is defined and populated with the particle data as described below, the mesh can be further refined into a nested hierarchy of rectangular grids of different sizes and levels
of refinement, following the implementation called “patch-based AMR”, originally described in \citet{Berger1984}. The algorithm, which has been developed for the RADAMESH radiative-transfer code \citep{Cantalupo2011}, is described in detail in section 3.1 of \citet{Cantalupo2011}. Because our goal is to compare to the uniform three-dimensional grids of MUSE and KCWI, we do not use the multi-mesh capabilities of P2C here. 
Currently, the gas attributes that can be mapped to the grid are their density, x-, y- \& z-velocity, temperature, emissivity and emissivity weighted velocities. 
The grid cells emissivity values are calculated from the luminosity of the particles. A particle's luminosity is obtained by integrating $\epsilon_{\mathrm{Ly}\alpha}$, as defined in Equation \ref{eq:LyaEm}, over the particle's volume. The particle's luminosity is then distributed over the grid according to a given smoothing kernel.
We choose to use the same smoothing kernel as is used in the EAGLE simulations: the $C^2$ kernel from \citet{Wendland_1995} \citep{team2017eagle}, 
but note that the actual choice of the smoothing kernel has little effect on scales larger than the smoothing kernel itself, which is usually the case for the majority of the quantities. 
The user can also choose to exclude gas above a given uniform density threshold or metallicity dependant threshold described in Equation \ref{eq:SFthr}, as usually done in cosmological simulation to define star-forming regions. A minimum temperature floor, e.g. due to photo-heating, can also be imposed.  

\subsubsection{Mock integral-field-spectroscopy observations}\label{subsec:mockCubes}
\begin{figure}
	\centering
	\includegraphics[width=0.95\columnwidth]{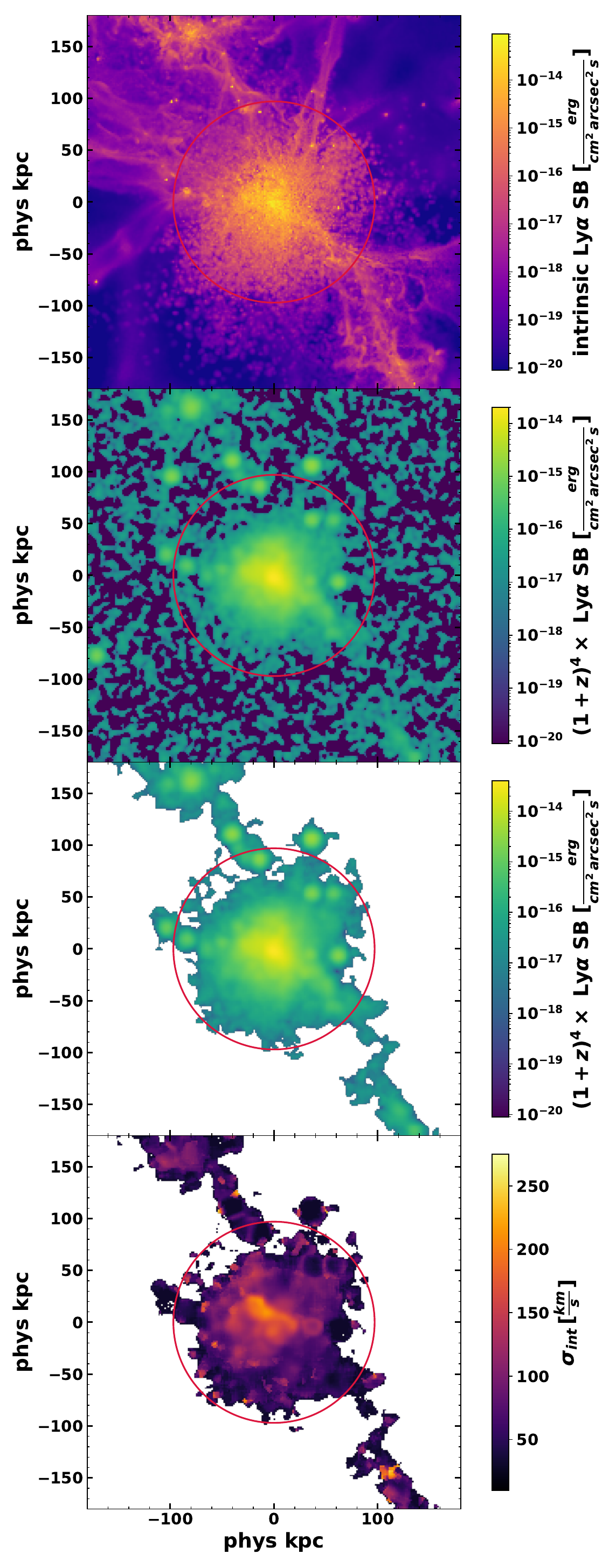}
    \caption{A visual representation of the conversion from simulation to velocity dispersion maps of a $1.9 \times 10^{12} M_{\odot}$ halo included in the analysis. From top to bottom: The intrinsic Ly$\alpha$ SB map, obtained by summing the emissivity grid output by \texttt{P2C} over a line of sight. The SB map of the mock cube generated from said grid. The SB map of the central Ly$\alpha$ nebula extracted using \texttt{CubEx}. Lastly, the intrinsic velocity dispersion map of the nebula. In each panel the halo's virial radius is marked with a red circle and the SB values in the top three panels are not corrected for cosmological dimming.}
    \label{fig:illMethods}
\end{figure}
Due to the complex, non-spherical morphology of the ``cold'' components in the CGM, the same structure observed from different directions can appear to have completely different morphologies. 
We take advantage of this effect to increase our sample size by generating three mock integral-field-spectroscopy observations (mock cubes) for each halo included in this analysis by using three perpendicular lines of sight.
The mock cubes are designed to be directly comparable to cubes obtained from the MUSE integral field spectrograph, meaning that the cubes have a spatial resolution of 0.2 arcsec ($\sim1.5$ physical kpc (pkpc) at both redshifts) and a spectral resolution of 1.25 \si{\angstrom}. Their side length is 3.084 cMpc and 3.478 cMpc at $z\sim3$ and $z\sim3.5$ respectively.

The first step in generating the mock cubes is to assign all the gas particles in the box centred on the halo centre to a 3D grid and calculate the emissivity in each cell using \texttt{P2C}.
The velocity of the cells is with respect to the bulk-velocity of the respective central halo and the velocity shift due to the Hubble-flow is accounted for. As mentioned in Section \ref{subsec:LyaEm}, we impose a photo-heating temperature floor of $5 \times 10^4 K$ on all the cells in the grid.
When building the mock cube for a given line of sight, the emissivity field and the emissivity weighted line-of-sight velocity field are used. Each cell in the emissivity field is assigned to a spectral layer based on the corresponding line-of-sight velocity cell. The value of the emissivity grid in that cell is then added to the cell in the mock cube with the same spatial coordinates projected along the line of sight and the corresponding spectral coordinate. 
The spectral coordinates are calculated by dividing the grid's line-of-sight velocity range into spectral layers of 1.25 \si{\angstrom} and assigning the cells to spectral layers based on their line-of-sight velocity with respect to the central halo. This corresponds to layers with a  width of 75 km s$^{-1}$ for redshift $z\sim3$ and 68 km s$^{-1}$ for redshift $z\sim3.5$.

Operating under the assumption of \textit{maximal fluorescence}, we treat the whole mock-cube simulation volume as ionised. In order to avoid over-ionising the gas at the largest distances we simply impose an upper fluorescent SB limit\begin{footnote}{This upper limit should be thought of as a safety net for extreme cases with negligible effect on the generated mock observations. As a reference point: The maximum number of voxels affected by this upper limit in one mock observation is roughly 30 out of 512$^2 \times$ 22.}\end{footnote}. The limit depends on the distance from the central ionising source as well as its ionising luminosity. It is given by
\begin{equation}\label{eq:SBlimit}
    \mathrm{SB}_{\mathrm{max}} = 2.25\mathrm{e}{-17} \, \Bigg(\frac{1 + 2.3}{1 + z}\Bigg)^4 \frac{1}{\mathrm{R}^2} \, \frac{\mathrm{erg}}{\mathrm{s} \,\, \mathrm{cm}^2 \, \mathrm{arcsec}^2 \, \si{\angstrom}} \, ,
\end{equation}
where R is the distance to the central ionising source in units of phys Mpc assuming an ionising luminosity comparable to the UM287 quasar \citep{Cantalupo2014, Cantalupo2019}. As the i-band magnitude of the quasar UM287 is comparable to that of bright quasars observed with MUSE this relation is also applicable to our mock observations.
To mimic typical seeing conditions we apply two dimensional Gaussian smoothing to each spectral layer individually, we additionally mimic the typical MUSE line spread function as reported in \citet{Bacon2017} by applying Gaussian smoothing along the spectral dimension.
We then add artificial noise to the mock cubes layer by layer which has a Gaussian distribution with a standard deviation of $\sigma = 5\times10^{-20}\,\, \mathrm{erg}/(\mathrm{s} \,\, \mathrm{cm}^2 \, \mathrm{arcsec}^2 \, \si{\angstrom})$. We note that this noise level typically corresponds to a time integration of more than 10 hours with MUSE, i.e. to a deep observation. However, as we demonstrate in Appendix \ref{app:noiseLevel}, the results presented here are not particularly affected by the chosen noise level as long as the nebula is detected with at least two wavelength layers per spaxel.

As the gas's emissivity depends on the square of its density and thus on the CGM clumping factor (see Equation \ref{eq:LyaEm}), there is a link between the signal-to-noise ratio (SNR) in an individual mock cube voxel and the density of the gas attributed to that voxel. 
In general, at a set distance from the halo centre, the average gas density and thus both the emissivity and SNR are lower for a halo with lower mass. 
The nebulae extracted from mock cubes around haloes with masses above $10^{12} \mathrm{M}_{\odot}$ are not affected by noise levels which are an order of magnitude higher than the one used here. However, haloes below  $10^{12} \mathrm{M}_{\odot}$ in EAGLE have average densities and clumping factors which would preclude their detection in shallow MUSE observations (a detailed comparison of mock and observed SB will be presented in Paper II). We stress that the SB normalisation, driven by the unknown gas clumping factor, is not important for the results presented here (see Appendix \ref{app:HeIIvelDisp}). Therefore, instead of increasing the SB normalisation, or the gas clumping factor, by an arbitrary value we have decided to keep the noise level low in order to increase the detectability of nebulae across a large mass range. 

\subsubsection{Detection and extraction of Ly\textalpha \hspace{1pt} nebulae} \label{subsec:detExLya}
\begin{table*}
    \centering
    \begin{tabular}{p{0.05\textwidth}||p{0.27\textwidth}|p{0.27\textwidth}|p{0.3\textwidth}}
        \toprule
        z & number of haloes & number of Ly$\alpha$ nebulae  & detection rate\\
        \midrule
        3.528 & 73 (NoAGN), 74 (RECAL), 639 (Ref) & 219 (NoAGN), 222 (RECAL), 1917 (Ref) & 100\% (NoAGN), 100\% (RECAL), 100\% (Ref) \\
        3.017 & 107 (NoAGN), 109 (RECAL), 941 (Ref) & 321 (NoAGN), 327 (RECAL), 2769 (Ref) & 100\% (NoAGN), 100\% (RECAL), 98\% (Ref)\\
        
        \bottomrule
    \end{tabular}
    \caption{The number of haloes and Ly$\alpha$ nebulae analysed. From left to right the columns show: the redshift of the simulation snapshots, the number of haloes within the halo mass range $10^{11.75} \mathrm{M}_{\odot}$ - $10^{13.25} \mathrm{M}_{\odot}$ in each simulation snapshot, the number of Ly$\alpha$ nebulae extracted from those haloes using three perpendicular lines of sight and the detection rate of Ly$\alpha$ nebulae in the mock observations.}
    \label{tab:extrNebInfo}
\end{table*}
In order to make our analysis as similar as possible to the actual observations, we detect and extract the Ly$\alpha$ nebulae from the mock cubes using \texttt{CubEx} from the \texttt{CubeExtractor} package \citep{Cantalupo2019} which has been widely used in the literature \citep{borisova, Marino2018, Marino2019, ArrigoniB2019, Langen2022}. In particular, we use the following parameters, which are very similar to the ones used in actual observations. For an object to be extracted we require that all voxels attributed to that object have a SNR value above 2.0 and that the object consists of at least 1000 voxels. If multiple objects extracted from the mock cube fulfil these criteria we choose the object with the largest number of voxels. We find that the object with the largest number of voxels always spatially coincides with the massive halo selected as the centre of the mock cube. If no object is extracted from the mock cube, that cube is discarded from our sample. Table \ref{tab:extrNebInfo} lists how many haloes are analysed at each redshift, how many nebulae are extracted and the Ly$\alpha$ nebulae detection rates. The number of nebulae is higher than the number of haloes as for each halo we generate three mock cubes using three perpendicular lines of sight.
As done in observations, we calculate the velocity dispersion as the second moment of the flux distribution using the segmentation mask generated by \texttt{CubEx} through the \texttt{Cube2Im} software, also part of the \texttt{CubeExtractor} package. We require that the detected emission occupies at least 2 spectral layers at a given spatial coordinate for the velocity dispersion to be calculated in that spaxel. We also apply a $3 \times 3$ spatial boxcar smoothing filter before generating the velocity dispersion maps. 
As we do not model any radiative transfer effects, the velocity dispersion calculated in this way directly traces the kinematics of the emitting gas. We therefore refer to it as the \textit{intrinsic} velocity dispersion, in order to differentiate this quantity from the observed velocity dispersion, which is subject to radiative transfer effects.

In Figure \ref{fig:illMethods}, we give an example of the process of converting the simulated haloes to velocity dispersion maps. 
Each panel depicts one stage of this conversion for a  $1.9 \times 10^{12} M_{\odot}$ halo included in this analysis.
In each panel the halo centre is located in the middle and its virial radius is marked with a red circle.
The top panel shows the Ly$\alpha$ emissivity as calculated by \texttt{P2C}, summed up over a given line of sight axis, resulting in an intrinsic Ly$\alpha$ SB map of the halo. The main halo and subhaloes are visible as well as the two major cosmic web filaments penetrating it. There are also numerous, more delicate filaments evident in emission. 
The second panel shows a surface brightness map generated from the mock cube containing the halo, using the same line of sight. The main halo, some subhaloes and the main filaments are still visible, however, the more tenuous filaments have now become undetectable under the noise. 
The third panel from the top contains the surface brightness map solely of the region attributed to the main Ly$\alpha$ nebula as extracted by \texttt{CubEx} using SNR = 2.0. Both major filaments are included in the detected emission. 
The bottom panel shows the intrinsic velocity dispersion map of the main nebula generated using \texttt{Cube2Im}. There is clearly a maximum in the region of the halo centre with the intrinsic velocity dispersion decreasing at larger distances. 

It is worth stressing that the mock Ly$\alpha$ nebulae obtained through our method exhibit morphologies which resemble comparable observed nebulae \citep{borisova, ArrigoniB2019, Fossati2021}, although they tend to be systematically dimmer. This may be due to the small scale clumpiness which is unresolved in the EAGLE/ENGINE simulations, as we will discuss in greater detail in Paper II.

\subsection{Kinematical analysis of the cold CGM} \label{sec:results}
In order to quantify the link between the kinematics of the cold ($<10^5K$) CGM and the mass of the host halo in the simulations, we compare the spherically averaged radial velocity profiles of the dark matter, the cold, Ly$\alpha$ emitting gas and the hot ($>10^5K$) gas. Building on the link revealed in the spherically averaged radial velocity profiles (Section \ref{subsec:radVelProf}), we investigate the evolution of the circularly averaged intrinsic velocity dispersion profiles with host halo mass, showing they are actually self-similar. We switch from the radial velocity to the intrinsic Ly$\alpha$ velocity dispersion, as the radial velocity is not an observable. 
With the link between the intrinsic velocity dispersion and the host halo mass established in Section \ref{subsec:velDispProf}, we discuss how we correct for the resonant broadening of the Ly$\alpha$ line in Section \ref{subsec:corrBroadeningLya}. Finally, in Section \ref{subsec:velDispAnnuli} we derive an analytical relation based on the self-similarity of the Ly$\alpha$ velocity dispersion profiles which can be used to constrain the host halo mass. 

\subsubsection{Radial velocity profiles}\label{subsec:radVelProf}
\begin{figure*}
	\centering
	\includegraphics[width=0.85\textwidth]{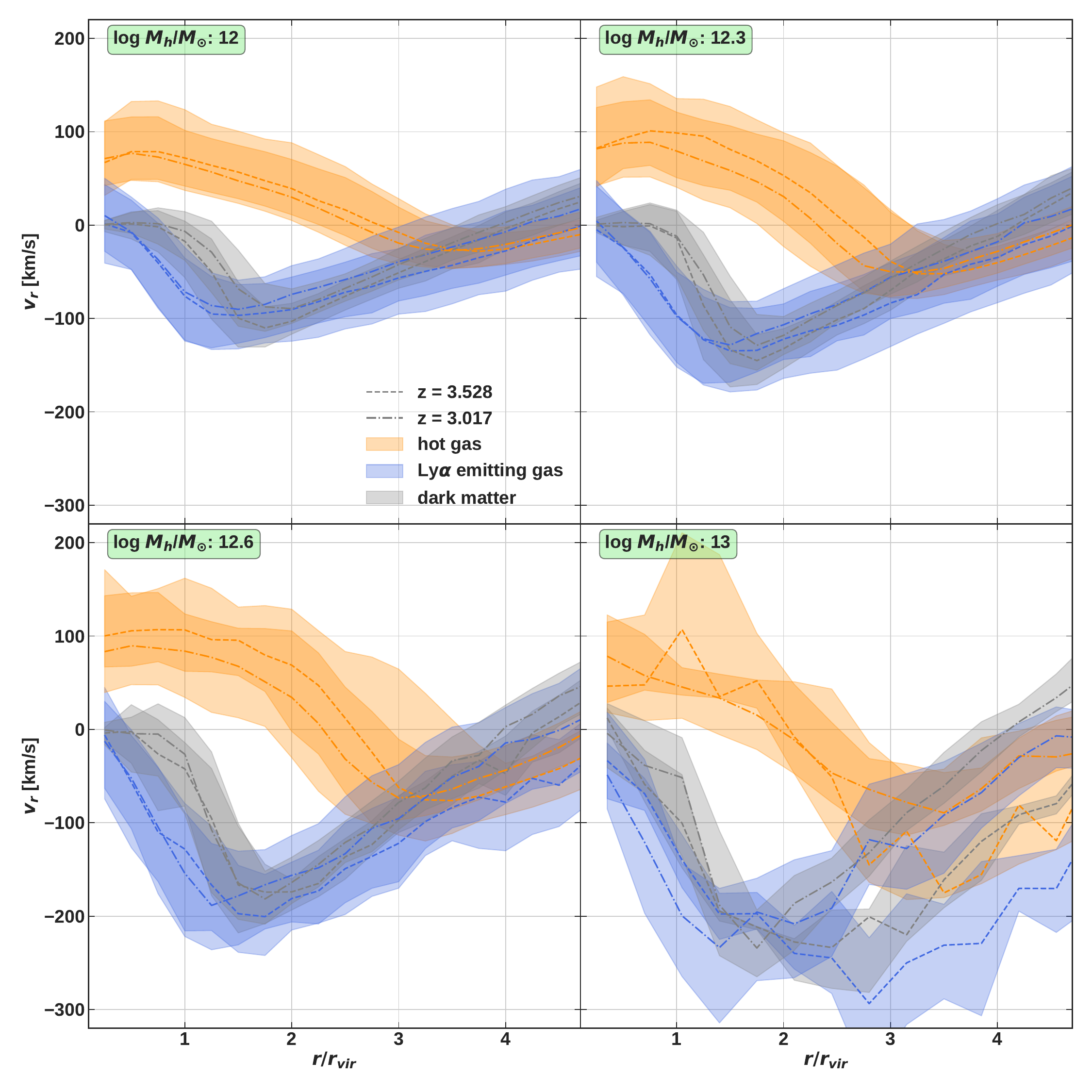}
    \caption{Spherically averaged radial velocity profiles of the dark matter (grey), the Ly$\alpha$ emitting gas (blue) and the hot ($> 10^5 K$) gas (orange) at redshifts $z = 3.528$ (dashed lines) and $z = 3.017$ (dash-dotted lines). Before stacking, distances from the center of the haloes for each individual profile are rescaled by the host halo's virial radius. The mass bins are centred on $10^{12} M_{\odot}$, $10^{12.3} M_{\odot}$, $10^{12.6} M_{\odot}$ and $10^{13} M_{\odot}$ as indicated in the labels within the panels. The three lower mass bins are 0.3 dex wide and the highest mass bin has an extent of 0.5 dex in order to increase statistics. The shaded areas indicate the range spanned by the 25th and 75th percentile for each radial profile. The radial velocity of the hot gas diverges from that of the Ly$\alpha$ emitting gas and the dark matter at $\sim 3 r_{vir}$. Remarkably, the radial velocities of the Ly$\alpha$ emitting gas and dark matter follow each other up until $\sim 1.5 r_{vir}$. We will use this result to derive a relation between the line-of-sight velocity dispersion of the Ly$\alpha$ emitting gas and halo mass as explained in Section \ref{subsec:velDispAnnuli}.
    }
    \label{fig:radVel}
\end{figure*}

We analyse the radial velocity of the gas and dark matter surrounding the haloes in the Ref and RECAL simulations at redshifts $z\sim3.5$ and $z\sim3$. To do this, we calculate the spherically averaged radial velocity profiles of the dark matter, the hot gas ($T > 10^5K$) and the Ly$\alpha$ emitting gas within five virial radii of each halo. For each simulation and redshift we divide the haloes into four mass bins and stack the radial profiles of each mass bin by calculating the median of all spherically averaged radial profiles in that mass bin. For the dark matter we calculate the radial velocity of each dark matter particle with respect to the halo's centre of gravity. 
For the Ly$\alpha$ emitting gas and the hot gas we calculate the radial velocity of the gas in each cell obtained with \texttt{P2C}. 
To calculate the profiles of the Ly$\alpha$ emitting gas we calculate the average radial velocity of the gas in all cells contained within a given spherical shell and use the Ly$\alpha$ emissivity of the cells as weights.
Figure \ref{fig:radVel} shows the radial velocity profiles at $z\sim3.5$ and $z\sim3$ for the four mass bins considered.

\begin{figure*}
	\centering
	\includegraphics[width=1.0\textwidth]{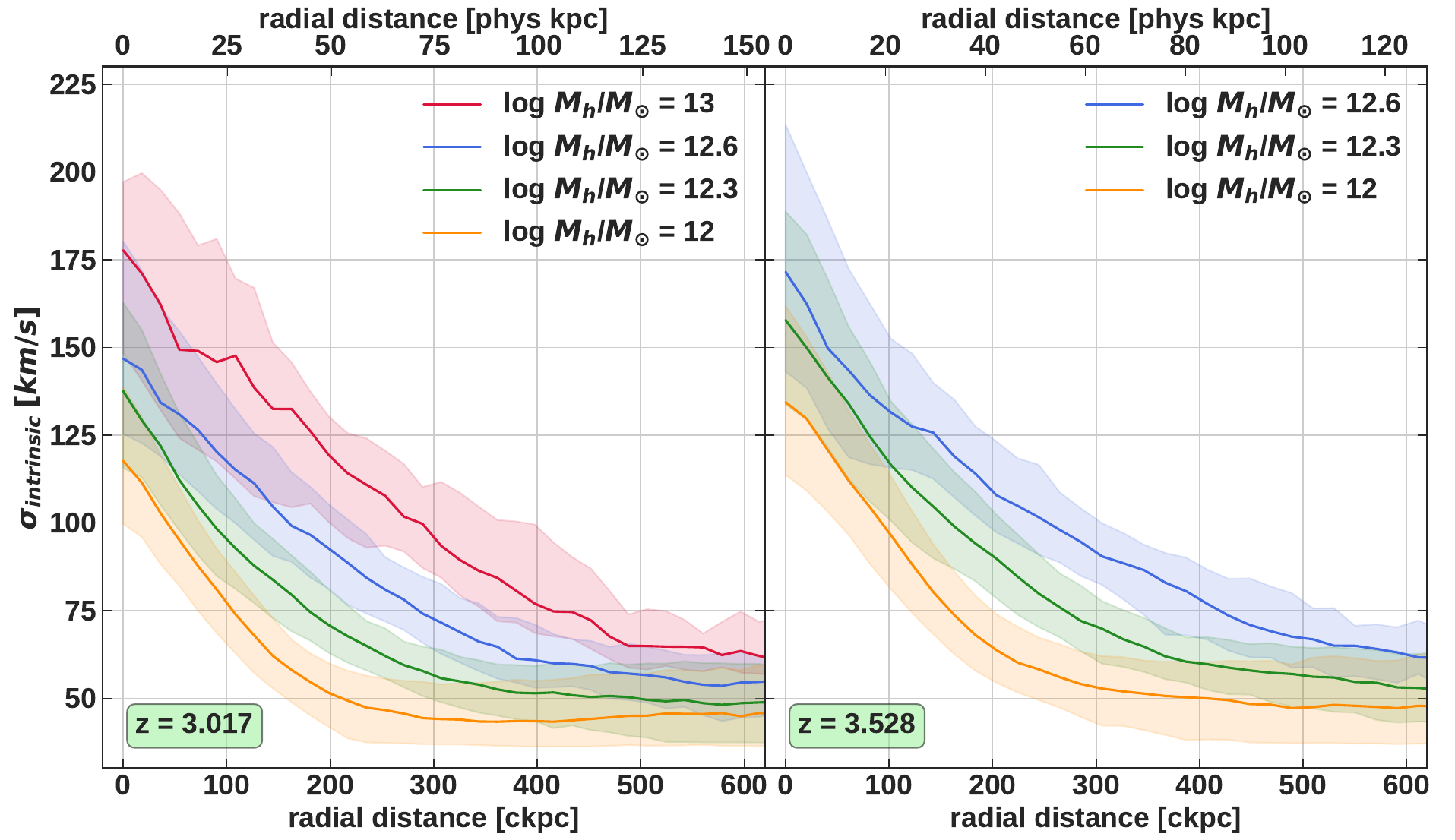}
    \caption{Circularly averaged intrinsic velocity dispersion profiles of the extracted Ly$\alpha$ nebulae hosted by haloes with masses of $10^{12 \pm 0.15} M_{\odot}$ (yellow), $10^{12.3 \pm 0.15} M_{\odot}$ (green),  $10^{12.6 \pm 0.15} M_{\odot}$ (blue) \&  $10^{13 \pm 0.25} M_{\odot}$ (red). The left panel shows the velocity dispersion profiles for Ly$\alpha$ nebulae at redshift $z = 3.017$ and the right panel shows the profiles for $z=3.528$. The region between the 25th and 75th percentile of the velocity dispersion for each mass bin is delimited by the shaded areas. The mass bin $10^{13 \pm 0.25} M_{\odot}$ is not included for $z = 3.528$ as there are only 2 haloes contained in that mass bin at that redshift. The maximum velocity dispersion value increases with halo mass analogously to the maximum radial inflow velocity. Additionally, the shape of the intrinsic velocity dispersion profiles becomes flatter/less concave with increasing halo mass.}
    \label{fig:velDispProf}

\end{figure*}
At a distance of roughly three virial radii and further away from the halo centre the radial velocity of the Ly$\alpha$ emitting gas is consistent with that of the hot gas and the dark matter for all halo masses, redshifts, and simulations.
The negative radial velocities of the gas and dark matter imply that all three are flowing towards the halo centre. The fact that the radial velocity values of both the hot and Ly$\alpha$ emitting gas are consistent with those of the dark matter suggests that, at these large distances, the gas is kinematically tracing the dark matter. In all four mass bins, at roughly three virial radii from the halo centre the average radial velocity of the hot gas begins to increase, diverging from that of the dark matter and Ly$\alpha$ emitting gas.
This divergence is likely caused by hot outflows which result in a positive (outflowing) radial velocity for the hot gas out to roughly two virial radii.
The outflows are driven by stellar \& AGN-feedback that simultaneously heat and expel the gas from the central galaxy and the halo in the EAGLE \& ENGINE simulations. 
However, not all the gas is heated and expelled as indicated by the fact the Ly$\alpha$ emitting gas, which effectively corresponds to the cold gas ($T<10^5$ K), continues to fall towards the halo centre, kinematically tracing the dark matter.

Both the dark matter and Ly$\alpha$ emitting gas are accelerated towards the halo centre until their radial inflow velocity abruptly starts to decrease. This decrease occurs between one and two virial radii with the decrease in the dark matter's radial inflow velocity occurring up to 0.5 virial radii before that of the Ly$\alpha$ emitting gas. The apparent radial deceleration of the dark matter is likely due to the co-existence of particles with positive and negative radial velocities associated with virialisation. The radial deceleration of the Ly$\alpha$ emitting gas could be explained by weak virial shocks that, despite decreasing the radial inflow velocity, do not heat it to virial temperature or other hydrodynamical interactions. We note that the dark matter's point of maximum infall velocity coincides with the halo's splashback radius \citep{Fillmore1984, Bertschinger1985, Diemer2014, More2015}. For both the dark matter and Ly$\alpha$ emitting gas the maximum radial infall velocity increases with increasing halo mass.

These results clearly indicate that the kinematics of the Ly$\alpha$ emitting gas are dominated by gravity up to 1.5-3 virial radii from the halo centre, at least within the assumptions made in the EAGLE and ENGINE models. 
Moreover, the fact that the radial velocity of the Ly$\alpha$ emitting gas is almost exclusively negative implies that Ly$\alpha$ emission traces the gas accreting into the halo.

As mentioned in Section \ref{subsec:CosmSims}, the incorporation of both the RECAL and NoAGN ENGINE simulations allows us to quantify the effect of the EAGLE AGN-feedback implementation on the kinematics of the gas. 
We find that the AGN-feedback has no effect on the radial velocity of the Ly$\alpha$ emitting gas and that stellar feedback is the main driver of the hot outflows as can be seen in Appendix \ref{app:AGNfeedbackEffect} in Figures \ref{fig:radVel_AGNvsNoAGNz3p528} \& \ref{fig:radVel_AGNvsNoAGNz3p017}. 

Our findings with regard to the radial velocity profiles are broadly consistent with similar analyses performed on different sets of cosmological simulations. For instance, the distance at which the radial velocity of the Ly$\alpha$ emitting gas begins increasing is consistent with the distance of $0.75 - 1.25 r_{vir}$ that \citet{Nelson2016} find.
The bi-modal behaviour of the hot and cold (Ly$\alpha$ emitting) gas has also been observed by \citet{Huscher2021} for an EAGLE zoom simulation of galaxy haloes at redshift $z\sim 2-3$ with masses of $\approx 10^{12}\mathrm{M}_{\odot}$. 

The profiles shown in Figure \ref{fig:radVel} clearly reveal a correlation between the maximum infall velocity and halo mass.
Similarly, there is evidence for a correlation between the point of deceleration of the accreting gas and the halo's virial radius. These results suggest that the maximum infall velocity and point of deceleration of the Ly$\alpha$ emitting gas could be used to determine the halo's mass, if they could be observed. 

\subsubsection{Velocity dispersion profiles} \label{subsec:velDispProf}
\begin{figure*}
	\centering
	\includegraphics[width=1.0\textwidth]{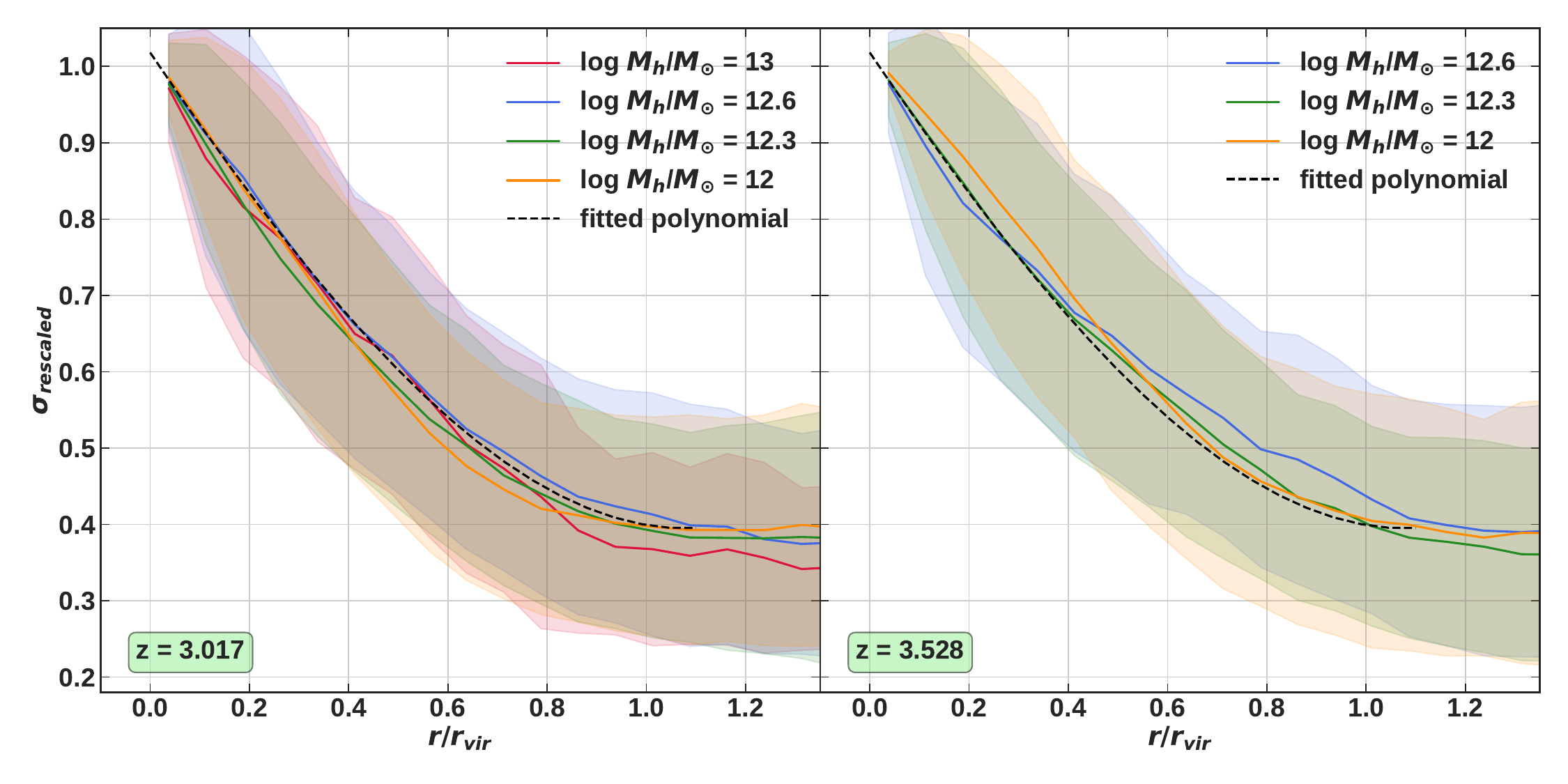}
    \caption{Circularly averaged profiles of the rescaled intrinsic velocity dispersion as a function of virial radius of the halo hosting the extracted Ly$\alpha$ nebulae. The average profiles of the mass bin $10^{12 \pm 0.15} M_{\odot}$ (yellow), $10^{12.3 \pm 0.15} M_{\odot}$ (green), $10^{12.6 \pm 0.15} M_{\odot}$ (blue) \& $10^{13 \pm 0.35} M_{\odot}$ (red) are shown with the shaded regions indicating the respective standard deviations. The left panel shows the profiles for redshift $z = 3.017$ and the right shows them for $z = 3.528$. The joint rescaling of the distance to the halo centre and the intrinsic velocity dispersion leads to all profiles of the different mass bins lying on top of each other, demonstrating the dependence on halo mass of both the intrinsic velocity dispersion values and the profiles shapes. The third degree polynomial fitted to all individual profiles is plotted in black in both panels. Its excellent agreement with each average profile further confirms the self-similarity of the rescaled velocity dispersion profiles.}
    \label{fig:rescVelDispProf}
\end{figure*}
As the radial velocity of the Ly$\alpha$ emitting gas is not a direct observable, we investigate whether there is any correlation present between the observable line-of-sight velocity dispersion and halo mass. Analogously to the radial velocity profiles, we calculate the circularly averaged velocity dispersion profile for each halo and each of its lines of sight in all three simulations at both redshifts. Then we divide the haloes into the same mass bins and stack the circularly averaged velocity dispersion profiles for each mass bin by calculating the median velocity dispersion profile.
The stacked velocity dispersion profiles are shown in Figure \ref{fig:velDispProf}. As explained in Section \ref{subsec:detExLya}, we remind the reader that the plots show the intrinsic velocity dispersion without taking radiative transfer effects into account.

Previous observations of Ly$\alpha$ nebulae have found average velocity dispersion values of $\sigma \approx 250$ km/s and higher \citep{ArrigoniBattaia2015, borisova, Cantalupo2019, Marino2019, Drake2022}. Our stacked profiles do not reach such high values and are more compatible with velocity dispersion values $\sigma \approx 100 \mathrm{ km/s}$ of HeII-1640$\mathrm{\AA}$ nebulae that are co-spatial with Ly$\alpha$ nebulae around AGNs \citep[][Travascio et al. in Prep.]{Marino2019}. 
The lower velocity dispersion values measured in the co-spatial HeII nebulae are due to the absence of resonant broadening effects. This implies that the intrinsic kinematics of the emitting gas in the CGM is traced by the emission from the HeII-1640$\mathrm{\AA}$ transition in observations, just as the intrinsic kinematics of the emitting gas in the simulations is traced by the mock Ly$\alpha$ emission.
The agreement between our simulated intrinsic velocity dispersion values and the observed ones indicates that although the CGM is likely unresolved in the simulations, its large scale kinematics are well reproduced.

Similarly to the maximum infall velocity, the maximum intrinsic velocity dispersion increases with halo mass at both redshifts for all three simulations. The shape of the profiles also becomes flatter/less concave with increasing halo mass. Both of these findings can be explained with the behaviour of the radial velocity profiles discussed in Section \ref{subsec:radVelProf}. In particular, due to the projection effects, higher infall velocities translate to a larger spread of velocities along the line of sight. 
The flattening is connected to the fact that the point where the radial inflow velocity of the gas starts to decrease is at the same distance in virial radii units from the halo centre. Due to $r_{vir}\propto M_{halo}^{(1/3)}$, this point is farther away from the halo centre for higher mass haloes. In particular, this point of deceleration corresponds to the region where the velocity dispersion profile transitions from constant to monotonically increasing.
Thus, the farther away this point is from the halo centre, the flatter the profile. 

In order to quantify in which way the flattening depends on the halo mass, we plot the intrinsic velocity dispersion as a function of r/r$_{vir}$ and rescale the profiles by their value in the the inner most radial bin in Figure \ref{fig:rescVelDispProf}. 
At both redshifts the rescaling leads to the profiles coming to lie on top of each other, independently of redshift and halo mass. This self-similarity with respect to halo mass of the rescaled velocity dispersion profiles is consistent with expectations from dark matter simulations, observational results pointing to the self-similarity of the cold CGM with respect to the virial radius \citep{Churchill2013, Churchill2013_2} and with the fact that the Ly$\alpha$ kinematics closely trace the dark matter kinematics in the EAGLE simulations (see Figure \ref{fig:radVel}). This result gives us the opportunity to constrain the virial radius, and thus the halo mass, from the shape of the Ly$\alpha$ velocity dispersion profile, which is an observable quantity.
In order to facilitate this task, we obtain an analytical relation between the rescaled velocity dispersion and the halo's virial radius by fitting a third degree polynomial to the rescaled velocity dispersion profiles of each individual nebula. 
The coefficients of the fitted polynomial
\begin{equation} \label{eq:3rdPoly}
    \sigma_{rescaled}(x) = ax^3 + bx^2 + cx + d 
\end{equation}
are $a = 0.168 \pm 0.016$, $b=0.174 \pm 0.03$, $c=-0.96 \pm 0.016$ and $d=1.018 \pm 0.001$, where $x$ is the projected distance to the centre of the nebula rescaled by the halo's virial radius: $r/r_{vir}$.
While performing the fit, only values at radial distances less than 1.25 $r/r_{vir}$ are considered in order to avoid including the plateau region of the profiles which could be affected by signal-to-noise and spectral resolution limitations. 
The analytical relation is plotted with a black, dashed line in both panels in Figure \ref{fig:rescVelDispProf}. The mean rescaled velocity dispersion profiles for each mass bin are also plotted in red, blue, green and orange with the standard deviation for each mass bin indicated by the shaded regions. The fact that the fitted analytical function is in good agreement with the mean profile of each halo mass bin indicates that this relation holds equally well for the whole mass range considered in this work. We stress that fitting the polynomial to the  median rescaled profiles would result in a very similar shape and thus would not significantly affect our results as presented in Section \ref{subsec:velDispAnnuli}.

In principle, applying this relation to observations could directly constrain the virial radius and thus halo mass. 
A direct comparison between the self-similar rescaled velocity dispersion profiles and observations is however impractical for the majority of observed nebulae given their limited signal-to-noise ratios. Moreover, the central bin used for the rescaling is typically dominated by the bright quasar Point-Spread-Function (PSF). For these reasons we introduce a new parametrisation in Section \ref{subsec:velDispAnnuli} which is based on the analytical relation presented in Figure \ref{fig:rescaledAnnuli} and mitigates these observational limitations. However, before proceeding further, the resonant broadening of the Ly$\alpha$ velocity profile must be taken into account, as discussed in the next section.

\subsubsection{Ly$\alpha$ spectral broadening} \label{subsec:corrBroadeningLya}
In the previous sections we generate Ly$\alpha$ velocity dispersion profiles under the assumption of maximal fluorescence and ignoring possible radiative transfer effects (``intrinsic'' velocity dispersion profiles). 
In this section, we first verify that the gas in the CGM is not highly self-shielded and then derive the effect of radiative transfer on the shape and normalisation of the observable velocity dispersion. We then compare our analytical expectations with observations.

The large Ly$\alpha$ absorption cross section ($\sigma_0\simeq5.9\times10^{-14}$ cm$^{2}$ at line center for T=10$^4$K) implies that even a highly ionised medium such as the CGM of quasars could have a high opacity to Ly$\alpha$ photons generated by recombinations.
Let us assume, following \citet{PezzulliCantalupo2019}, that the CGM is composed of a hot component (at about the virial temperature) and a cold component at a temperature fixed by quasar photoionisation in the form of clouds (of arbitrary shape) with a typical size $l$ and average density $n$. The average neutral density ($<n_{\mathrm{HI}}>$) of these clouds at a distance $r$ from the quasar will be given by
\begin{eqnarray}
<n_{\mathrm{HI}}> \simeq \frac{<n_{\mathrm{H}}>^2\ \alpha(T)\ C_l}{\Gamma_{i}} = \nonumber \\ 
4.8\times10^{-6}  C_l \ T_4^{-0.75} \left[\frac{<n_{\mathrm{H}}>}{\mathrm{cm^{-3}}}\right]^2 \left[\frac{r}{30\ \mathrm{kpc}}\right]^{2} \,\mathrm{cm^{-3}} , 
\end{eqnarray}
under the plausible assumption that $\Gamma_i\gg n\cdot \alpha(T)$, 
where $\Gamma_i\simeq10^{-7}(r/30\mathrm{kpc})^{-2} \mathrm{s^{-1}}$ is the photoionisation rate of bright quasars in MUSE surveys such as \citet{borisova} (see \citet{PezzulliCantalupo2019} for details), $T_4\equiv T/(10^4\mathrm{K})$, $\alpha(T)=4.8\times10^{-13} T_4^{-0.75} \mathrm{cm}^3 \mathrm{s}^{-1}$ is the hydrogen case A recombination coefficient\begin{footnote}
    {The case A recombination coefficient $\alpha(T)$ quoted in this section is taken from \citet{PezzulliCantalupo2019} for the sake of consistency with their derivation. However, although this approximation differs slightly from the $\alpha(T)$ defined in Section \ref{subsec:LyaEm}, both calculations of the Ly$\alpha$ emissivity $\epsilon_{\mathrm{Ly}\alpha}$ in Section \ref{subsec:LyaEm} and of the average neutral fraction $x_{\mathrm{HI}}$ this Section are insensitive to the exact approximation of $\alpha(T)$ used.}
\end{footnote}, and $C_l$ is the clumping factor over scales $l$ previously introduced in Section \ref{subsubsec:degSBCl}. 

The neutral hydrogen column density ($N_{\mathrm{HI}}$) of such clouds is thus
\begin{eqnarray}\label{NHI_1}
    N_\mathrm{HI}=\ <n_{\mathrm{HI}}>  \simeq \nonumber  \\
    1.5\times10^{15} C_l \ T_4^{-0.75} \left[\frac{n}{\mathrm{cm}^{-3}}\right]^2 \left[\frac{r}{30\ \mathrm{kpc}}\right]^{2}\left[\frac{l}{100\mathrm{pc}}\right]\ \mathrm{cm^{-2}}\ .
\end{eqnarray}
Under the assumption that the observed Ly$\alpha$ SB is produced by recombination radiation, \citet{PezzulliCantalupo2019} have derived the following radial average density profiles for clouds at a distance $r$ from quasars in the \citet{borisova} sample at $z\sim3.2$ (which we assume in the following as a reference redshift)
\begin{equation}\label{nprof}
    n(r)=1.2\ C_l^{-1/2} T_4^{0.48} \left[\frac{f_v}{10^{-3}}\right]^{-0.5}  \left[\frac{r}{30\ \mathrm{kpc}}\right]^{-1.25} \mathrm{cm^{-3}} ,
\end{equation}
where $f_v$ is the volume filling factor occupied by cold clouds with average a density $n$ on scales of $l$.

The cloud size is connected to the volume filling factor through this relation
\begin{equation}\label{size_eq}
    l=r_{\mathrm{vir}}\cdot f_v \cdot f_c^{-1}\simeq
    74 \left[\frac{f_v}{10^{-3}}\right] f_c^{-1} M_{12}^{1/3}\ \mathrm{pc},
\end{equation}
where $r_{\mathrm{vir}}$ is the halo's virial radius, $f_c$ is the covering factor as seen by the quasar, or, equivalently, the average number of clouds with size $l$ between the quasar and the virial radius (assumed here to be the far ``edge'' of the CGM for simplicity), and $M_{12}$ is the total halo mass in units of $10^{12}$ M$_{\odot}$.

Inserting Equations \ref{nprof} and \ref{size_eq} into Equation \ref{NHI_1}, we finally obtain
\begin{equation}\label{NHI_fin}
    N_\mathrm{HI}\simeq 1.6\times10^{15} \left[\frac{r}{30\ \mathrm{kpc}}\right]^{-0.5} T_4^{0.21} f_c^{-1} M_{12}^{1/3}\ \mathrm{cm^{-2}},
\end{equation}
or, equivalently, in terms of line centre optical depth to the Ly$\alpha$ radiation ($\tau_0$)
\begin{equation}\label{tau_0}
    \tau_0\simeq 94.4 \left[\frac{r}{30\ \mathrm{kpc}}\right]^{-0.5} T_4^{0.21} f_c^{-1} M_{12}^{1/3},
\end{equation}
where we have used the relation $\sigma_0\simeq5.9\times10^{-14} T_4^{-1/2}$ for the line center cross section and assumed the \emph{internal} velocity dispersion of the clouds, or equivalently, the gas velocity dispersion on scales of $l$, to be dominated by the thermal broadening due to quasar photoheating. 

Equation \ref{NHI_1} implies that both $N_\mathrm{HI}$ and $\tau_0$ depend linearly on the size of the clouds $l$, which is currently unknown. 
As demonstrated in Equations \ref{size_eq} - \ref{tau_0} the unknown cloud size $l$ can be recast in terms of the covering factor $f_c$ along the quasar line-of-sight. Doing this, one can characterise the local $N_\mathrm{HI}$ and $\tau_0$ as inversely proportional to a global value of $f_c$ and mildly proportional to the other parameters (distance from the quasar, cloud temperature and halo mass). 
In principle, the covering factor can be measured by looking at quasar absorption spectra.
In practice, however, it could be challenging to identify and count absorption lines in a very narrow velocity window, which from the results presented above should be of the order of a few hundred km/s without a very precise measurement of the quasar systemic redshift. 

However, we can provide some useful limits on the value of $f_c$ from the following considerations. 
The smoothness of observed SB maps indicates that the individual clouds must have sizes smaller than the spatial resolution element of the MUSE observations, which is typically 5 kpc. Additionally, observations in both absorption and emission have placed upper limits on the individual cloud sizes of $l\lesssim 20-500$ pc \citep{Crighton2015,ArrigoniBattaia2015}.
Using eq. \ref{size_eq} and assuming $f_v=10^{-3}$, this would imply $f_c>0.15$ and thus $N_\mathrm{HI}<10^{16}$ cm$^{-2}$. Such column densities and their associated optical depths are below the HI self-shielding limit to the ionizing radiation, implying that these clouds produce Ly$\alpha$ photons from recombination efficiently, as initially assumed. We discuss how our results would be effected by the presence of neutral gas, due to a smaller ionisation cone for instance, in Section \ref{subsec:ProsCons}. 
On the other hand, as long as $f_c<100$ (and thus $\tau_0>1$), the associated absorption lines to these clouds in the spectra of observed quasars should be easily measurable. 

Let us consider, as an example, the observed $dN/dz$ of observed absorption line systems in the presence of the quasar ``proximity effect'' and overdensity of matter in their associated haloes. In particular, we can define the expected number of absorption line systems at a distance $r$ from the quasar as
\begin{equation}
    \frac{dN}{dz}=N_0 (1+z)^{\gamma} [1+\frac{\Gamma_i(r)}{\Gamma_{\mathrm{UVB}}}]^{-(\beta-1)}\, \delta^{\, (\beta-1)} \times \Delta(r),
\end{equation}
where $N_0(=6.1)$ and $\gamma(=2.47)$ are observationally derived parameters which describe the number density of absorption line systems away from quasars in a given column density range ($13.64\lessapprox \mathrm{log} N_{\mathrm{HI}} \lessapprox 17$, corresponding to the $f_c$ limits discussed above). 
The parameter $\beta(=1.6)$ is also observationally constrained and describes the column density distribution function which is consistent with a power law with index $-\beta$  \cite[see e.g.][and references therein]{2009Meiksin} and $\Gamma_{\mathrm{UVB}}$ is the photoionisation rate in absence of radiation from the quasar (i.e., only due to the cosmic UV background). 
$\Delta(r)$ is a factor which accounts for the increase in number density of systems of a given column density (in absence of quasar radiation) at a distance $r$ from the quasar. The factor $\delta^{\, (\beta-1)}$ quantifies the fact that the internal density of these clouds could increase due to compression from the hot component of the CGM, which in turn would lead to an increase of the recombination rate, thus counteracting the aforementioned ``proximity effect''.
Substituting the numerical values, assuming $\Gamma_{\mathrm{UVB}}=10^{-12}$ s$^{-1}$, $z=3.5$, $\delta = 1$ and using $r=r_{\mathrm{vir}}$, we obtain 
\begin{equation} \label{eq:NumClouds}
    f_c\simeq 0.5\Delta z\times \Delta(r_{\mathrm{vir}})\simeq 0.75 \frac{\Delta \mathrm{v}(\Delta z)}{100 \mathrm{km}\ \mathrm{s^{-1}}}\frac{\Delta({r_{\mathrm{vir}}})}{1000}. 
\end{equation}
As the value of $\delta$ is unknown, we set it to the lower limit of 1, but note that $\delta$ could contribute to a higher covering factor. 
We also stress that the actual value of $\Delta(r_{\mathrm{vir}})$ is unknown and likely larger than the canonical value of 200 given by gravity alone since the physics that determines the properties of clouds in the CGM of a massive halo could be different to the physics for the generic cloud population in the IGM.
However, it is interesting to consider that unless $\Delta(r_{\mathrm{vir}})$ is extremely large, we obtain $f_c$ values which are of the order of unity which would imply $\tau_0>>1$, in addition to $\tau_0<10^4$, as derived above. 

Once produced within the clouds, the Ly$\alpha$ photons thus cannot directly escape from the interior regions of the cloud and will be absorbed and re-emitted by atoms \emph{within} the cloud (thus experiencing negligible spatial diffusion compared to CGM scales) until their frequency is sufficiently far away from the line center. Unfortunately, there are no analytical solutions to predict the emerging spectral shape of the Ly$\alpha$ photons at such values of $\tau_0$. However, we do expect that the emerging spectrum would have a significant depletion of Ly$\alpha$ photons at the line center with respect to the ``intrinsic'' spectrum. As a reference, a pure absorbing screen with $N_{\mathrm{HI}}\simeq10^{16}$ cm$^{-2}$ (which is on the flat or ``logarithmic'' part of the Ly$\alpha$ equivalent width curve of growth) would produce an absorption line with a FWHM of about four times the value of the Doppler parameter \citep[e.g.][]{2009Meiksin}. 

The emerging spectrum from an individual cloud would then be significantly broader than the ``intrinsic one''. However, the amount of broadening for each individual cloud would be rather insensitive to the actual value of $\tau_0$ as long as it is within the range $10\lessapprox \tau_0 \lessapprox 10^4$, which is the case for the clouds discussed above. In particular, given the very weak dependency on distance from the quasar of Equation \ref{tau_0}, $\tau_0$ variations would be of the order of a few. This would result in very similar broadening independent on cloud distance from the quasar. 

The resulting integrated spectrum would however depend on the number of clouds encountered by the Ly$\alpha$ photons before escaping the CGM and thus on $f_c$. If $f_c<1$, broadening would only happen within the individual clouds. The emerging spectrum would therefore be a convolution of the clouds' velocity dispersion due to their bulk motion and the individual (constant) broadening within individual clouds. The expected effect is thus a flattening of the Ly$\alpha$ observed velocity dispersion profile with respect to the ``intrinsic'' one. As discussed below, this would imply that the halo masses derived from Ly$\alpha$ velocity dispersion profiles should be considered as upper limits. On the other hand, if $f_c>1$, the broadening of the emerging spectrum would also depend on the bulk velocity dispersion between different clouds providing a possible mechanism to produce a constant relative broadening. A detailed calculation of the magnitude of the spectral broadening would require detailed radiative transfer simulations. However, radiative transfer effects do not change our results as long as the \emph{shape} of the velocity dispersion profile is not significantly affected by resonant broadening. 

In order to empirically understand whether solely the normalisation and not the shape of the velocity dispersion profile is changed, observations of a non-resonant line, such HeII-1640$\mathrm{\AA}$ (i.e., HeII H$\alpha$) are needed. 
This line is particularly useful for at least three reasons: i) it is a primordial element like hydrogen, thus we expect that both are distributed in the CGM in a similar fashion, ii) its transition wavelength places it in an observable range from the ground with MUSE, iii) for fully doubly ionised Helium, its flux is expected to be relatively bright, i.e. about one third of the Ly$\alpha$ flux for recombination radiation. 
Unfortunately, as discussed in \citet{Cantalupo2019}, HeII emission is typically much fainter than expected for fully, doubly ionised Helium in quasar nebulae suggesting the presence of very dense gas in the CGM. We will return to this point in Paper II. From an observational point of view, this makes HeII nebulae more challenging to detect and indeed they are rarely found in the literature. 

To overcome this limitation we have re-examined several MUSE medium-deep ($\sim4$h) observations of quasars and optimised the analysis to specifically search for HeII emission. Our optimised methodology, applied to 26 of the 28 quasar nebulae in the MAGG sample \citep{Fossati2021}, where two nebulae are excluded due to being gravitationally lensed and associated with multiple quasars respectively,
allowed us to discover 14 individually detected HeII nebulae\begin{footnote}{We note that the fraction of detected nebulae is mostly determined in the MAGG sample by the data noise properties and in particular by the presence of bright sky lines at the expected HeII wavelengths combined with the intrinsic faintness of the line.}\end{footnote}. 
The data analysis and results will be presented in detail in Travascio et al. (in Prep.). These HeII nebulae have SB values which are typically about 10 - 20 times fainter than the Ly$\alpha$ SB at the same spatial location, consistent with previous results at lower redshifts \citep[e.g.][]{Cantalupo2019}. The faintness of the emission only allows us to probe CGM kinematics, when HeII is detected, up to about 200 ckpc from the quasars.  
The dispersion typically reaches values between 100 and 150 km/s and in three cases values above 200 km/s. 
Interestingly, these values are very consistent with the intrinsic velocity dispersion values shown in Figure \ref{fig:velDispProf}. 
In order to reduce the noise associated with the fainter HeII emission, we compare the median velocity dispersion profiles of both HeII and Ly$\alpha$ emission (using the same subset of sources) instead of the individual profiles. The ratio of these median velocity dispersion profiles is found to be consistent with a constant value of 5.66$\pm$0.68 (where the errors indicate the 25th and 75th percentiles) at every radius at which HeII is detected, i.e. up to 200 ckpc.

In light of the discussion above, this result suggests, in the case of $f_c<1$, that the Ly$\alpha$ velocity dispersion profile is not significantly flattened with respect to the intrinsic one, or, in the case of $f_c>1$, that the radiative transfer effects due to multiple clouds along the line of sight do not change the \emph{shape} of the Ly$\alpha$ velocity dispersion profiles. In both cases, this would imply that the Ly$\alpha$ velocity dispersion profiles can be used as direct tracers of the halo kinematics.
In the remainder of this work, we will assume for simplicity that this result holds at all radii and for all nebulae. 

Finally, we stress that, even in the case of $f_c<1$, the velocity dispersion measured in emission as discussed here, would still be a good representation of the overall kinematics of the system, instead of the velocity dispersion within individual clouds. This is the case as long as multiple clouds at different radial distances are present within the spatial resolution element. 
Indeed, assuming $f_v\simeq10^{-3}$ would imply $l\simeq370$ pc for a $M_{12}=1$ halo\begin{footnote}{In reference to the work of \citet{PezzulliCantalupo2019}, we note that this value for $f_v$ would be compatible with a cosmological baryon fraction within haloes, quasar photo-heating and Ly$\alpha$ emission produced by recombination radiation, as long as $C_l>10$. A value of $C_l$ different than 1 would imply that some regions of the emitting clouds should be out of pressure-equilibrium with the ambient medium. We will return to this point extensively in Paper II.}\end{footnote}. The spatial resolution element used in the observations and in the production of mock cubes is of the order of 5 kpc (determined by the seeing). Assuming for the sake of simplicity that the typical line of sight length through the halo is about the same size of the virial radius, we obtain that at least  $\sim18$ clouds should be contributing to each individual spatial resolution element if $f_c>0.1$. Although this number is sensitive to the actual (unknown) value of $f_v$, our order of magnitude estimate is useful in conveying that even a covering below unity plausibly results in a significant number of clouds along the line of sight within the spatial resolution element. 

\subsection{Constraining quasar halo masses} \label{subsec:velDispAnnuli}
In the previous sections, we have seen that the shape of the normalised Ly$\alpha$ ``intrinsic'' velocity dispersion profiles is self-similar if represented in units of $r/r_{vir}$ (see Figure \ref{fig:rescVelDispProf}) and can be described by an analytical function (see Equation \ref{eq:3rdPoly}). Moreover, through analytical considerations and by comparing to HeII emission observations (for a sub-sample of nebulae), we have shown that this result should also apply to the \emph{observed} Ly$\alpha$ velocity profiles, which differ from the ``intrinsic'' ones by a renormalisation factor which is independent of radius.  

Taking advantage of these results, we present an analytical relation based on the self similarity of the rescaled velocity dispersion profiles, which can be used to constrain the quasar halo masses using the observed Ly$\alpha$ velocity profile \emph{shape} as a function of comoving radial distance from the quasar. To this end, we introduce a new variable representing
the ratio of the median velocity dispersion in two concentric annuli
\begin{equation}
    \eta^{140-200}_{40-100} \equiv \frac{\sigma_{140-200}}{\sigma_{40-100}} \ ,
\end{equation}
where $\sigma_{40-100}$ is the \emph{median} velocity dispersion of spaxels within the annulus at 40 to 100 ckpc from the quasar, and $\sigma_{140-200}$ is the \emph{median} velocity dispersion within the annulus 140 to 200 ckpc.  
These annuli have been carefully chosen in order to avoid the central regions (affected by quasar PSF subtraction in observations) and to maximise the velocity dispersion variations across the relevant spatial scales as shown in Figure \ref{fig:velDispProf}.  
Note that a ``flatter'' velocity dispersion profile as a function of comoving distance (corresponding to larger halo masses) would have a higher $\eta^{140-200}_{40-100}$ than a steeper profile (corresponding to lower halo masses). At the same time, $\eta^{140-200}_{40-100}$ is independent of any radiative transfer broadening effects, as long as these effects are mostly independent of radius (as argued in the previous sections). 

\begin{figure}
	\centering
	\includegraphics[width=\columnwidth]{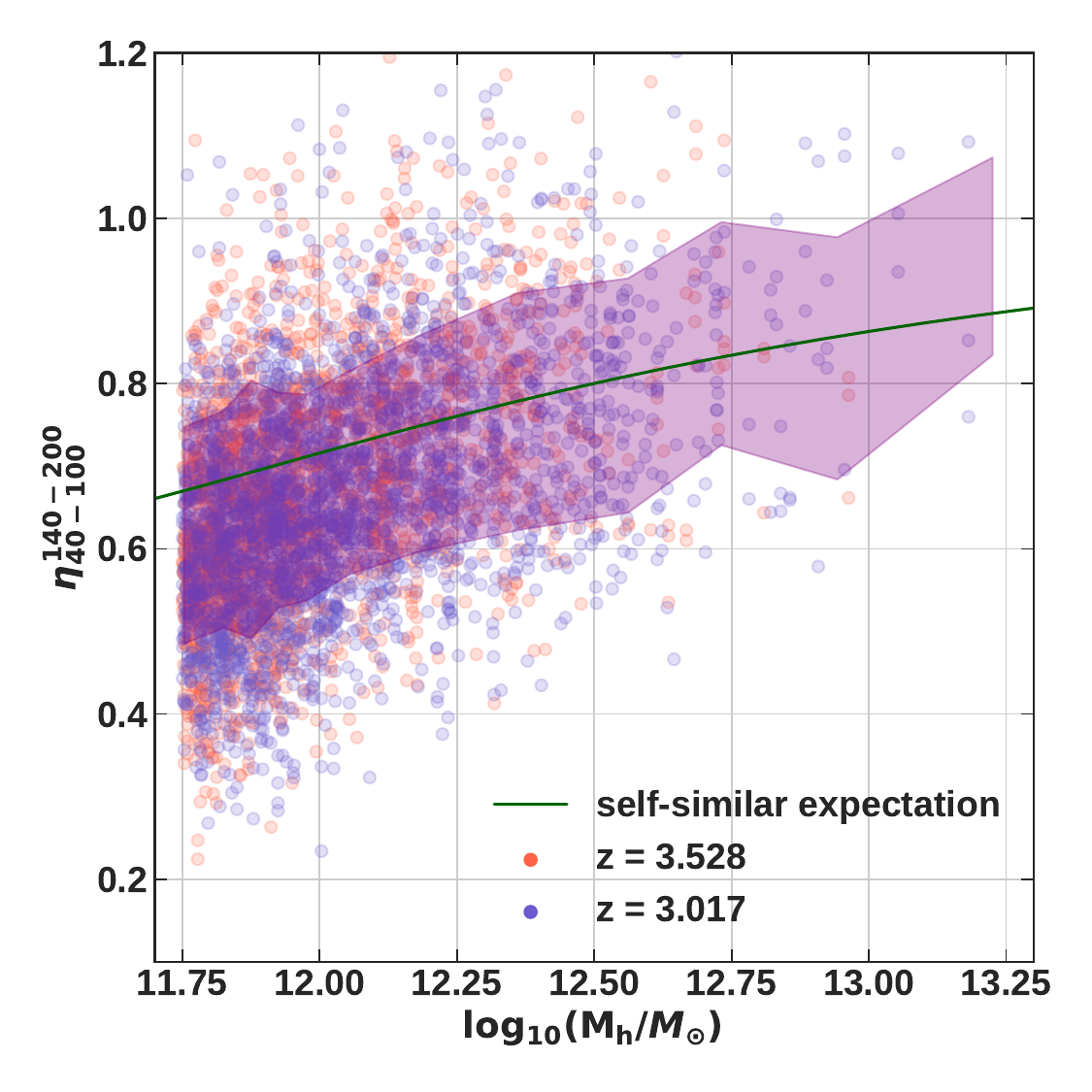}
    \caption{The ratio of the median velocity dispersion values in the outer (140 - 200 ckpc) and inner (40 - 100 ckpc) annuli from all three simulations and both redshifts as a function of halo mass. The blue dots refer to  $\eta^{140-200}_{40-100}$ at $z \sim3$, the red dots to $z\sim3.5$ and the purple shaded region indicates the standard deviation of the individual $\eta^{140-200}_{40-100}$ values as a function of halo mass. The analytical relation based on the self-similarity of the rescaled velocity dispersion profiles given in Equation \ref{eq:etaAnalytical} is plotted in green.
    The values of $\eta^{140-200}_{40-100}$ do not vary with resolution, redshift or feedback implementation, which is why we combine all six simulation snapshots. 
    Despite the significant scatter there is a clear correlation with halo mass as expected from our analytical relation.}
    \label{fig:rescaledAnnuli}
\end{figure}

As the function for $\sigma_{rescaled}$ given in Equation \ref{eq:3rdPoly} is monotonic, the median $\sigma_{rescaled}$ within an annulus is simply the value of $\sigma_{rescaled}$ at the radius where half the surface of the annulus is reached. 
Hence, $\eta^{140-200}_{40-100}$ can be written as
\begin{equation}
    \eta^{140-200}_{40-100} = \frac{\sigma(r_{out}/r_{vir})}{\sigma(r_{in}/r_{vir})}.
\end{equation}
We have dropped the subscript $rescaled$ from $\sigma$ for the sake of readability, $r_{out} \sim 172$ ckpc and $r_{in} \sim 76$ ckpc refer to the radii at which the median $\sigma$ is reached in the two annuli and all radii are in comoving units. With the relation
\begin{equation} \label{eq:rvir}
    r_{vir} = \Bigg( \frac{M_h \: G}{100 \: H_0^2 \: \Omega_m} \Bigg)^{1/3},
\end{equation}
which is redshift independent for comoving units within our redshift range of interest\footnote{In comoving units the mass of a halo can be written as $M_h = 100\: r_{vir}^3 \; H_0^2 \:G^{-1} (\Omega_m + \Omega_{\Lambda}(1 +z)^{-3})$. As $\Omega_m$ dominates $\Omega_{\Lambda}(1 +z)^{-3}$ at $z>2$ one can neglect the second term in the brackets, resulting in the relation quoted in Equation \ref{eq:rvir}.}, the velocity dispersion ratio $\eta^{140-200}_{40-100}$ can be rewritten as
\begin{equation}\label{eq:etaAnalytical}
    \eta^{140-200}_{40-100} =  \frac{a \: P \: r_{out}^3 \: \mu^{3} + b \: P^{2/3} \: r_{out}^2 \: \mu^{2} + c \: P^{1/3} \: r_{out} \: \mu + d}{a \: P \: r_{in}^3 \: \mu^{3} + b \: P^{2/3} \: r_{in}^2 \: \mu^{2} + c \: P^{1/3} \: r_{in} \: \mu + d}.
\end{equation}
Where $\mu \equiv M_h^{-1/3}$, $P = (100 \: H_0^2 \: \Omega_m)/G = \num{3.197e4}$ M$_{\odot}$ ckpc$^{-3}$, $G$ is the gravitational constant, the coefficients $a$, $b$, $c$ and $d$ are those of the fitted polynomial in Equation \ref{eq:3rdPoly} and $M_h$ is in units of M$_{\odot}$.

In Figure \ref{fig:rescaledAnnuli} we plot the analytically derived $\eta^{140-200}_{40-100}$ given in Equation \ref{eq:etaAnalytical} as a function of halo mass (through its relation with the virial radius) in green. We plot the $\eta^{140-200}_{40-100}$ of the individual mock Ly$\alpha$ nebulae to confirm that their $\eta^{140-200}_{40-100}$ scales with halo mass as predicted by Equation \ref{eq:etaAnalytical}. Additionally, we indicate the region within the standard deviation of the simulated nebulae's $\eta^{140-200}_{40-100}$ as a function of halo mass with the purple shaded region.
We emphasise that the analytical relation is based on the self-similarity of the shape of the velocity dispersion profiles which are independent of mass and redshift if represented in units of the virial radius. The large scatter of the individual $\eta^{140-200}_{40-100}$ values in Figure \ref{fig:rescaledAnnuli} is consistent with the scatter in the individual velocity dispersion profiles presented in Figure \ref{fig:rescVelDispProf} and reflects the systematic effects related to morphological and physical properties of the haloes as discussed in detail in Section \ref{subsec:ProsCons}. As such, we remind the reader that the analytical $\eta^{140-200}_{40-100}$ versus halo mass function presented here should be seen as a relation valid for a population of haloes rather than for single quasars.
The blue dots refer to redshift $z\sim3$ and the red dots to redshift $z\sim3.5$. 
As expected, the values of $\eta^{140-200}_{40-100}$ appear to be largely independent of redshift within the range explored here.
Moreover, we have also verified that they are independent of simulation resolution and AGN-feedback implementation.

By varying the Ly$\alpha$ emissivity of each particle (by a constant factor, for simplicity), we have also verified that the analytical relation presented above is independent of the surface brightness normalisation as long as emission is detectable. We refer to Figure \ref{fig:HeIIvelDisp} in Appendix \ref{app:HeIIvelDisp} for more details. Moreover, we have also found that $\eta^{140-200}_{40-100}$ is largely independent of the spectral resolution of the mock cubes as long as it is high enough to resolve the typical width of the ``intrinsic'' velocity dispersion value (which is typically the case for MUSE within the considered spatial annuli). 
Because the velocity dispersion should decreases with increasing distance from the halo centre, the values of $\eta^{140-200}_{40-100}$ are expected to be lower than 1, as is the case for the analytical $\eta^{140-200}_{40-100}$ plotted in Figure \ref{fig:rescaledAnnuli}.
However, as is evident from Figure \ref{fig:rescaledAnnuli} there are some individual instances of $\eta^{140-200}_{40-100}$ that have values larger than 1, implying a larger velocity dispersion at larger distances from the quasar. This can be attributed, e.g., 
to the superposition along the line of sight of multiple halos separated by hundreds of kpc or a few Mpc which are, however, spectrally blended with the quasar Ly$\alpha$ nebula. These projection effects are expected to happen, in a statistical sense, both in our mock observations and in real data (see \citet{Cantalupo2019} for a discussion of one of these possible cases associated with the Slug Nebula). 

We propose using the function given in Equation \ref{eq:etaAnalytical} to constrain the mass of haloes hosting observed Ly$\alpha$ nebulae powered by bright quasars based on their measured $\eta^{140-200}_{40-100}$. 
Theoretically, the $\eta^{140-200}_{40-100}$ value of a sample with a given mass distribution can be expressed as
\begin{equation}
    \eta^{140-200}_{40-100} =  \frac{\int_{0}^{\infty}\eta(M_h)\,P(M_h)\:n(M_h) \: dM_h}{\int_{0}^{\infty}P(M_h) \: n(M_h)\: dM_h} ,
\end{equation}
where $P(M_h)$ is the mass dependent probability of a halo hosting a bright quasar and is $n(M_h)$ is the halo mass function. 
Following the standard procedure used in quasar clustering studies, e.g. \citet{Eftekharzadeh2015}, we use two possible functional shapes for $P(M_h)$: i) a delta function (which thus defines a ``characteristic" halo mass) and, ii) a step function (which defines a ``minimum" halo mass) as described in detail below. We stress, however, that any shape of $P(M_h)$ can in principle be used in combination with our method. 
Thus, we define the characteristic halo mass $M_h$ of a sample of Ly$\alpha$ nebulae as the halo mass for which the analytical $\eta^{140-200}_{40-100}$ in Equation \ref{eq:etaAnalytical} corresponds to the measured $\eta^{140-200}_{40-100}$ of the sample. 
Additionally, a minimum halo mass $M_{min}$ can be derived by assuming that $P(M_h)$ is a step function for which $P(M_h)=1$ for $M_h>M_{min}$ and zero otherwise. In the latter case, we interpret the measured $\eta^{140-200}_{40-100}$ as the mean  $\eta^{140-200}_{40-100}$ of all haloes with masses above $M_{min}$, weighted by the halo mass function \citep[see Equation 9,][]{Eftekharzadeh2015}.

In the following sections, we apply our method to a subset of observed MUSE Ly$\alpha$ nebulae around bright quasars at $3<z<4$ and provide constraints on the characteristic and minimum halo masses of the samples based on the CGM kinematics.

\section{Application of Mass estimation method} \label{sec:Application}
As a first application of the method presented above, we use the analytical relation given in Equation \ref{eq:etaAnalytical} to constrain the characteristic mass of haloes hosting bright quasars and surrounding Ly$\alpha$ nebulae presented in the first MUSE GTO survey around bright quasars (MUSE Quasar Nebulae snapshot survey, or MQN) \citep{borisova} and those included in the MAGG sample \citep{Lofthouse2020, Fossati2021}. 
These nebulae are extended enough and have sufficiently high SNR to be excellent candidates for our mass estimation method.

\subsection{The observed Ly$\alpha$ nebula samples} \label{subsec:ObSamples}
The Ly$\alpha$ nebulae sample presented in \citet{borisova} is comprised of two sub-samples, observed during the two different MUSE GTO programs: 094.A-0396, 095.A-0708, 096.A-0345 PI: S. Lilly \& 094.A- 0131, 095.A-0200, 096.A-0222 PI: J. Schaye. We solely consider the first sub-sample (094.A-0396, 095.A-0708, 096.A-0345), which consists of 12 of radio-quiet quasars within the redshift range  $z \approx 3.0 - 3.3$. For the sake of brevity we refer to this sub-sample as the MQN $z\sim3.1$ sample. 
We note that a handful of the nebulae in the higher redshift sub-sample (094.A- 0131, 095.A-0200, 096.A-0222) are included in the MAGG sample.  
Relevant for this analysis is that the observed quasars are some of the brightest known radio quiet quasars within the redshift range considered. 

We calculate the velocity dispersion ratio  $\eta^{140 - 200}_{40-100}$ for 10 of the 12 Ly$\alpha$ nebulae using the velocity dispersion maps obtained as discussed in \citet{borisova} and presented in Figure 7 of that work\begin{footnote}{It is worth noting that although Figure 7 in \citet{borisova} shows the FWHM, they calculate the FWHM by using the relation $2.35\times\sigma$. Therefore these values could be directly used as our method requires sigma modulo any constant multiplicative factor.}\end{footnote}. 
When calculating the velocity dispersion ratio  $\eta^{140 - 200}_{40-100}$ for the extracted nebulae, we require that at least 20\% of spaxels in the outer annulus have a velocity dispersion measurement. Due to this requirement, one nebula (MQN17) is excluded from our analysis. We further exclude another nebula (MQN07) as its SB peak does not spatially coincide with the quasar, possibly suggesting that the quasar could be hosted by a companion or satellite galaxy not located at the halo center. This is different than our mock observations for which, by construction, the quasar is placed at the center of mass of the host halo. As a consequence, all distances relevant for our empirical relations are calculated with respect to the center of the halo, making our mass estimation method likely unsuitable for observed nebulae with clear displacements between quasar position and the SB spatial peak. This issue could be solved by changing the definition of halo centre in observed nebulae to the SB spatial peak. For simplicity however, in this first analysis, we have excluded the peculiar nebula MQN07.

The MAGG sample is introduced and described in detail in \citet{Lofthouse2020} \& \citet{Fossati2021}, here we briefly summarise the aspects relevant to this work. The sample consists of 28 quasars for which archival high-resolution (R$\gtrsim$ 30 000) spectroscopy is available,
with m$_r$ < 19 AB mag covering a redshift range of $z \approx 3.2 - 4.5$. The selection criteria for the quasars require that these are observable from the VLT with low airmass, and have at least one intervening strong hydrogen absorption line system at $z > 3.05$ with $N_{\mathrm{HI}}>10^{17} \mathrm{cm}^{-2}$.
The original extraction and detection of Ly$\alpha$ nebulae in \citet{Fossati2021} is performed with \texttt{CubEx} imposing a SNR threshold of 2.0 and a minimum number of connected voxels of 1000.
We note that one of the Ly$\alpha$ nebulae is excluded from the analysis in \citet{Fossati2021}, due to it being strongly lensed and thus exhibiting an irregular morphology. The sample can be split into a high and low redshift sub-samples with median redshifts of $z \sim 4.1$ and $z \sim 3.5$ respectively.  We apply our mass estimation method to each redshift sub-sample separately and to the  whole sample combined with the MQN sample in order to investigate any potential redshift evolution of the typical mass of halo hosting a quasar. 

\subsection{Characteristic quasar halo masses as a function of redshift} \label{subsec:haloMassEstObs}
\begin{figure}
	\includegraphics[width=0.98\columnwidth]{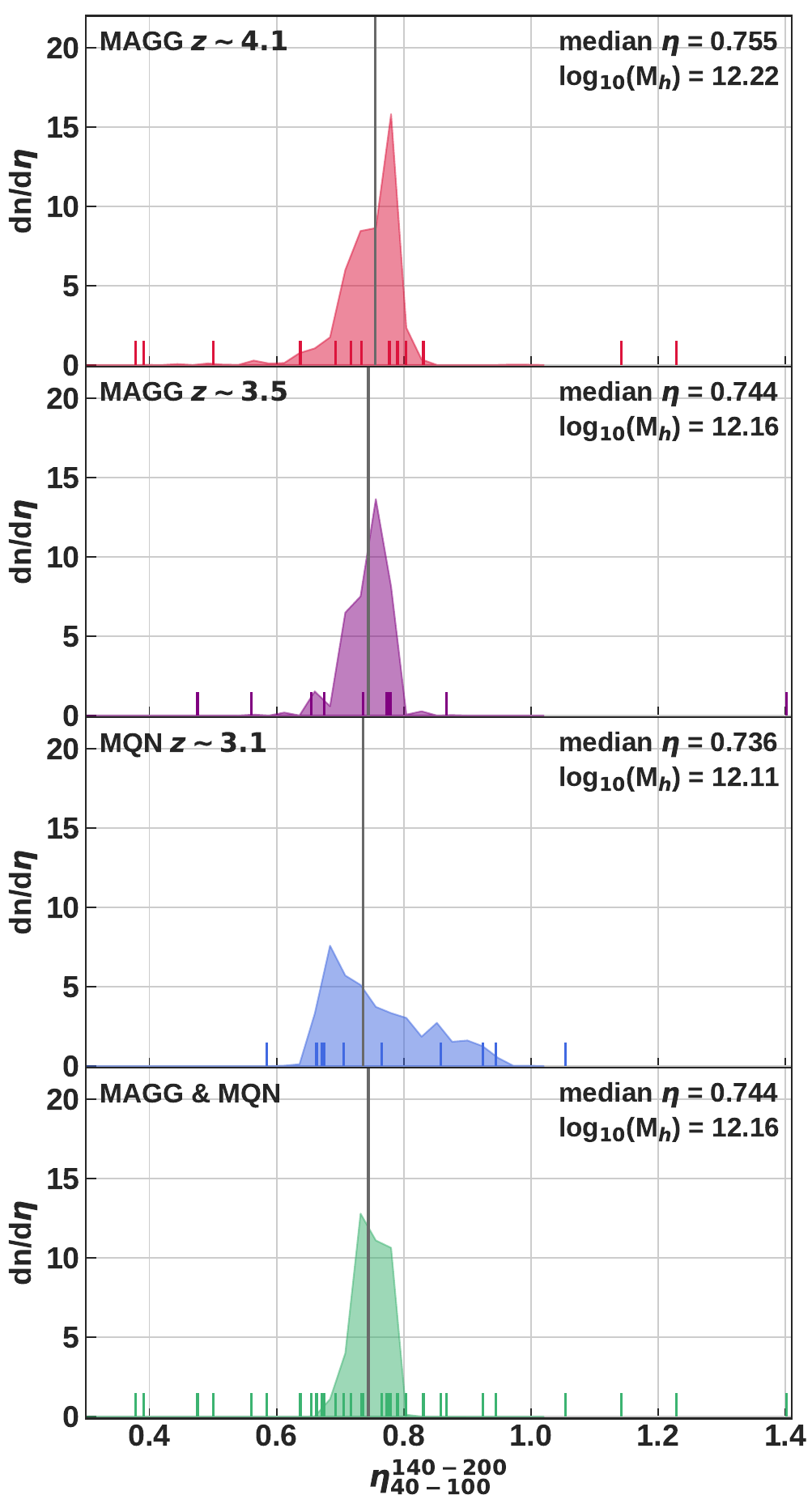}
    \caption{Starting from the top, the panels refer to the MAGG $z\sim4.1$, $z\sim3.5$, MQN $z\sim3.5$ and to all three samples combined. In each panel the vertical ticks indicate the $\eta^{140-200}_{40-100}$ of the individual Ly$\alpha$ nebulae included in the analysis and the shaded region refers to the distribution of medians of the randomly redrawn samples normalised such that the integral of the area equals one. The samples respective median $\eta^{140-200}_{40-100}$ are marked with a vertical, solid line.
    The characteristic halo mass for each sample as obtained using our analytical relation (Equation \ref{eq:etaAnalytical}) is quoted in each panel.}
    \label{fig:massEst}
\end{figure}
We calculate the median $\eta^{140-200}_{40-100}$ of the three samples described in Section \ref{subsec:ObSamples}.
This results in the following values: 0.755, 0.744 and 0.736 for the MAGG $z\sim4.1$, $z\sim3.5$ and MQN $z\sim3.1$ samples respectively. 
In Figure \ref{fig:massEst} we indicate the individual $\eta^{140-200}_{40-100}$ values of each Ly$\alpha$ nebula with vertical ticks. The top three panels refer to the three samples separately and the bottom panel refers to the combination of all three samples. 
In each panel the median of the samples is marked with a solid vertical grey line. 
We use the median of each $\eta^{140-200}_{40-100}$ distribution and Equation \ref{eq:etaAnalytical} to obtain a characteristic quasar halo mass for each sample as outlined in Section \ref{subsec:velDispAnnuli}. 
We estimate the uncertainties of the characteristic halo mass by means of a bootstrap estimate. We randomly re-draw a population of $\eta^{140-200}_{40-100}$ values from each observed sample 10000 times and calculate the median of each random population. The shaded region in each panel of Figure \ref{fig:massEst} indicates the distribution of these medians, normalised so that the integral of the distributions is one.
We calculate the standard deviation of these 10000 median $\eta^{140-200}_{40-100}$ values and quote this standard deviation as the uncertainty of our mass estimates.
The decadic logarithms of the obtained quasar halo mass estimates in units of M$_{\odot}$ are thus $12.22^{+0.28}_{-0.25}$, $12.16^{+0.18}_{-0.17}$ and $12.11^{+0.47}_{-0.42}$ for the MAGG $z\sim4.1$, $z\sim3.5$ and MQN $z\sim3.1$ samples respectively.
We use the median $\eta^{140-200}_{40-100}$ of the samples instead of the mean to limit our sensitivity to outliers. 
We stress, however, that using the mean instead of the median would result in halo mass estimates that are consistent to the above values within their errorbars. In particular,  
calculating the characteristic masses from the mean would result in halo mass estimates with decadic logarithms of 
$12.15^{+0.28}_{-0.25}$, $12.28^{+0.19}_{-0.18}$ and $12.40^{+0.59}_{-0.44}$ for the three samples (in order of decreasing redshift).
We compare the measured $\eta^{140-200}_{40-100}$ values of the individual Ly$\alpha$ nebulae in the MAGG $z\sim3.5$ sample to the number of Ly$\alpha$ emitters (LAEs) in each MUSE field as reported by \citet{Fossati2021} and find evidence for a weak correlation (see Appendix $\mathrm{\ref{app:checkClustering}}$), which strengthens our results.

Despite the relative broadness of the observed $\eta^{140-200}_{40-100}$ distributions, the median values of $\eta^{140-200}_{40-100}$ are very much consistent with each other, independent of redshift.
Combining all three redshift ranges results in an almost symmetrical distribution of $\eta^{140-200}_{40-100}$ around a median of 0.744. We therefore obtain a characteristic halo mass with a decadic logarithm of $12.16^{+0.14}_{-0.13}$ ($12.26^{+0.15}_{-0.14}$ if we use the mean $\eta^{140-200}_{40-100}$ of the combined samples instead).
In addition to calculating the characteristic halo masses of the observed samples, we also calculate the minimum halo mass following the procedure described in Section \ref{subsec:velDispAnnuli} and \citet{Eftekharzadeh2015}. The decadic logarithms of the minimum halo masses obtained in this way are: 12.04, 11.94, 11.87 and 11.94 respectively for the MAGG $z\sim4.1$, $z\sim3.5$, MQN $z\sim3.1$ samples and for all samples combined.

Characteristic halo masses as large as $10^{13}\mathrm{M}_{\odot}$ are thus clearly outside our $1\sigma$ confidence interval at $3.1<z<4.5$. This result, combined with the other literature measurements presented in Figure \ref{fig:QSOclustering}, has important implications for the derivation of the CGM physical properties, such as emitting gas densities and clumpiness as will be discussed in detail in Paper II (see also discussion in \citet{PezzulliCantalupo2019}). Moreover, this result has important implications when compared to other quasar host halo mass measurements, e.g. from clustering as discussed in Section \ref{subsec:discHaloMassEst}. 


\section{Discussion} \label{sec:Discussion}

With the help of cosmological simulations and Ly$\alpha$ nebulae observations at $z>3$, we have shown that it is possible to derive new constraints on quasar halo masses from the CGM kinematics that are complementary to quasar clustering measurements. Despite the model uncertainties and statistical limitations due to the size of the observed samples used in this work, these constraints have the potential to provide a new view of the host haloes of quasars, their environment and thus of CGM properties. 
In Section \ref{subsec:ProsCons} we give an overview of the limitations of the halo mass estimation method presented here. In particular, among the possible model uncertainties, we discuss the effect of the AGN feedback implementation in EAGLE on our results in Section \ref{subsec:effAGNfeedback}. Finally, in Section \ref{subsec:discHaloMassEst} we put our method into a broader context by comparing it and the derived mass estimates to other estimates and methods. 

\subsection{Limitations of our method} \label{subsec:ProsCons}
As is evident from Figures \ref{fig:rescVelDispProf} and \ref{fig:rescaledAnnuli}, there is a relatively large scatter of the simulated velocity dispersion profiles and ratios $\eta^{140-200}_{40-100}$.
Such a large scatter stems both from systematic effects related to morphological and physical properties of the haloes and, partly, from the noise associated with the (mock) measurements themselves. 
Indeed, realistic halo shapes are far from simple spheres and radial inflows may occur along a few filaments coming from various directions \citep{Bond1996, Keres2005}. As such, the same halo as seen from different directions could have different radial profiles of the line of sight velocity dispersion as measured in emission. In addition, realistic haloes do not live in isolation: projection effects due to multiple haloes along the line of sight (as discussed in Section \ref{subsec:velDispAnnuli}) can artificially increase the velocity dispersion. Because these effects are aleatory and, in principle, independent of the quasar host halo mass, the error on the median halo mass of a sample of quasars could be decreased by increasing the sample size.
As a first application, we have used a relatively limited sample of MUSE Ly$\alpha$ nebulae around quasars for which we had access to the velocity dispersion maps. A much larger dataset is already available and in principle our analysis could be repeated using a larger sample. Moreover, the ubiquity of Ly$\alpha$ nebulae around quasars at all redshift explored so far ($z>2$) provide the opportunity to extend the redshift range and plan for dedicated surveys targeting a much larger number of quasars.

Observational limitations are related to the necessary SNR required to measure the line of sight velocity dispersion at a given distance from the quasar. Because the method presented here is based on the value of $\eta^{140-200}_{40-100}$, observable emission should extend to at least 200 ckpc from the quasar. While this is the case for the majority of bright quasar nebulae discovered so far, fainter quasars and quasars at higher redshifts more often have smaller detectable nebulae \citep{Farina2019, Mackenzie2021} at a given SB sensitivity level. As long as these quasars are ``illuminating'' gas on large scales, this issue can be solved by increasing the exposure time of the observations.

We have proposed the use of the Ly$\alpha$ emission as a tracer of the shape of the velocity dispersion profile given the brightness of this line which makes it easily detectable on large scales even in short exposure times. However, as discussed in detail in section \ref{subsec:corrBroadeningLya}, this is a resonant line which may suffer from radiative transfer broadening. We have argued, based on analytical considerations and on complementary HeII-H$\alpha$ observations for a subsample of quasar nebulae, that the Ly$\alpha$ broadening with respect to the ``intrinsic'' velocity dispersion should be independent of distance, making our method based on $\eta^{140-200}_{40-100}$ independent of Ly$\alpha$ broadening. In the model presented in Section \ref{subsec:corrBroadeningLya}, a series of simplified assumptions have been made, for instance, that the quasar ionising radiation is isotropic and that the whole CGM is thus ionised. 
However, some observations suggest that this is not the case, at least for quasars which are around 2-3 mag fainter than the sample studied here \citep{Prochaska2013}, and that part of the transverse direction is not ionised by the quasar. While we have verified that this does not affect our result based on the ``intrinsic'' velocity dispersion (for an opening angle at least as large as $60^\circ$), this could have an effect on the broadening of the Ly$\alpha$ emission produced on the far side of the quasar halo. For ionising cones as expected from the AGN unification model \citep[e.g.][]{denBrok2020}, this would possibly imply a larger broadening at larger distances from the quasar. In this case, our estimate should be considered an \emph{upper limit} on the quasar halo mass. 

Lastly, it is worth stressing that our results are based on the fact that the kinematics of the cold emitting gas is \textit{on average} determined by gravity only and thus by the host halo mass to a high degree, as derived and demonstrated using the EAGLE and ENGINE simulations. 
However, different assumptions, e.g., concerning galaxy feedback, or missing physics in these simulations could lead to a different result, i.e. to different cold gas kinematic patterns and radial velocity dispersion profiles. 
For instance, if galactic feedback is preferentially increasing the velocity dispersion of the emitting gas closer to the center of the haloes, then the actual values of $\eta^{140-200}_{40-100}$ would be \emph{lower} than predicted from pure gravitational effects. This would imply that the actual halo mass could be higher than predicted by from our model. This scenario can however be excluded by looking at the \emph{absolute} value of the velocity dispersion in the inner annulus used in our analysis as measured through a non-resonant line, such as HeII-H$\alpha$ emission. In particular, the values measured in the MAGG sample by Travascio et al. (in Prep.) and discussed in Section \ref{subsec:velDispProf} are of the order of 100 km/s which is very much consistent (or slightly below) the intrinsic velocity dispersion in the mock observations presented here (see Figure \ref{subsec:velDispProf}). Future H$\alpha$ observations, e.g. with JWST, will test this hypothesis. With the data available up until now we then find it very implausible that the actual masses of the quasar host haloes are \emph{significantly larger} than what is found in this work. 

\subsection{The impact of AGN feedback on CGM kinematics }\label{subsec:effAGNfeedback}
Though not the main focus of this work, we take advantage of the different implementation of feedback in the EAGLE simulations to verify the possible effect of AGN feedback on the overall CGM kinematics (and thus also on our halo mass estimate). 
In particular, in this section, we compare the radial velocity profiles obtained from the NoAGN and RECAL simulations (see Table \ref{tab:simProp}).  
For haloes in the three highest mass bins contained in these two simulations the inclusion of AGN-feedback leads to higher maximum outflow velocities of the hot gas in the radial velocity profiles. The radial velocity profiles of the Ly$\alpha$ emitting gas, however, are not affected by the inclusion of AGN-feedback and there is no significant difference between the radial profiles obtained from the NoAGN and the RECAL simulations at either redshift as can be seen in Figures \ref{fig:radVel_AGNvsNoAGNz3p528} and \ref{fig:radVel_AGNvsNoAGNz3p017} in Appendix \ref{app:AGNfeedbackEffect}. 
This is likely due to the fact that, as mentioned in Section \ref{subsec:radVelProf}, the Ly$\alpha$ emitting gas roughly corresponds to the cool gas ($\lessapprox 10^5 K$). The low temperature of the gas implies that it has not been significantly heated by feedback processes. Indeed, the hot gas ejected from the galaxy due to the AGN-feedback (as implemented in EAGLE) is expected to take the path of least resistance and therefore flows out of the galaxy via the regions with lower densities, without interacting with the cold and dense accreting filaments (see also \citet{vanDeVoort2011}). As such, the presence of AGN feedback, at least as implemented in EAGLE, has little effect on our halo mass estimates based on $\eta^{140-200}_{40-100}$.

We stress however that AGN feedback implementation is still highly debated in the literature. While works based on EAGLE-like simulations, e.g. \citet{Rahmati2015}, or other simulations such as SIMBA \citep{Dave2019} typically found that the HI distribution in the CGM, traced via mean Ly$\alpha$ flux fluctuation profiles at high redshift, is not affected by the presence of AGN-feedback \citep[e.g.][]{Sorini2020}, other works, such as \citet{Costa2022}, suggest that AGN feedback can also have an effect on the cold CGM component, at least at very high redshift ($z>6$). From an observational view-point, the situation is also not clear. While extended emission around radio-galaxies and radio-loud quasar has often been associated with outflows \citep{VillarMartin1999, Silva2018} given the large velocity dispersion values ($>600$ km/s) found in non-resonant emission lines, the large majority of radio-quiet nebulae discovered around quasars at $z>2$ present relatively quiet kinematics consistent with gravitational motion only, e.g., in HeII-H$\alpha$
as we have discussed in Section \ref{subsec:velDispProf} (see also \citet{Cantalupo2017} for a review). The discriminating factor could be the presence of a radio-jet on CGM scales which certainly has the potential to inject energy and momentum, possibly also affecting the cold CGM component. It could thus be interesting to apply our methodology to a sample of radio-loud quasars and galaxies in the future and compare the \emph{intrinsic} velocity dispersion profiles to search for AGN feedback effects, assuming, e.g., that radio-loud quasars live in similar haloes with respect to their radio-quiet counterparts. Unfortunately, radio-loud AGN are much rarer and thus a specific sample should be built for this purpose. We note that, especially for radio-galaxies, a non-resonant line (e.g., hydrogen or helium H$\alpha$) is necessary for this analysis since Ly$\alpha$ broadening could be different for these systems. This would be the case, for instance, if the ionisation cone of these AGN is oriented along the plane of the sky rather than along our line-of-sight as for quasars. 
 
\subsection{Comparison to other halo mass estimation methods} \label{subsec:discHaloMassEst}
\begin{figure*}
	\includegraphics[width=1.0\textwidth]{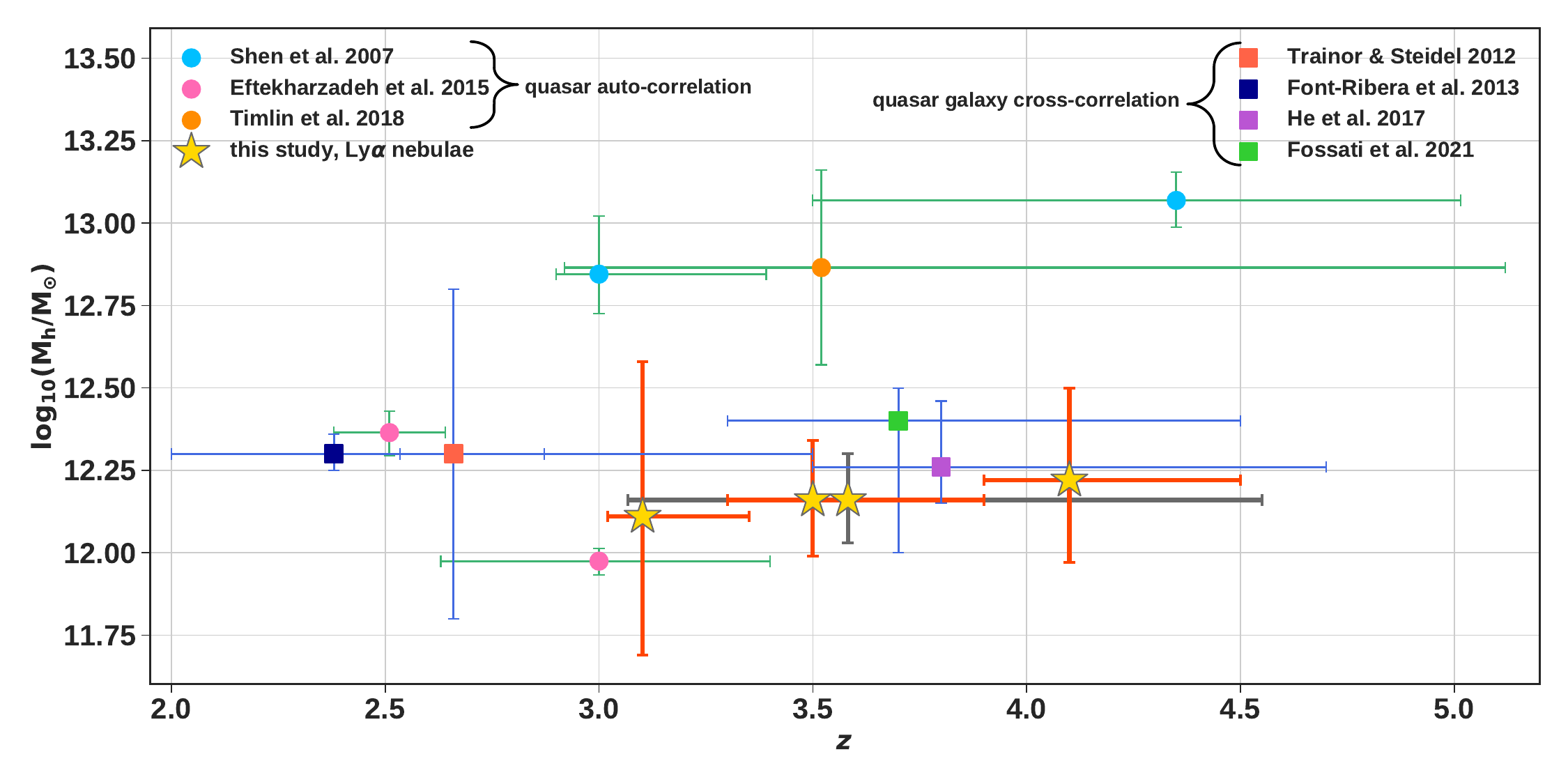}
    \caption{Comparison of our quasar host halo median mass estimates for the MQN $z\sim3.1$, $z\sim3.5$, $z\sim4.1$ MAGG samples (yellow stars with red error bars) and those three samples combined (yellow stars with grey error bars) with  quasar halo mass estimates obtained from QSO-clustering studies (circles with green error bars) and galaxy-quasar clustering studies (squares with blue error bars), as reported in the legend (see Section \ref{subsec:haloMassEstObs}). All horizontal errorbars denote the redshift range of the quasars included in the different studies. The vertical errorbars denote the halo mass range for each sample. For some studies (\citet{He2017}, \citet{Timlin2018} \& \citet{Fossati2021}), only a halo mass range and not a characteristic mass are given. In this case, we centre the errorbars on the averaged halo mass within the quoted mass range for visualisation purposes. The characteristic quasar halo mass and mass range of the quasar sample presented in \citet{Shen2007} \& \citet{FontRibera2013} are taken from \citet{Eftekharzadeh2015}.  At $z\sim3$ our measurements are consistent with the quasar auto-correlation constraints presented in \citet{Eftekharzadeh2015}, which conflict with those presented in \citet{Shen2007}. Our method, which is independent of clustering studies, thus suggest that (bright) quasar are typically hosted by $10^{12.16}\ M_{\odot}$ haloes, independent of redshift in the range $3<z<4$. See section \ref{subsec:ProsCons} for a detailed discussion of the strength and limitation of our methods and the possible implication of this results. }
    \label{fig:QSOclustering}
\end{figure*}

Typical methods currently used to estimate the mass of the haloes hosting quasars
are based on clustering and include, in particular, quasar auto-correlation (QSO-clustering studies) \citep{Shen2007, daAngela2008, Shen2009, Eftekharzadeh2015, Timlin2018} or quasar-galaxy cross-correlation studies \citep{Trainor2012, FontRibera2013, He2017}. In some recent works, kinematics of galaxies and some simplified treatment of Ly$\alpha$ emission kinematics have also been used to put some constraints for quasars at $z>2$ \citep{Lau2018, ArrigoniB2019, Fossati2021}. In this section, we compare these results to those obtained by applying the method developed here to the MAGG and MQN samples as detailed in Section \ref{subsec:haloMassEstObs}.

We stress that, due to the differences in sample sizes and redshifts it is not trivial to compare our mass estimates with those obtained measuring the quasar auto-correlation or quasar-galaxy cross-correlation functions. 
Furthermore, these differences and additional caveats, such as the modelling of nonlinear scales and the inclusion of Poisson shot-noise, also make it non-trivial to compare the results obtained using correlation functions with each other. 
Despite this, we give a qualitative comparison of the quasar halo mass estimates obtained in this study and those from QSO-clustering studies and quasar-galaxy cross-correlation studies in Figure \ref{fig:QSOclustering}. Our results  and their uncertainties for the redshifts $z\sim3.1$ (MQN sample), $z\sim 3.5$ \& $z\sim4.1$ (MAGG sample) are indicated with yellow stars and red error bars. The characteristic mass and associated uncertainties for the three samples combined is plotted with a yellow star and grey error bars. Mass estimates and their uncertainties obtained with QSO-clustering are plotted with circles and green error bars, the results from quasar-galaxy cross-correlation studies are indicated with squares and blue error bars. The colours of the individual markers refer to each individual piece of work.

Our inferred halo mass is lower than the mass estimate of quasar hosting haloes obtained from QSO clustering studies performed at comparable redshifts by \citet{Shen2007} ($z\sim3$ and $z\sim4.3$, blue circle and green error bars) \& \citet{Timlin2018} ($z\sim 3.5$, orange circle and green error bars). 
\citet{Eftekharzadeh2015} (pink circles and green error bars) use a far larger sample, which they divide into three redshift sub-samples, with the two higher redshift bins being most compatible with the quasars used in our analysis. Their mass estimate for the $z\sim3$ bin is lower than our mass estimate, while their mass estimate for the $z\sim2.5$ redshift bin is consistent with our results. The fact that clustering studies do not include Poisson shot-noise and only use a linear bias relation while fitting the correlation function also on (mildly) nonlinear scales could lead to an overestimation of the bias and hence of the haloes masses. The first effect is mitigated with increasing sample size and could explain why \citet{Shen2007} \& \citet{Timlin2018} find higher halo masses.

Our mass estimates are in very good agreement with those obtained from quasar-galaxy cross-correlation reported in Figure \ref{fig:QSOclustering}. 
These include the work of \citet{Trainor2012} (dark orange square and blue error bars), in which they calculate the quasar-galaxy cross-correlation function for a sample of 15 hyper-luminous quasars and the surrounding galaxies that lie within 4.2 h$^{-1}$cMpc at redshifts $2.5 < z < 2.9$, as a part of the Keck Baryonic Structure Survey (KBSS) \citep{Steidel2014}. Comparing the observed cross-correlation function to the galaxy-quasar cross-correlation function of simulated halo populations while varying masses of both the simulated quasar haloes and surrounding galaxy haloes they find that the hyper-luminous quasars are hosted by haloes with a median mass of $10^{12.3\pm0.5} M_{\odot}$.
\citet{FontRibera2013} ($z\sim2.38$, dark blue squares and blue error bars) measure the cross-correlation function of quasars from the Baryon Oscillation Spectroscopic Survey (BOSS) \citep{Dawson2013} and the Ly$\alpha$ forest absorption in redshift space. They measure bias factors consistent with auto-correlation measurements at comparable redshifts.
\citet{He2017} ($z\sim3.8$, purple square and blue error bars) use a combination of two quasar samples totalling 1'243 quasars and 25'790 bright $z\sim4$ Lyman break galaxies to calculate a quasar-galaxy cross-correlation function and derive the bias by comparing the measured clustering strength to that of the underlying dark matter based on linear structure formation theory. From this bias they derive a halo mass range of $10^{12.15} M_\odot$ -  $10^{12.46} M_\odot$.

Our estimate is also consistent with that obtained by \citet{Fossati2021} for the same MAGG sample used in this work. In particular, these authors calculate the quasar-galaxy cross-correlation function for the MAGG sample's 28 quasars and 113 LAEs. However, this measurement is not used to constrain the quasar halo masses as the small field of view of the MUSE instrument prevents the inclusion of larger scales, thus hampering the conversion of the cross-correlation function into a bias and quasar halo mass. 
Instead, they constrain the quasar halo masses in the MAGG sample by calculating the overdensity of LAEs around the MAGG quasars and comparing this overdensity to that of galaxies around the 15 hyper-luminous quasars from the KBSS calculated in \citet{Trainor2012}. As the two galaxy overdensity values are consistent with one another, \citet{Fossati2021} deduce that the quasar halo masses in the MAGG sample are likely also consistent with those calculated in \citet{Trainor2012} using the quasar-galaxy cross-correlation function. 
\citet{Fossati2021} further evaluate the velocity offset of the LAEs with respect to the nebulae's redshift. They compare the kinematic dispersion of these galaxies with that of galaxies in the same line of sight velocity window centred on haloes with a mass of M$_h \sim 10^{12.4} - 10^{12.6}$ M$_{\odot}$ from the Millennium simulation \citep{Springel2005}, finding that the observed kinematic dispersion is consistent with a typical halo mass of $10^{12.5}$ M$_{\odot}$.
Based on these two comparisons, \citet{Fossati2021} derive a quasar halo mass estimate of $10^{12} M_\odot$ -  $10^{12.5} M_\odot$. Although this estimate is not strictly based on the quasar-galaxy cross-correlation function, we still include it in our comparison in Figure \ref{fig:QSOclustering} for the sake of completeness. It is plotted with a green square and blue error bars.

We note that combining the halo mass estimates of the MQN sample at $z\sim3.1$ and the MAGG sample with that of the KBSS would imply a negligible evolution of bright quasar hosting halo masses from $z\sim2.7$ to $z\sim3.7$.  This is consistent with a negligible halo mass evolution within the MAGG sample from $z\sim3.5$ to $z\sim4.1$ already mentioned in Section \ref{subsec:haloMassEstObs} and could have significant implications for our understanding of quasar formation and evolution. However, this interpretation needs additional confirmation as additional effects, such as selection biases affecting both samples, could be responsible for the apparent non-evolution of the mass of haloes hosting bright quasars. 

Finally, it is interesting to compare our method to the analysis performed by \citet{ArrigoniB2019} on their sample of 61 quasar Ly$\alpha$ nebulae as a part of the MUSEUM survey. In particular, they calculate the spatial average of the Ly$\alpha$ velocity dispersion over the whole area in which the emission is detected ($<\sigma_\mathrm{Ly\alpha}>$) quoting a value of $\sigma_{Ly\alpha}< 400$ km s$^{-1}$ (and typical values around 250 km s$^{-1}$ according to their Figure 11). 
Assuming a NFW profile \citep{NFW1997} and a concentration of $c~\sim3.7$ at $z\sim3$, they estimate that a halo with a mass of $10^{12.5}\mathrm{M}_{\odot}$ should have a maximum circular velocity of $v_{circ}^{max} = 360 \mathrm{km\ s^{-1}}$. Further assuming that the velocity dispersion obeys $\sqrt{2}\,\sigma_{1D\,rms} = v_{circ}^{max}$ \citep{Tormen1997}, they infer $\sigma_{1D\,rms}\simeq250 \mathrm{km\ s^{-1}}$, which they note is similar to their $<\sigma_\mathrm{Ly\alpha}>$, implying a similar halo mass to the one obtained using our method.
We note that, given that the MUSEUM survey is shallower in exposure time with respect to MAGG (40 minutes vs a typical exposure time of 4 hours) and the MQN sample (1 hour exposure time), nebulae in this survey are typically detected up to smaller distances from the quasars with respect to these other surveys. The median maximum extent of the nebulae in the MUSEUM survey is indeed around 170 ckpc, thus the $<\sigma_\mathrm{Ly\alpha}>$ typically represent only the inner regions of the nebulae. 
However, because the area changes for each nebula and because of the large spatial averaging it is not easy to translate their $<\sigma_\mathrm{Ly\alpha}>$ in any of the quantities used in this work, such as the median velocity dispersion in our inner and outer annuli (which are defined at fixed comoving distances from the quasars). 
Given the large differences between our method and the approach used by \citet{ArrigoniB2019}, it therefore possible that this agreement is partially coincidental.
We have seen that Ly$\alpha$ broadening significantly increases the observed line widths compared to non-resonant emission, such as HeII-H$\alpha$, making it non-trivial to directly convert the absolute Ly$\alpha$ velocity dispersion values to a measurement of the gas's intrinsic kinematics. 
\citet{Farina2019} also apply this analysis to the 12 Ly$\alpha$ nebulae at $z>5.7$ detected by them and measure an average 1D rms velocity dispersion of $\sigma_{1D\,rms}\simeq340 \pm 125 \mathrm{km\ s^{-1}}$, which is consistent with the gravitational motions in a $10^{12.5}\mathrm{M}_{\odot}$ halo at $z=6$. Although the caveats mentioned above also apply to the comparison of our results with those obtained by \citet{Farina2019}, taken together these results support a link between the Ly$\alpha$ kinematics and dark matter halo mass even at higher redshifts.

\section{Summary \& Conclusions} \label{sec:SumConcl}

The discovery of ubiquitous Ly$\alpha$ emission from the CGM and IGM around quasars at $z>2$ gives us the unique opportunity to constrain the physical properties of gas around galaxies directly through emission, provided that the quasar host halo mass is known \citep[e.g.][]{PezzulliCantalupo2019}. Unfortunately, current constraints on quasar host halo masses given by clustering studies \citep[e.g. using SDSS][]{Shen2007, Eftekharzadeh2015} exhibit significant discrepancies at $z\simeq3$ (see Figure \ref{fig:QSOclustering}), the redshift at which most Ly$\alpha$ nebulae have been discovered so far.  

We develop a new method to constrain quasar halo masses based on the kinematics of the cold (T $< 10^5$ K), Ly$\alpha$ emitting CGM. By using the cosmological simulations EAGLE and ENGINE \citep{Schaye2015}, we first show in Section \ref{subsec:radVelProf} that the kinematics of cold Ly$\alpha$ emitting gas in the CGM of massive haloes ($\mathrm{M}_h > 10^{11.75} \mathrm{M}_{\odot}$) should directly depend on the total halo mass, rather than galactic feedback over the scales of interests for Ly$\alpha$ nebulae observations. In particular, we find that the radial velocity profiles of cold emitting gas (Ly$\alpha$-emissivity-weighted) closely follow the radial velocity profiles of the dark matter, at least in the EAGLE and ENGINE simulations (Figure \ref{fig:radVel}) at distances between roughly 1.5 and 5 virial radii from the centres of haloes in the mass range $10^{11.75} \mathrm{M}_{\odot}$ - $10^{13.25}\mathrm{M}_{\odot}$. At distances below 1.5 virial radii from the halo centre the Ly$\alpha$ emitting gas is predominantly inflowing.

With the aim of exploiting this result, we generate mock MUSE-like observations of Ly$\alpha$ emission from massive haloes in EAGLE and ENGINE under the assumption of maximally fluorescent emission due to bright quasar ionisation (see Sections \ref{subsec:LyaEm} \& \ref{subsec:mockCubes}), including sky background noise, atmospheric smoothing and the finite MUSE spectral resolution. The mock observations are then analysed with the same software as is used for MUSE observations, producing first and second moment maps of the flux distribution (Section \ref{subsec:detExLya}). These maps are then used to generate the \emph{intrinsic} velocity dispersion profiles of the Ly$\alpha$ emission (i.e., without considering the effect of resonant broadening) as a function of projected distance from the quasar, a quantity which can be directly compared to observations. We find that, once rescaled by the virial radius and normalised to the value of the central velocity dispersion, these profiles become self-similar (Figure \ref{fig:rescVelDispProf}), demonstrating that they could be used to derive a constraint on the virial radius of the associated halo and thus on its total mass. 

Taking advantage of this self-similarity, we define a new variable which can be directly measured in observations and used to derive the halo mass: the velocity dispersion ratio $\eta^{140-200}_{40-100}$. This variable represents the ratio of the median velocity dispersion values in two concentric annuli (40-100 and 140-200 ckpc). These annuli have been carefully selected to maximise the $\eta^{140-200}_{40-100}$ variation across the expected halo mass range associated with quasars \citep[$10^{12} \mathrm{M}_{\odot}$ and $10^{13} \mathrm{M}_{\odot}$][]{Shen2007, Eftekharzadeh2015, Timlin2018}, while at the same time excluding the inner regions typically affected by the quasar PSF in observations. 
More importantly, in Section \ref{subsec:corrBroadeningLya}, we show that the value of $\eta^{140-200}_{40-100}$ is unaffected by the radiative transfer spectral broadening associated to resonant Ly$\alpha$ emission. This result is obtained by comparing the observed HeII-H$\alpha$ velocity dispersion values to the Ly$\alpha$ velocity dispersion values for a subset of quasars for which both measurements are available (see Section \ref{subsec:corrBroadeningLya}). 

As a first application of our new methodology, we apply our analytical relation based on the self-similarity of the velocity dispersion profiles to 37 Ly$\alpha$ nebulae observed at $3<z<4.5$ as part of the MAGG \citep{Fossati2021} and MQN \citep{borisova} surveys. 
As is typically done in clustering analysis studies, we derive a characteristic halo mass and a minimum halo mass for each of our quasar samples, obtaining the following characteristic masses in units of solar masses with decadic logarithms of: $12.11^{+0.47}_{-0.42}$, $12.16^{+0.18}_{-0.17}$, $12.22^{+0.28}_{-0.25}$ and the following minimum halo masses: 11.87, 11.94, 12.04 at $z\sim3.1$, $z\sim3.5$ and, $z\sim4.1$ respectively (Section \ref{subsec:haloMassEstObs}).
Given the fact that these mass estimates are consistent which each other at different redshifts, we also obtain a combined constraint by combining all quasars in our samples, resulting in a characteristic halo mass with a decadic logarithm of $12.16^{+0.14}_{-0.13}$ in units of solar masses and a minimum halo mass of $11.94$.
We then compare our results to other estimates of quasar host halo masses obtained through clustering studies at similar redshifts (see Figure \ref{fig:QSOclustering}) finding good agreement with quasar-galaxy cross-correlation studies and intermediate values between the two discrepant quasar auto-correlation clustering measurements of \citet{Eftekharzadeh2015} and \citet{Shen2007} at z$\sim3$. We stress that our method is based on the kinematics of cold emitting gas in the CGM of quasars, thereby providing an independent estimate with respect to these studies. Despite the relatively large errorbars due to the limited number of nebulae used in this study, our results suggest no significant redshift evolution of the (bright) quasar host halo masses across the explored redshift range.
Combining our studies with cross-correlation clustering results and with the auto-correlation clustering measurements of \citet{Eftekharzadeh2015} at $z\sim2.5$ would consistently give a characteristic halo mass of $\sim10^{12.2} \mathrm{M}_{\odot}$ and thus disfavour masses around $\sim10^{13} \mathrm{M}_{\odot}$, such as those suggested by \citet{Shen2007}, with a high level of confidence.

The uncertainties associated with our measurement can be significantly improved in the future by taking advantage of the large sample of quasar Ly$\alpha$ nebulae discovered so far \citep{Cai2019, Farina2019, Mackenzie2021}, which have not been included in the present study for the sake of brevity. Moreover, ongoing and future observations of H$\alpha$ emission from quasar nebulae at z$>2$, e.g. with JWST or from the ground at some particular redshift \citep[e.g.][]{Langen2022} will further reduce possible uncertainties associated with the Ly$\alpha$ line radiative transfer, allowing, for instance, to use the \emph{absolute} value of the velocity dispersion as an additional constraint together with the \emph{shape} of the velocity dispersion profile. Additionally, our method could be applied to different subsamples of quasars, e.g., as a function of their UV luminosity or radio-loudness, in order to disentangle possible environmental effects associated with different quasar sub-samples. 

In the context of the RePhyNe project, whose main goal is to constrain and resolve the physics of the cold component of the CGM, the results presented here provide a possible resolution to the discrepancy previously found in the literature concerning the host halo masses of quasars at $z\sim3$. This allows us to break several degeneracies between, e.g., implied cold gas densities and CGM clumpiness from the Ly$\alpha$ emission and halo masses. In particular, based on the analytical model presented in \citet{PezzulliCantalupo2019}, a host halo mass of $\sim10^{12.2} \mathrm{M}_{\odot}$ (or lower) for quasar Ly$\alpha$ nebulae would imply high densities in the CGM which cannot be easily confined by the thermal pressure of the hot virialised gas. This would imply, e.g., the presence of broad gas density distributions in the CGM of quasars, as suggested by \citet{Cantalupo2019}, or, alternatively, a significant contribution to the Ly$\alpha$ emission from mechanisms differing from recombination radiation (a hypothesis that will be directly tested by JWST H$\alpha$ emission observations and has so far been excluded by current observations from the ground, \citep[e.g.][]{Langen2022, Leibler2018}). In the second paper of this series, we will take advantage of the results presented here to put strong constraints on the physical properties of cold emitting gas in the CGM of high redshift galaxies (hosting an AGN) by comparing analytical and numerical models to the observed Ly$\alpha$ Surface Brightness profiles. 

\section*{Acknowledgements}
We would like to thank the referee for their comments and suggestions, which were useful in improving this work.
This project has received funding from the European Research Council (ERC) under the European Union’s Horizon 2020 research and innovation programme (grant agreement Nos. 757535 and 864361) and by Fondazione Cariplo (grant Nos. 2018-2329 and 2020-0902). 
We also acknowledge the Virgo Consortium for making their simulation data available. This work used the DiRAC@Durham facility managed by the Institute for Computational Cosmology on behalf of the STFC DiRAC HPC Facility \url{www.dirac.ac.uk}. The equipment was funded by BEIS capital funding via STFC capital grants ST/P002293/1, ST/R002371/1 and ST/S002502/1, Durham University and STFC operations grant ST/R000832/1. DiRAC is part of the National e-Infrastructure. The {\sc eagle} simulations were performed using the DiRAC-2 facility at Durham, managed by the ICC, and the PRACE facility Curie based in France at TGCC, CEA, Bruyèresle-Châtel. Gabriele Pezzulli further acknowledges support from the Netherlands Research School for Astronomy (Nederlandse Onderzoekschool Voor Astronomie, NOVA), Project 10.1.5.18. This work was supported by STFC grant ST/T000244/1. 
\section*{Data Availability}
The simulations used in this work are available on the EAGLE database \url{http://icc.dur.ac.uk/Eagle/database.php} and the observational data can accessed on the ESO Science Archive Facility \url{http://archive.eso.org/cms.html}. As for the codes, \texttt{P2C} is available at \url{https://gitlab.com/sdebeer/P2C}, the \texttt{CubeExtractor} package can be requested from Sebastiano Cantalupo and all other codes are available from Stephanie de Beer upon reasonable request.


\bibliographystyle{mnras}
\bibliography{sources} 

\begin{thebibliography}{}
\makeatletter
\relax
\def\mn@urlcharsother{\let\do\@makeother \do\$\do\&\do\#\do\^\do\_\do\%\do\~}
\def\mn@doi{\begingroup\mn@urlcharsother \@ifnextchar [ {\mn@doi@}
  {\mn@doi@[]}}
\def\mn@doi@[#1]#2{\def\@tempa{#1}\ifx\@tempa\@empty \href
  {http://dx.doi.org/#2} {doi:#2}\else \href {http://dx.doi.org/#2} {#1}\fi
  \endgroup}
\def\mn@eprint#1#2{\mn@eprint@#1:#2::\@nil}
\def\mn@eprint@arXiv#1{\href {http://arxiv.org/abs/#1} {{\tt arXiv:#1}}}
\def\mn@eprint@dblp#1{\href {http://dblp.uni-trier.de/rec/bibtex/#1.xml}
  {dblp:#1}}
\def\mn@eprint@#1:#2:#3:#4\@nil{\def\@tempa {#1}\def\@tempb {#2}\def\@tempc
  {#3}\ifx \@tempc \@empty \let \@tempc \@tempb \let \@tempb \@tempa \fi \ifx
  \@tempb \@empty \def\@tempb {arXiv}\fi \@ifundefined
  {mn@eprint@\@tempb}{\@tempb:\@tempc}{\expandafter \expandafter \csname
  mn@eprint@\@tempb\endcsname \expandafter{\@tempc}}}

\bibitem[\protect\citeauthoryear{{Adams} et~al.,}{{Adams}
  et~al.}{2021}]{Chombo}
{Adams} M.,  et~al., 2021, Technical Report LBNL-6616E, { Chombo Software
  Package for AMR Applications - Design Document}.
Lawrence Berkeley National Laboratory

\bibitem[\protect\citeauthoryear{{Agertz}, {Teyssier}  \& {Moore}}{{Agertz}
  et~al.}{2009}]{Agertz2009}
{Agertz} O.,  {Teyssier} R.,   {Moore} B.,  2009, \mn@doi [\mnras]
  {10.1111/j.1745-3933.2009.00685.x}, \href
  {https://ui.adsabs.harvard.edu/abs/2009MNRAS.397L..64A} {397, L64}

\bibitem[\protect\citeauthoryear{{Arrigoni Battaia}, {Hennawi}, {Prochaska}  \&
  {Cantalupo}}{{Arrigoni Battaia} et~al.}{2015}]{ArrigoniBattaia2015}
{Arrigoni Battaia} F.,  {Hennawi} J.~F.,  {Prochaska} J.~X.,   {Cantalupo} S.,
  2015, \mn@doi [\apj] {10.1088/0004-637X/809/2/163}, \href
  {https://ui.adsabs.harvard.edu/abs/2015ApJ...809..163A} {809, 163}

\bibitem[\protect\citeauthoryear{{Arrigoni Battaia}, {Hennawi}, {Prochaska},
  {O{\~n}orbe}, {Farina}, {Cantalupo}  \& {Lusso}}{{Arrigoni Battaia}
  et~al.}{2019}]{ArrigoniB2019}
{Arrigoni Battaia} F.,  {Hennawi} J.~F.,  {Prochaska} J.~X.,  {O{\~n}orbe} J.,
  {Farina} E.~P.,  {Cantalupo} S.,   {Lusso} E.,  2019, \mn@doi [\mnras]
  {10.1093/mnras/sty2827}, \href
  {https://ui.adsabs.harvard.edu/abs/2019MNRAS.482.3162A} {482, 3162}

\bibitem[\protect\citeauthoryear{{Augustin}, {P{\'e}roux}, {Hamanowicz},
  {Kulkarni}, {Rahmani}  \& {Zanella}}{{Augustin} et~al.}{2021}]{Augustin2021}
{Augustin} R.,  {P{\'e}roux} C.,  {Hamanowicz} A.,  {Kulkarni} V.,  {Rahmani}
  H.,   {Zanella} A.,  2021, \mn@doi [\mnras] {10.1093/mnras/stab1673}, \href
  {https://ui.adsabs.harvard.edu/abs/2021MNRAS.505.6195A} {505, 6195}

\bibitem[\protect\citeauthoryear{{Bacon} et~al.,}{{Bacon}
  et~al.}{2010}]{Bacon2010}
{Bacon} R.,  et~al., 2010, in {McLean} I.~S.,  {Ramsay} S.~K.,   {Takami} H.,
  eds,  Society of Photo-Optical Instrumentation Engineers (SPIE) Conference
  Series Vol. 7735, Ground-based and Airborne Instrumentation for Astronomy
  III. p. 773508, \mn@doi{10.1117/12.856027}

\bibitem[\protect\citeauthoryear{{Bacon} et~al.,}{{Bacon}
  et~al.}{2017}]{Bacon2017}
{Bacon} R.,  et~al., 2017, \mn@doi [\aap] {10.1051/0004-6361/201730833}, \href
  {https://ui.adsabs.harvard.edu/abs/2017A&A...608A...1B} {608, A1}

\bibitem[\protect\citeauthoryear{Bacon et~al.,}{Bacon et~al.}{2021}]{Bacon2021}
Bacon R.,  et~al., 2021, \mn@doi [{\aa}p]
  {10.1051/0004-6361/20203988710.48550/arXiv.2102.05516}, 647, A107

\bibitem[\protect\citeauthoryear{{Berger} \& {Oliger}}{{Berger} \&
  {Oliger}}{1984}]{Berger1984}
{Berger} M.~J.,  {Oliger} J.,  1984, \mn@doi [Journal of Computational Physics]
  {10.1016/0021-9991(84)90073-1}, \href
  {https://ui.adsabs.harvard.edu/abs/1984JCoPh..53..484B} {53, 484}

\bibitem[\protect\citeauthoryear{{Bertschinger}}{{Bertschinger}}{1985}]{Bertschinger1985}
{Bertschinger} E.,  1985, \mn@doi [\apjs] {10.1086/191028}, \href
  {https://ui.adsabs.harvard.edu/abs/1985ApJS...58...39B} {58, 39}

\bibitem[\protect\citeauthoryear{{Bielby}, {Crighton}, {Fumagalli}, {Morris},
  {Stott}, {Tejos}  \& {Cantalupo}}{{Bielby} et~al.}{2017}]{Bielby2017}
{Bielby} R.,  {Crighton} N.~H.~M.,  {Fumagalli} M.,  {Morris} S.~L.,  {Stott}
  J.~P.,  {Tejos} N.,   {Cantalupo} S.,  2017, \mn@doi [\mnras]
  {10.1093/mnras/stx528}, \href
  {https://ui.adsabs.harvard.edu/abs/2017MNRAS.468.1373B} {468, 1373}

\bibitem[\protect\citeauthoryear{{Bolton}, {Puchwein}, {Sijacki}, {Haehnelt},
  {Kim}, {Meiksin}, {Regan}  \& {Viel}}{{Bolton} et~al.}{2017}]{Bolton2017}
{Bolton} J.~S.,  {Puchwein} E.,  {Sijacki} D.,  {Haehnelt} M.~G.,  {Kim} T.-S.,
   {Meiksin} A.,  {Regan} J.~A.,   {Viel} M.,  2017, \mn@doi [\mnras]
  {10.1093/mnras/stw2397}, \href
  {https://ui.adsabs.harvard.edu/abs/2017MNRAS.464..897B} {464, 897}

\bibitem[\protect\citeauthoryear{{Bond}, {Kofman}  \& {Pogosyan}}{{Bond}
  et~al.}{1996}]{Bond1996}
{Bond} J.~R.,  {Kofman} L.,   {Pogosyan} D.,  1996, \mn@doi [\nat]
  {10.1038/380603a0}, \href
  {https://ui.adsabs.harvard.edu/abs/1996Natur.380..603B} {380, 603}

\bibitem[\protect\citeauthoryear{{Bordoloi} et~al.,}{{Bordoloi}
  et~al.}{2022}]{Bordoloi2022}
{Bordoloi} R.,  et~al., 2022, \mn@doi [\nat] {10.1038/s41586-022-04616-1},
  \href {https://ui.adsabs.harvard.edu/abs/2022Natur.606...59B} {606, 59}

\bibitem[\protect\citeauthoryear{{Borisova} et~al.,}{{Borisova}
  et~al.}{2016}]{borisova}
{Borisova} E.,  et~al., 2016, \mn@doi [\apj] {10.3847/0004-637X/831/1/39},
  \href {https://ui.adsabs.harvard.edu/abs/2016ApJ...831...39B} {831, 39}

\bibitem[\protect\citeauthoryear{{Bunker}, {Smith}, {Spinrad}, {Stern}  \&
  {Warren}}{{Bunker} et~al.}{2003}]{Bunker2003}
{Bunker} A.,  {Smith} J.,  {Spinrad} H.,  {Stern} D.,   {Warren} S.,  2003,
  \mn@doi [\apss] {10.1023/A:1024038312479}, \href
  {https://ui.adsabs.harvard.edu/abs/2003Ap&SS.284..357B} {284, 357}

\bibitem[\protect\citeauthoryear{{Cai} et~al.,}{{Cai} et~al.}{2019}]{Cai2019}
{Cai} Z.,  et~al., 2019, \mn@doi [\apjs] {10.3847/1538-4365/ab4796}, \href
  {https://ui.adsabs.harvard.edu/abs/2019ApJS..245...23C} {245, 23}

\bibitem[\protect\citeauthoryear{{Cantalupo}}{{Cantalupo}}{2017}]{Cantalupo2017}
{Cantalupo} S.,  2017, in {Fox} A.,  {Dav{\'e}} R.,  eds,  Astrophysics and
  Space Science Library Vol. 430, Gas Accretion onto Galaxies. p.~195
  (\mn@eprint {arXiv} {1612.00491}), \mn@doi{10.1007/978-3-319-52512-9_9}

\bibitem[\protect\citeauthoryear{{Cantalupo} \& {Porciani}}{{Cantalupo} \&
  {Porciani}}{2011}]{Cantalupo2011}
{Cantalupo} S.,  {Porciani} C.,  2011, \mn@doi [\mnras]
  {10.1111/j.1365-2966.2010.17799.x}, \href
  {https://ui.adsabs.harvard.edu/abs/2011MNRAS.411.1678C} {411, 1678}

\bibitem[\protect\citeauthoryear{{Cantalupo}, {Porciani}, {Lilly}  \&
  {Miniati}}{{Cantalupo} et~al.}{2005}]{Cantalupo2005}
{Cantalupo} S.,  {Porciani} C.,  {Lilly} S.~J.,   {Miniati} F.,  2005, \mn@doi
  [\apj] {10.1086/430758}, \href
  {https://ui.adsabs.harvard.edu/abs/2005ApJ...628...61C} {628, 61}

\bibitem[\protect\citeauthoryear{{Cantalupo}, {Porciani}  \&
  {Lilly}}{{Cantalupo} et~al.}{2008}]{Cantalupo2008}
{Cantalupo} S.,  {Porciani} C.,   {Lilly} S.~J.,  2008, \mn@doi [\apj]
  {10.1086/523298}, \href
  {https://ui.adsabs.harvard.edu/abs/2008ApJ...672...48C} {672, 48}

\bibitem[\protect\citeauthoryear{{Cantalupo}, {Lilly}  \&
  {Haehnelt}}{{Cantalupo} et~al.}{2012}]{Cantalupo2012}
{Cantalupo} S.,  {Lilly} S.~J.,   {Haehnelt} M.~G.,  2012, \mn@doi [\mnras]
  {10.1111/j.1365-2966.2012.21529.x}, \href
  {https://ui.adsabs.harvard.edu/abs/2012MNRAS.425.1992C} {425, 1992}

\bibitem[\protect\citeauthoryear{{Cantalupo}, {Arrigoni-Battaia}, {Prochaska},
  {Hennawi}  \& {Madau}}{{Cantalupo} et~al.}{2014}]{Cantalupo2014}
{Cantalupo} S.,  {Arrigoni-Battaia} F.,  {Prochaska} J.~X.,  {Hennawi} J.~F.,
  {Madau} P.,  2014, \mn@doi [\nat] {10.1038/nature12898}, \href
  {https://ui.adsabs.harvard.edu/abs/2014Natur.506...63C} {506, 63}

\bibitem[\protect\citeauthoryear{{Cantalupo} et~al.,}{{Cantalupo}
  et~al.}{2019}]{Cantalupo2019}
{Cantalupo} S.,  et~al., 2019, \mn@doi [\mnras] {10.1093/mnras/sty3481}, \href
  {https://ui.adsabs.harvard.edu/abs/2019MNRAS.483.5188C} {483, 5188}

\bibitem[\protect\citeauthoryear{Childs et~al.,}{Childs et~al.}{2012}]{VisIt}
Childs H.,  et~al., 2012, in , High Performance Visualization--Enabling
  Extreme-Scale Scientific Insight.
pp 357--372, \mn@doi{10.1201/b12985}

\bibitem[\protect\citeauthoryear{{Churchill}, {Nielsen}, {Kacprzak}  \&
  {Trujillo-Gomez}}{{Churchill} et~al.}{2013a}]{Churchill2013}
{Churchill} C.~W.,  {Nielsen} N.~M.,  {Kacprzak} G.~G.,   {Trujillo-Gomez} S.,
  2013a, \mn@doi [\apjl] {10.1088/2041-8205/763/2/L42}, \href
  {https://ui.adsabs.harvard.edu/abs/2013ApJ...763L..42C} {763, L42}

\bibitem[\protect\citeauthoryear{{Churchill}, {Trujillo-Gomez}, {Nielsen}  \&
  {Kacprzak}}{{Churchill} et~al.}{2013b}]{Churchill2013_2}
{Churchill} C.~W.,  {Trujillo-Gomez} S.,  {Nielsen} N.~M.,   {Kacprzak} G.~G.,
  2013b, \mn@doi [\apj] {10.1088/0004-637X/779/1/87}, \href
  {https://ui.adsabs.harvard.edu/abs/2013ApJ...779...87C} {779, 87}

\bibitem[\protect\citeauthoryear{{Corlies}, {Peeples}, {Tumlinson}, {O'Shea},
  {Lehner}, {Howk}, {O'Meara}  \& {Smith}}{{Corlies}
  et~al.}{2020}]{Corlies2020}
{Corlies} L.,  {Peeples} M.~S.,  {Tumlinson} J.,  {O'Shea} B.~W.,  {Lehner} N.,
   {Howk} J.~C.,  {O'Meara} J.~M.,   {Smith} B.~D.,  2020, \mn@doi [\apj]
  {10.3847/1538-4357/ab9310}, \href
  {https://ui.adsabs.harvard.edu/abs/2020ApJ...896..125C} {896, 125}

\bibitem[\protect\citeauthoryear{{Costa}, {Arrigoni Battaia}, {Farina},
  {Keating}, {Rosdahl}  \& {Kimm}}{{Costa} et~al.}{2022}]{Costa2022}
{Costa} T.,  {Arrigoni Battaia} F.,  {Farina} E.~P.,  {Keating} L.~C.,
  {Rosdahl} J.,   {Kimm} T.,  2022, \mn@doi [\mnras] {10.1093/mnras/stac2432},
  \href {https://ui.adsabs.harvard.edu/abs/2022MNRAS.517.1767C} {517, 1767}

\bibitem[\protect\citeauthoryear{{Crain}, {Eke}, {Frenk}, {Jenkins},
  {McCarthy}, {Navarro}  \& {Pearce}}{{Crain} et~al.}{2007}]{Crain2007}
{Crain} R.~A.,  {Eke} V.~R.,  {Frenk} C.~S.,  {Jenkins} A.,  {McCarthy} I.~G.,
  {Navarro} J.~F.,   {Pearce} F.~R.,  2007, \mn@doi [\mnras]
  {10.1111/j.1365-2966.2007.11598.x}, \href
  {https://ui.adsabs.harvard.edu/abs/2007MNRAS.377...41C} {377, 41}

\bibitem[\protect\citeauthoryear{{Crain} et~al.,}{{Crain}
  et~al.}{2015}]{Crain_2015}
{Crain} R.~A.,  et~al., 2015, \mn@doi [\mnras] {10.1093/mnras/stv725}, \href
  {https://ui.adsabs.harvard.edu/abs/2015MNRAS.450.1937C} {450, 1937}

\bibitem[\protect\citeauthoryear{{Crighton}, {Hennawi}, {Simcoe}, {Cooksey},
  {Murphy}, {Fumagalli}, {Prochaska}  \& {Shanks}}{{Crighton}
  et~al.}{2015}]{Crighton2015}
{Crighton} N. H.~M.,  {Hennawi} J.~F.,  {Simcoe} R.~A.,  {Cooksey} K.~L.,
  {Murphy} M.~T.,  {Fumagalli} M.,  {Prochaska} J.~X.,   {Shanks} T.,  2015,
  \mn@doi [\mnras] {10.1093/mnras/stu2088}, \href
  {https://ui.adsabs.harvard.edu/abs/2015MNRAS.446...18C} {446, 18}

\bibitem[\protect\citeauthoryear{{Dav{\'e}}, {Angl{\'e}s-Alc{\'a}zar},
  {Narayanan}, {Li}, {Rafieferantsoa}  \& {Appleby}}{{Dav{\'e}}
  et~al.}{2019}]{Dave2019}
{Dav{\'e}} R.,  {Angl{\'e}s-Alc{\'a}zar} D.,  {Narayanan} D.,  {Li} Q.,
  {Rafieferantsoa} M.~H.,   {Appleby} S.,  2019, \mn@doi [\mnras]
  {10.1093/mnras/stz937}, \href
  {https://ui.adsabs.harvard.edu/abs/2019MNRAS.486.2827D} {486, 2827}

\bibitem[\protect\citeauthoryear{{Dawson} et~al.,}{{Dawson}
  et~al.}{2013}]{Dawson2013}
{Dawson} K.~S.,  et~al., 2013, \mn@doi [\aj] {10.1088/0004-6256/145/1/10},
  \href {https://ui.adsabs.harvard.edu/abs/2013AJ....145...10D} {145, 10}

\bibitem[\protect\citeauthoryear{{Dekel} et~al.,}{{Dekel}
  et~al.}{2009}]{Dekel2009}
{Dekel} A.,  et~al., 2009, \mn@doi [\nat] {10.1038/nature07648}, \href
  {https://ui.adsabs.harvard.edu/abs/2009Natur.457..451D} {457, 451}

\bibitem[\protect\citeauthoryear{{Diemer} \& {Kravtsov}}{{Diemer} \&
  {Kravtsov}}{2014}]{Diemer2014}
{Diemer} B.,  {Kravtsov} A.~V.,  2014, \mn@doi [\apj]
  {10.1088/0004-637X/789/1/1}, \href
  {https://ui.adsabs.harvard.edu/abs/2014ApJ...789....1D} {789, 1}

\bibitem[\protect\citeauthoryear{{Dijkstra}, {Haiman}  \& {Spaans}}{{Dijkstra}
  et~al.}{2006}]{Dijkstra2006}
{Dijkstra} M.,  {Haiman} Z.,   {Spaans} M.,  2006, \mn@doi [\apj]
  {10.1086/506244}, \href
  {https://ui.adsabs.harvard.edu/abs/2006ApJ...649...37D} {649, 37}

\bibitem[\protect\citeauthoryear{{Drake}, {Farina}, {Neeleman}, {Walter},
  {Venemans}, {Banados}, {Mazzucchelli}  \& {Decarli}}{{Drake}
  et~al.}{2019}]{Drake2019}
{Drake} A.~B.,  {Farina} E.~P.,  {Neeleman} M.,  {Walter} F.,  {Venemans} B.,
  {Banados} E.,  {Mazzucchelli} C.,   {Decarli} R.,  2019, \mn@doi [\apj]
  {10.3847/1538-4357/ab2984}, \href
  {https://ui.adsabs.harvard.edu/abs/2019ApJ...881..131D} {881, 131}

\bibitem[\protect\citeauthoryear{{Drake} et~al.,}{{Drake}
  et~al.}{2022}]{Drake2022}
{Drake} A.~B.,  et~al., 2022, \mn@doi [\apj] {10.3847/1538-4357/ac5043}, \href
  {https://ui.adsabs.harvard.edu/abs/2022ApJ...929...86D} {929, 86}

\bibitem[\protect\citeauthoryear{{Dutta} et~al.,}{{Dutta}
  et~al.}{2020}]{Dutta2020}
{Dutta} R.,  et~al., 2020, \mn@doi [\mnras] {10.1093/mnras/staa3147}, \href
  {https://ui.adsabs.harvard.edu/abs/2020MNRAS.499.5022D} {499, 5022}

\bibitem[\protect\citeauthoryear{{Dutta} et~al.,}{{Dutta}
  et~al.}{2021}]{Dutta2021}
{Dutta} R.,  et~al., 2021, \mn@doi [\mnras] {10.1093/mnras/stab2752}, \href
  {https://ui.adsabs.harvard.edu/abs/2021MNRAS.508.4573D} {508, 4573}

\bibitem[\protect\citeauthoryear{EAGLE-Team}{EAGLE-Team}{2017}]{team2017eagle}
EAGLE-Team 2017, The EAGLE simulations of galaxy formation: Public release of
  particle data (\mn@eprint {arXiv} {1706.09899})

\bibitem[\protect\citeauthoryear{{Eftekharzadeh} et~al.,}{{Eftekharzadeh}
  et~al.}{2015}]{Eftekharzadeh2015}
{Eftekharzadeh} S.,  et~al., 2015, \mn@doi [\mnras] {10.1093/mnras/stv1763},
  \href {https://ui.adsabs.harvard.edu/abs/2015MNRAS.453.2779E} {453, 2779}

\bibitem[\protect\citeauthoryear{{Ellison}, {Ibata}, {Pettini}, {Lewis},
  {Aracil}, {Petitjean}  \& {Srianand}}{{Ellison} et~al.}{2004}]{Ellision2004}
{Ellison} S.~L.,  {Ibata} R.,  {Pettini} M.,  {Lewis} G.~F.,  {Aracil} B.,
  {Petitjean} P.,   {Srianand} R.,  2004, \mn@doi [\aap]
  {10.1051/0004-6361:20034003}, \href
  {https://ui.adsabs.harvard.edu/abs/2004A&A...414...79E} {414, 79}

\bibitem[\protect\citeauthoryear{{Fardal}, {Katz}, {Gardner}, {Hernquist},
  {Weinberg}  \& {Dav{\'e}}}{{Fardal} et~al.}{2001}]{Fardal2001}
{Fardal} M.~A.,  {Katz} N.,  {Gardner} J.~P.,  {Hernquist} L.,  {Weinberg}
  D.~H.,   {Dav{\'e}} R.,  2001, \mn@doi [\apj] {10.1086/323519}, \href
  {https://ui.adsabs.harvard.edu/abs/2001ApJ...562..605F} {562, 605}

\bibitem[\protect\citeauthoryear{{Farina} et~al.,}{{Farina}
  et~al.}{2017}]{Farina2017}
{Farina} E.~P.,  et~al., 2017, \mn@doi [\apj] {10.3847/1538-4357/aa8df4}, \href
  {https://ui.adsabs.harvard.edu/abs/2017ApJ...848...78F} {848, 78}

\bibitem[\protect\citeauthoryear{{Farina} et~al.,}{{Farina}
  et~al.}{2019}]{Farina2019}
{Farina} E.~P.,  et~al., 2019, \mn@doi [\apj] {10.3847/1538-4357/ab5847}, \href
  {https://ui.adsabs.harvard.edu/abs/2019ApJ...887..196F} {887, 196}

\bibitem[\protect\citeauthoryear{{Fernandez-Figueroa}
  et~al.,}{{Fernandez-Figueroa} et~al.}{2022}]{Fernandez2022}
{Fernandez-Figueroa} A.,  et~al., 2022, \mn@doi [\mnras]
  {10.1093/mnras/stac2851}, \href
  {https://ui.adsabs.harvard.edu/abs/2022MNRAS.517.2214F} {517, 2214}

\bibitem[\protect\citeauthoryear{{Fielding} et~al.,}{{Fielding}
  et~al.}{2020}]{Fielding2020}
{Fielding} D.~B.,  et~al., 2020, \mn@doi [\apj] {10.3847/1538-4357/abbc6d},
  \href {https://ui.adsabs.harvard.edu/abs/2020ApJ...903...32F} {903, 32}

\bibitem[\protect\citeauthoryear{{Fillmore} \& {Goldreich}}{{Fillmore} \&
  {Goldreich}}{1984}]{Fillmore1984}
{Fillmore} J.~A.,  {Goldreich} P.,  1984, \mn@doi [\apj] {10.1086/162070},
  \href {https://ui.adsabs.harvard.edu/abs/1984ApJ...281....1F} {281, 1}

\bibitem[\protect\citeauthoryear{{Font-Ribera} et~al.,}{{Font-Ribera}
  et~al.}{2013}]{FontRibera2013}
{Font-Ribera} A.,  et~al., 2013, \mn@doi [\jcap]
  {10.1088/1475-7516/2013/05/018}, \href
  {https://ui.adsabs.harvard.edu/abs/2013JCAP...05..018F} {2013, 018}

\bibitem[\protect\citeauthoryear{{Fossati} et~al.,}{{Fossati}
  et~al.}{2021}]{Fossati2021}
{Fossati} M.,  et~al., 2021, \mn@doi [\mnras] {10.1093/mnras/stab660}, \href
  {https://ui.adsabs.harvard.edu/abs/2021MNRAS.503.3044F} {503, 3044}

\bibitem[\protect\citeauthoryear{{Fumagalli}, {O'Meara}, {Prochaska}  \&
  {Worseck}}{{Fumagalli} et~al.}{2013}]{Fumagalli2013}
{Fumagalli} M.,  {O'Meara} J.~M.,  {Prochaska} J.~X.,   {Worseck} G.,  2013,
  \mn@doi [\apj] {10.1088/0004-637X/775/1/78}, \href
  {https://ui.adsabs.harvard.edu/abs/2013ApJ...775...78F} {775, 78}

\bibitem[\protect\citeauthoryear{{Garc{\'\i}a-Vergara}, {Hennawi}, {Barrientos}
   \& {Rix}}{{Garc{\'\i}a-Vergara} et~al.}{2017}]{garcia2017}
{Garc{\'\i}a-Vergara} C.,  {Hennawi} J.~F.,  {Barrientos} L.~F.,   {Rix} H.-W.,
   2017, \mn@doi [\apj] {10.3847/1538-4357/aa8b69}, \href
  {https://ui.adsabs.harvard.edu/abs/2017ApJ...848....7G} {848, 7}

\bibitem[\protect\citeauthoryear{{Gronke}, {Oh}, {Ji}  \& {Norman}}{{Gronke}
  et~al.}{2022}]{Gronke2022}
{Gronke} M.,  {Oh} S.~P.,  {Ji} S.,   {Norman} C.,  2022, \mn@doi [\mnras]
  {10.1093/mnras/stab3351}, \href
  {https://ui.adsabs.harvard.edu/abs/2022MNRAS.511..859G} {511, 859}

\bibitem[\protect\citeauthoryear{{Guo} et~al.,}{{Guo} et~al.}{2020}]{Guo2020}
{Guo} Y.,  et~al., 2020, \mn@doi [\apj] {10.3847/1538-4357/ab9b7f}, \href
  {https://ui.adsabs.harvard.edu/abs/2020ApJ...898...26G} {898, 26}

\bibitem[\protect\citeauthoryear{{Haiman} \& {Rees}}{{Haiman} \&
  {Rees}}{2001}]{Haiman2001}
{Haiman} Z.,  {Rees} M.~J.,  2001, \mn@doi [\apj] {10.1086/321567}, \href
  {https://ui.adsabs.harvard.edu/abs/2001ApJ...556...87H} {556, 87}

\bibitem[\protect\citeauthoryear{{Haiman}, {Spaans}  \& {Quataert}}{{Haiman}
  et~al.}{2000}]{Haiman2000}
{Haiman} Z.,  {Spaans} M.,   {Quataert} E.,  2000, \mn@doi [\apjl]
  {10.1086/312754}, \href
  {https://ui.adsabs.harvard.edu/abs/2000ApJ...537L...5H} {537, L5}

\bibitem[\protect\citeauthoryear{He et~al.,}{He et~al.}{2017}]{He2017}
He W.,  et~al., 2017, \mn@doi [Publications of the Astronomical Society of
  Japan] {10.1093/pasj/psx129}, 70

\bibitem[\protect\citeauthoryear{{Hennawi} \& {Prochaska}}{{Hennawi} \&
  {Prochaska}}{2013}]{Hennawi2013}
{Hennawi} J.~F.,  {Prochaska} J.~X.,  2013, \mn@doi [\apj]
  {10.1088/0004-637X/766/1/58}, \href
  {https://ui.adsabs.harvard.edu/abs/2013ApJ...766...58H} {766, 58}

\bibitem[\protect\citeauthoryear{{Hennawi} et~al.,}{{Hennawi}
  et~al.}{2006}]{Hennawi2006}
{Hennawi} J.~F.,  et~al., 2006, \mn@doi [\apj] {10.1086/507069}, \href
  {https://ui.adsabs.harvard.edu/abs/2006ApJ...651...61H} {651, 61}

\bibitem[\protect\citeauthoryear{{Hui} \& {Gnedin}}{{Hui} \&
  {Gnedin}}{1997}]{HuiGnedin1997}
{Hui} L.,  {Gnedin} N.~Y.,  1997, \mn@doi [\mnras] {10.1093/mnras/292.1.27},
  \href {https://ui.adsabs.harvard.edu/abs/1997MNRAS.292...27H} {292, 27}

\bibitem[\protect\citeauthoryear{{Hummels} et~al.,}{{Hummels}
  et~al.}{2019}]{Hummels2019}
{Hummels} C.~B.,  et~al., 2019, \mn@doi [\apj] {10.3847/1538-4357/ab378f},
  \href {https://ui.adsabs.harvard.edu/abs/2019ApJ...882..156H} {882, 156}

\bibitem[\protect\citeauthoryear{{Huscher}, {Oppenheimer}, {Lonardi}, {Crain},
  {Richings}  \& {Schaye}}{{Huscher} et~al.}{2021}]{Huscher2021}
{Huscher} E.,  {Oppenheimer} B.~D.,  {Lonardi} A.,  {Crain} R.~A.,  {Richings}
  A.~J.,   {Schaye} J.,  2021, \mn@doi [\mnras] {10.1093/mnras/staa3203}, \href
  {https://ui.adsabs.harvard.edu/abs/2021MNRAS.500.1476H} {500, 1476}

\bibitem[\protect\citeauthoryear{{Joung}, {Bryan}  \& {Putman}}{{Joung}
  et~al.}{2012}]{Joung2012}
{Joung} M.~R.,  {Bryan} G.~L.,   {Putman} M.~E.,  2012, \mn@doi [\apj]
  {10.1088/0004-637X/745/2/148}, \href
  {https://ui.adsabs.harvard.edu/abs/2012ApJ...745..148J} {745, 148}

\bibitem[\protect\citeauthoryear{{Kere{\v{s}}}, {Katz}, {Weinberg}  \&
  {Dav{\'e}}}{{Kere{\v{s}}} et~al.}{2005}]{Keres2005}
{Kere{\v{s}}} D.,  {Katz} N.,  {Weinberg} D.~H.,   {Dav{\'e}} R.,  2005,
  \mn@doi [\mnras] {10.1111/j.1365-2966.2005.09451.x}, \href
  {https://ui.adsabs.harvard.edu/abs/2005MNRAS.363....2K} {363, 2}

\bibitem[\protect\citeauthoryear{{Kere{\v{s}}}, {Katz}, {Fardal}, {Dav{\'e}}
  \& {Weinberg}}{{Kere{\v{s}}} et~al.}{2009}]{Keres2009}
{Kere{\v{s}}} D.,  {Katz} N.,  {Fardal} M.,  {Dav{\'e}} R.,   {Weinberg} D.~H.,
   2009, \mn@doi [\mnras] {10.1111/j.1365-2966.2009.14541.x}, \href
  {https://ui.adsabs.harvard.edu/abs/2009MNRAS.395..160K} {395, 160}

\bibitem[\protect\citeauthoryear{{Kulier}, {Padilla}, {Schaye}, {Crain},
  {Schaller}, {Bower}, {Theuns}  \& {Paillas}}{{Kulier}
  et~al.}{2019}]{Kulier2019}
{Kulier} A.,  {Padilla} N.,  {Schaye} J.,  {Crain} R.~A.,  {Schaller} M.,
  {Bower} R.~G.,  {Theuns} T.,   {Paillas} E.,  2019, \mn@doi [\mnras]
  {10.1093/mnras/sty2914}, \href
  {https://ui.adsabs.harvard.edu/abs/2019MNRAS.482.3261K} {482, 3261}

\bibitem[\protect\citeauthoryear{{Langen}, {Cantalupo}, {Steidel}, {Chen},
  {Pezzulli}  \& {Gallego}}{{Langen} et~al.}{2023}]{Langen2022}
{Langen} V.,  {Cantalupo} S.,  {Steidel} C.~C.,  {Chen} Y.,  {Pezzulli} G.,
  {Gallego} S.~G.,  2023, \mn@doi [\mnras] {10.1093/mnras/stac3205}, \href
  {https://ui.adsabs.harvard.edu/abs/2023MNRAS.519.5099L} {519, 5099}

\bibitem[\protect\citeauthoryear{{Lau}, {Prochaska}  \& {Hennawi}}{{Lau}
  et~al.}{2018}]{Lau2018}
{Lau} M.~W.,  {Prochaska} J.~X.,   {Hennawi} J.~F.,  2018, \mn@doi [\apj]
  {10.3847/1538-4357/aab78e}, \href
  {https://ui.adsabs.harvard.edu/abs/2018ApJ...857..126L} {857, 126}

\bibitem[\protect\citeauthoryear{{Leclercq} et~al.,}{{Leclercq}
  et~al.}{2017}]{Leclercq2017}
{Leclercq} F.,  et~al., 2017, \mn@doi [\aap] {10.1051/0004-6361/201731480},
  \href {https://ui.adsabs.harvard.edu/abs/2017A&A...608A...8L} {608, A8}

\bibitem[\protect\citeauthoryear{{Leibler}, {Cantalupo}, {Holden}  \&
  {Madau}}{{Leibler} et~al.}{2018}]{Leibler2018}
{Leibler} C.~N.,  {Cantalupo} S.,  {Holden} B.~P.,   {Madau} P.,  2018, \mn@doi
  [\mnras] {10.1093/mnras/sty1764}, \href
  {https://ui.adsabs.harvard.edu/abs/2018MNRAS.480.2094L} {480, 2094}

\bibitem[\protect\citeauthoryear{{Li} \& {Tonnesen}}{{Li} \&
  {Tonnesen}}{2020}]{Li2020}
{Li} M.,  {Tonnesen} S.,  2020, \mn@doi [\apj] {10.3847/1538-4357/ab9f9f},
  \href {https://ui.adsabs.harvard.edu/abs/2020ApJ...898..148L} {898, 148}

\bibitem[\protect\citeauthoryear{{Lofthouse} et~al.,}{{Lofthouse}
  et~al.}{2020}]{Lofthouse2020}
{Lofthouse} E.~K.,  et~al., 2020, \mn@doi [\mnras] {10.1093/mnras/stz3066},
  \href {https://ui.adsabs.harvard.edu/abs/2020MNRAS.491.2057L} {491, 2057}

\bibitem[\protect\citeauthoryear{{Lofthouse} et~al.,}{{Lofthouse}
  et~al.}{2023}]{Lofthouse2023}
{Lofthouse} E.~K.,  et~al., 2023, \mn@doi [\mnras] {10.1093/mnras/stac3089},
  \href {https://ui.adsabs.harvard.edu/abs/2023MNRAS.518..305L} {518, 305}

\bibitem[\protect\citeauthoryear{{Lopez} et~al.,}{{Lopez}
  et~al.}{2018}]{Lopez2018}
{Lopez} S.,  et~al., 2018, \mn@doi [\nat] {10.1038/nature25436}, \href
  {https://ui.adsabs.harvard.edu/abs/2018Natur.554..493L} {554, 493}

\bibitem[\protect\citeauthoryear{{Lusso}, {Worseck}, {Hennawi}, {Prochaska},
  {Vignali}, {Stern}  \& {O'Meara}}{{Lusso} et~al.}{2015}]{Lusso2015}
{Lusso} E.,  {Worseck} G.,  {Hennawi} J.~F.,  {Prochaska} J.~X.,  {Vignali} C.,
   {Stern} J.,   {O'Meara} J.~M.,  2015, \mn@doi [\mnras]
  {10.1093/mnras/stv516}, \href
  {https://ui.adsabs.harvard.edu/abs/2015MNRAS.449.4204L} {449, 4204}

\bibitem[\protect\citeauthoryear{{Mackenzie} et~al.,}{{Mackenzie}
  et~al.}{2021}]{Mackenzie2021}
{Mackenzie} R.,  et~al., 2021, \mn@doi [\mnras] {10.1093/mnras/staa3277}, \href
  {https://ui.adsabs.harvard.edu/abs/2021MNRAS.502..494M} {502, 494}

\bibitem[\protect\citeauthoryear{{Mandelker}, {Padnos}, {Dekel}, {Birnboim},
  {Burkert}, {Krumholz}  \& {Steinberg}}{{Mandelker}
  et~al.}{2016}]{Mandelker2016}
{Mandelker} N.,  {Padnos} D.,  {Dekel} A.,  {Birnboim} Y.,  {Burkert} A.,
  {Krumholz} M.~R.,   {Steinberg} E.,  2016, \mn@doi [\mnras]
  {10.1093/mnras/stw2267}, \href
  {https://ui.adsabs.harvard.edu/abs/2016MNRAS.463.3921M} {463, 3921}

\bibitem[\protect\citeauthoryear{{Mandelker}, {van Dokkum}, {Brodie}, {van den
  Bosch}  \& {Ceverino}}{{Mandelker} et~al.}{2018}]{Mandelker2018}
{Mandelker} N.,  {van Dokkum} P.~G.,  {Brodie} J.~P.,  {van den Bosch} F.~C.,
  {Ceverino} D.,  2018, \mn@doi [\apj] {10.3847/1538-4357/aaca98}, \href
  {https://ui.adsabs.harvard.edu/abs/2018ApJ...861..148M} {861, 148}

\bibitem[\protect\citeauthoryear{{Mandelker}, {Nagai}, {Aung}, {Dekel},
  {Padnos}  \& {Birnboim}}{{Mandelker} et~al.}{2019}]{Mandelker2019}
{Mandelker} N.,  {Nagai} D.,  {Aung} H.,  {Dekel} A.,  {Padnos} D.,
  {Birnboim} Y.,  2019, \mn@doi [\mnras] {10.1093/mnras/stz012}, \href
  {https://ui.adsabs.harvard.edu/abs/2019MNRAS.484.1100M} {484, 1100}

\bibitem[\protect\citeauthoryear{{Marino} et~al.,}{{Marino}
  et~al.}{2018}]{Marino2018}
{Marino} R.~A.,  et~al., 2018, \mn@doi [\apj] {10.3847/1538-4357/aab6aa}, \href
  {https://ui.adsabs.harvard.edu/abs/2018ApJ...859...53M} {859, 53}

\bibitem[\protect\citeauthoryear{{Marino} et~al.,}{{Marino}
  et~al.}{2019}]{Marino2019}
{Marino} R.~A.,  et~al., 2019, \mn@doi [\apj] {10.3847/1538-4357/ab2881}, \href
  {https://ui.adsabs.harvard.edu/abs/2019ApJ...880...47M} {880, 47}

\bibitem[\protect\citeauthoryear{{Martin}, {Moore}, {Morrissey}, {Matuszewski},
  {Rahman}, {Adkins}  \& {Epps}}{{Martin} et~al.}{2010}]{Martin2010}
{Martin} C.,  {Moore} A.,  {Morrissey} P.,  {Matuszewski} M.,  {Rahman} S.,
  {Adkins} S.,   {Epps} H.,  2010, in {McLean} I.~S.,  {Ramsay} S.~K.,
  {Takami} H.,  eds,  Society of Photo-Optical Instrumentation Engineers (SPIE)
  Conference Series Vol. 7735, Ground-based and Airborne Instrumentation for
  Astronomy III. p. 77350M, \mn@doi{10.1117/12.858227}

\bibitem[\protect\citeauthoryear{{McAlpine} et~al.,}{{McAlpine}
  et~al.}{2016}]{McAlpine_2016}
{McAlpine} S.,  et~al., 2016, \mn@doi [Astronomy and Computing]
  {10.1016/j.ascom.2016.02.004}, \href
  {https://ui.adsabs.harvard.edu/abs/2016A&C....15...72M} {15, 72}

\bibitem[\protect\citeauthoryear{{Meiksin}}{{Meiksin}}{2009}]{2009Meiksin}
{Meiksin} A.~A.,  2009, \mn@doi [Reviews of Modern Physics]
  {10.1103/RevModPhys.81.1405}, \href
  {https://ui.adsabs.harvard.edu/abs/2009RvMP...81.1405M} {81, 1405}

\bibitem[\protect\citeauthoryear{{Monier}, {Turnshek}  \& {Lupie}}{{Monier}
  et~al.}{1998}]{Monier1998}
{Monier} E.~M.,  {Turnshek} D.~A.,   {Lupie} O.~L.,  1998, \mn@doi [\apj]
  {10.1086/305372}, \href
  {https://ui.adsabs.harvard.edu/abs/1998ApJ...496..177M} {496, 177}

\bibitem[\protect\citeauthoryear{{More}, {Diemer}  \& {Kravtsov}}{{More}
  et~al.}{2015}]{More2015}
{More} S.,  {Diemer} B.,   {Kravtsov} A.~V.,  2015, \mn@doi [\apj]
  {10.1088/0004-637X/810/1/36}, \href
  {https://ui.adsabs.harvard.edu/abs/2015ApJ...810...36M} {810, 36}

\bibitem[\protect\citeauthoryear{{Morrissey} et~al.,}{{Morrissey}
  et~al.}{2018}]{Morrissey2018}
{Morrissey} P.,  et~al., 2018, \mn@doi [\apj] {10.3847/1538-4357/aad597}, \href
  {https://ui.adsabs.harvard.edu/abs/2018ApJ...864...93M} {864, 93}

\bibitem[\protect\citeauthoryear{{Mortensen}, {Keerthi Vasan}, {Jones},
  {Faucher-Gigu{\`e}re}, {Sanders}, {Ellis}, {Leethochawalit}  \&
  {Stark}}{{Mortensen} et~al.}{2021}]{Mortensen2021}
{Mortensen} K.,  {Keerthi Vasan} G.~C.,  {Jones} T.,  {Faucher-Gigu{\`e}re}
  C.-A.,  {Sanders} R.~L.,  {Ellis} R.~S.,  {Leethochawalit} N.,   {Stark}
  D.~P.,  2021, \mn@doi [\apj] {10.3847/1538-4357/abfa11}, \href
  {https://ui.adsabs.harvard.edu/abs/2021ApJ...914...92M} {914, 92}

\bibitem[\protect\citeauthoryear{{Navarro}, {Frenk}  \& {White}}{{Navarro}
  et~al.}{1997}]{NFW1997}
{Navarro} J.~F.,  {Frenk} C.~S.,   {White} S. D.~M.,  1997, \mn@doi [\apj]
  {10.1086/304888}, \href
  {https://ui.adsabs.harvard.edu/abs/1997ApJ...490..493N} {490, 493}

\bibitem[\protect\citeauthoryear{{Nelson}, {Genel}, {Pillepich},
  {Vogelsberger}, {Springel}  \& {Hernquist}}{{Nelson}
  et~al.}{2016}]{Nelson2016}
{Nelson} D.,  {Genel} S.,  {Pillepich} A.,  {Vogelsberger} M.,  {Springel} V.,
   {Hernquist} L.,  2016, \mn@doi [\mnras] {10.1093/mnras/stw1191}, \href
  {https://ui.adsabs.harvard.edu/abs/2016MNRAS.460.2881N} {460, 2881}

\bibitem[\protect\citeauthoryear{{Osterbrock} \& {Ferland}}{{Osterbrock} \&
  {Ferland}}{2006}]{Osterbrock2006}
{Osterbrock} D.~E.,  {Ferland} G.~J.,  2006, {Astrophysics of gaseous nebulae
  and active galactic nuclei}

\bibitem[\protect\citeauthoryear{{Peeples} et~al.,}{{Peeples}
  et~al.}{2019}]{Peeples2019}
{Peeples} M.~S.,  et~al., 2019, \mn@doi [\apj] {10.3847/1538-4357/ab0654},
  \href {https://ui.adsabs.harvard.edu/abs/2019ApJ...873..129P} {873, 129}

\bibitem[\protect\citeauthoryear{{Pezzulli} \& {Cantalupo}}{{Pezzulli} \&
  {Cantalupo}}{2019}]{PezzulliCantalupo2019}
{Pezzulli} G.,  {Cantalupo} S.,  2019, \mn@doi [\mnras] {10.1093/mnras/stz906},
  \href {https://ui.adsabs.harvard.edu/abs/2019MNRAS.486.1489P} {486, 1489}

\bibitem[\protect\citeauthoryear{{Planck Collaboration} et~al.,}{{Planck
  Collaboration} et~al.}{2014}]{Planck2014}
{Planck Collaboration} et~al., 2014, \mn@doi [\aap]
  {10.1051/0004-6361/201321529}, \href
  {https://ui.adsabs.harvard.edu/abs/2014A&A...571A...1P} {571, A1}

\bibitem[\protect\citeauthoryear{{Prochaska} et~al.,}{{Prochaska}
  et~al.}{2013}]{Prochaska2013}
{Prochaska} J.~X.,  et~al., 2013, \mn@doi [\apj] {10.1088/0004-637X/776/2/136},
  \href {https://ui.adsabs.harvard.edu/abs/2013ApJ...776..136P} {776, 136}

\bibitem[\protect\citeauthoryear{{Rahmati}, {Schaye}, {Bower}, {Crain},
  {Furlong}, {Schaller}  \& {Theuns}}{{Rahmati} et~al.}{2015}]{Rahmati2015}
{Rahmati} A.,  {Schaye} J.,  {Bower} R.~G.,  {Crain} R.~A.,  {Furlong} M.,
  {Schaller} M.,   {Theuns} T.,  2015, \mn@doi [\mnras]
  {10.1093/mnras/stv1414}, \href
  {https://ui.adsabs.harvard.edu/abs/2015MNRAS.452.2034R} {452, 2034}

\bibitem[\protect\citeauthoryear{{Rauch}, {Sargent}, {Barlow}  \&
  {Carswell}}{{Rauch} et~al.}{2001}]{Rauch2001}
{Rauch} M.,  {Sargent} W. L.~W.,  {Barlow} T.~A.,   {Carswell} R.~F.,  2001,
  \mn@doi [\apj] {10.1086/323523}, \href
  {https://ui.adsabs.harvard.edu/abs/2001ApJ...562...76R} {562, 76}

\bibitem[\protect\citeauthoryear{{Rees} \& {Ostriker}}{{Rees} \&
  {Ostriker}}{1977}]{Rees1977}
{Rees} M.~J.,  {Ostriker} J.~P.,  1977, \mn@doi [\mnras]
  {10.1093/mnras/179.4.541}, \href
  {https://ui.adsabs.harvard.edu/abs/1977MNRAS.179..541R} {179, 541}

\bibitem[\protect\citeauthoryear{{Rosdahl} \& {Blaizot}}{{Rosdahl} \&
  {Blaizot}}{2012}]{Rosdahl2012}
{Rosdahl} J.,  {Blaizot} J.,  2012, \mn@doi [\mnras]
  {10.1111/j.1365-2966.2012.20883.x}, \href
  {https://ui.adsabs.harvard.edu/abs/2012MNRAS.423..344R} {423, 344}

\bibitem[\protect\citeauthoryear{{Rubin}, {Prochaska}, {Koo}, {Phillips}  \&
  {Weiner}}{{Rubin} et~al.}{2010}]{Rubin2010}
{Rubin} K. H.~R.,  {Prochaska} J.~X.,  {Koo} D.~C.,  {Phillips} A.~C.,
  {Weiner} B.~J.,  2010, \mn@doi [\apj] {10.1088/0004-637X/712/1/574}, \href
  {https://ui.adsabs.harvard.edu/abs/2010ApJ...712..574R} {712, 574}

\bibitem[\protect\citeauthoryear{{Rubin} et~al.,}{{Rubin}
  et~al.}{2018}]{Rubin2018}
{Rubin} K. H.~R.,  et~al., 2018, \mn@doi [\apj] {10.3847/1538-4357/aaaeb7},
  \href {https://ui.adsabs.harvard.edu/abs/2018ApJ...859..146R} {859, 146}

\bibitem[\protect\citeauthoryear{{Rudie} et~al.,}{{Rudie}
  et~al.}{2012}]{Rudie2012}
{Rudie} G.~C.,  et~al., 2012, \mn@doi [\apj] {10.1088/0004-637X/750/1/67},
  \href {https://ui.adsabs.harvard.edu/abs/2012ApJ...750...67R} {750, 67}

\bibitem[\protect\citeauthoryear{{Schaye} et~al.,}{{Schaye}
  et~al.}{2015}]{Schaye2015}
{Schaye} J.,  et~al., 2015, \mn@doi [\mnras] {10.1093/mnras/stu2058}, \href
  {https://ui.adsabs.harvard.edu/abs/2015MNRAS.446..521S} {446, 521}

\bibitem[\protect\citeauthoryear{{Shen} et~al.,}{{Shen}
  et~al.}{2007}]{Shen2007}
{Shen} Y.,  et~al., 2007, \mn@doi [\aj] {10.1086/513517}, \href
  {https://ui.adsabs.harvard.edu/abs/2007AJ....133.2222S} {133, 2222}

\bibitem[\protect\citeauthoryear{{Shen} et~al.,}{{Shen}
  et~al.}{2009}]{Shen2009}
{Shen} Y.,  et~al., 2009, \mn@doi [\apj] {10.1088/0004-637X/697/2/1656}, \href
  {https://ui.adsabs.harvard.edu/abs/2009ApJ...697.1656S} {697, 1656}

\bibitem[\protect\citeauthoryear{{Silk}}{{Silk}}{1977}]{Silk1977}
{Silk} J.,  1977, \mn@doi [\apj] {10.1086/154972}, \href
  {https://ui.adsabs.harvard.edu/abs/1977ApJ...211..638S} {211, 638}

\bibitem[\protect\citeauthoryear{{Silva} et~al.,}{{Silva}
  et~al.}{2018}]{Silva2018}
{Silva} M.,  et~al., 2018, \mn@doi [\mnras] {10.1093/mnras/stx3019}, \href
  {https://ui.adsabs.harvard.edu/abs/2018MNRAS.474.3649S} {474, 3649}

\bibitem[\protect\citeauthoryear{{Smette}, {Surdej}, {Shaver}, {Foltz},
  {Chaffee}, {Weymann}, {Williams}  \& {Magain}}{{Smette}
  et~al.}{1992}]{Smette1992}
{Smette} A.,  {Surdej} J.,  {Shaver} P.~A.,  {Foltz} C.~B.,  {Chaffee} F.~H.,
  {Weymann} R.~J.,  {Williams} R.~E.,   {Magain} P.,  1992, \mn@doi [\apj]
  {10.1086/171187}, \href
  {https://ui.adsabs.harvard.edu/abs/1992ApJ...389...39S} {389, 39}

\bibitem[\protect\citeauthoryear{{Sorini}, {Dav{\'e}}  \&
  {Angl{\'e}s-Alc{\'a}zar}}{{Sorini} et~al.}{2020}]{Sorini2020}
{Sorini} D.,  {Dav{\'e}} R.,   {Angl{\'e}s-Alc{\'a}zar} D.,  2020, \mn@doi
  [\mnras] {10.1093/mnras/staa2937}, \href
  {https://ui.adsabs.harvard.edu/abs/2020MNRAS.499.2760S} {499, 2760}

\bibitem[\protect\citeauthoryear{{Springel}}{{Springel}}{2010}]{Springel2010}
{Springel} V.,  2010, \mn@doi [\mnras] {10.1111/j.1365-2966.2009.15715.x},
  \href {https://ui.adsabs.harvard.edu/abs/2010MNRAS.401..791S} {401, 791}

\bibitem[\protect\citeauthoryear{{Springel} et~al.,}{{Springel}
  et~al.}{2005}]{Springel2005}
{Springel} V.,  et~al., 2005, \mn@doi [\nat] {10.1038/nature03597}, \href
  {https://ui.adsabs.harvard.edu/abs/2005Natur.435..629S} {435, 629}

\bibitem[\protect\citeauthoryear{{Steidel}, {Erb}, {Shapley}, {Pettini},
  {Reddy}, {Bogosavljevi{\'c}}, {Rudie}  \& {Rakic}}{{Steidel}
  et~al.}{2010}]{Steidel2010}
{Steidel} C.~C.,  {Erb} D.~K.,  {Shapley} A.~E.,  {Pettini} M.,  {Reddy} N.,
  {Bogosavljevi{\'c}} M.,  {Rudie} G.~C.,   {Rakic} O.,  2010, \mn@doi [\apj]
  {10.1088/0004-637X/717/1/289}, \href
  {https://ui.adsabs.harvard.edu/abs/2010ApJ...717..289S} {717, 289}

\bibitem[\protect\citeauthoryear{{Steidel}, {Bogosavljevi{\'c}}, {Shapley},
  {Kollmeier}, {Reddy}, {Erb}  \& {Pettini}}{{Steidel}
  et~al.}{2011}]{Steidel2011}
{Steidel} C.~C.,  {Bogosavljevi{\'c}} M.,  {Shapley} A.~E.,  {Kollmeier} J.~A.,
   {Reddy} N.~A.,  {Erb} D.~K.,   {Pettini} M.,  2011, \mn@doi [\apj]
  {10.1088/0004-637X/736/2/160}, \href
  {https://ui.adsabs.harvard.edu/abs/2011ApJ...736..160S} {736, 160}

\bibitem[\protect\citeauthoryear{{Steidel} et~al.,}{{Steidel}
  et~al.}{2014}]{Steidel2014}
{Steidel} C.~C.,  et~al., 2014, \mn@doi [\apj] {10.1088/0004-637X/795/2/165},
  \href {https://ui.adsabs.harvard.edu/abs/2014ApJ...795..165S} {795, 165}

\bibitem[\protect\citeauthoryear{{Tejos} et~al.,}{{Tejos}
  et~al.}{2021}]{Tejos2021}
{Tejos} N.,  et~al., 2021, \mn@doi [\mnras] {10.1093/mnras/stab2147}, \href
  {https://ui.adsabs.harvard.edu/abs/2021MNRAS.507..663T} {507, 663}

\bibitem[\protect\citeauthoryear{{Timlin} et~al.,}{{Timlin}
  et~al.}{2018}]{Timlin2018}
{Timlin} J.~D.,  et~al., 2018, \mn@doi [\apj] {10.3847/1538-4357/aab9ac}, \href
  {https://ui.adsabs.harvard.edu/abs/2018ApJ...859...20T} {859, 20}

\bibitem[\protect\citeauthoryear{{Tormen}, {Bouchet}  \& {White}}{{Tormen}
  et~al.}{1997}]{Tormen1997}
{Tormen} G.,  {Bouchet} F.~R.,   {White} S. D.~M.,  1997, \mn@doi [\mnras]
  {10.1093/mnras/286.4.865}, \href
  {https://ui.adsabs.harvard.edu/abs/1997MNRAS.286..865T} {286, 865}

\bibitem[\protect\citeauthoryear{{Trainor} \& {Steidel}}{{Trainor} \&
  {Steidel}}{2012}]{Trainor2012}
{Trainor} R.~F.,  {Steidel} C.~C.,  2012, \mn@doi [\apj]
  {10.1088/0004-637X/752/1/39}, \href
  {https://ui.adsabs.harvard.edu/abs/2012ApJ...752...39T} {752, 39}

\bibitem[\protect\citeauthoryear{{Travascio} et~al.,}{{Travascio}
  et~al.}{2020}]{Travascio2020}
{Travascio} A.,  et~al., 2020, \mn@doi [\aap] {10.1051/0004-6361/201936197},
  \href {https://ui.adsabs.harvard.edu/abs/2020A&A...635A.157T} {635, A157}

\bibitem[\protect\citeauthoryear{{Turner}, {Schaye}, {Steidel}, {Rudie}  \&
  {Strom}}{{Turner} et~al.}{2014}]{Turner2014}
{Turner} M.~L.,  {Schaye} J.,  {Steidel} C.~C.,  {Rudie} G.~C.,   {Strom}
  A.~L.,  2014, \mn@doi [\mnras] {10.1093/mnras/stu1801}, \href
  {https://ui.adsabs.harvard.edu/abs/2014MNRAS.445..794T} {445, 794}

\bibitem[\protect\citeauthoryear{{Umehata} et~al.,}{{Umehata}
  et~al.}{2019}]{umehata}
{Umehata} H.,  et~al., 2019, \mn@doi [Science] {10.1126/science.aaw5949}, \href
  {https://ui.adsabs.harvard.edu/abs/2019Sci...366...97U} {366, 97}

\bibitem[\protect\citeauthoryear{{Villar-Mart{\'\i}n}, {Binette}  \&
  {Fosbury}}{{Villar-Mart{\'\i}n} et~al.}{1999}]{VillarMartin1999}
{Villar-Mart{\'\i}n} M.,  {Binette} L.,   {Fosbury} R.~A.~E.,  1999, \mn@doi
  [\aap] {10.48550/arXiv.astro-ph/9903122}, \href
  {https://ui.adsabs.harvard.edu/abs/1999A&A...346....7V} {346, 7}

\bibitem[\protect\citeauthoryear{{Vossberg}, {Cantalupo}  \&
  {Pezzulli}}{{Vossberg} et~al.}{2019}]{Vossberg2019}
{Vossberg} A.-C.~E.,  {Cantalupo} S.,   {Pezzulli} G.,  2019, \mn@doi [\mnras]
  {10.1093/mnras/stz2276}, \href
  {https://ui.adsabs.harvard.edu/abs/2019MNRAS.489.2130V} {489, 2130}

\bibitem[\protect\citeauthoryear{Wendland}{Wendland}{1995}]{Wendland_1995}
Wendland H.,  1995, \mn@doi [Advances in Computational Mathematics]
  {10.1007/BF02123482}, 4, 389

\bibitem[\protect\citeauthoryear{{White} \& {Frenk}}{{White} \&
  {Frenk}}{1991}]{White1991}
{White} S. D.~M.,  {Frenk} C.~S.,  1991, \mn@doi [\apj] {10.1086/170483}, \href
  {https://ui.adsabs.harvard.edu/abs/1991ApJ...379...52W} {379, 52}

\bibitem[\protect\citeauthoryear{{Wisotzki} et~al.,}{{Wisotzki}
  et~al.}{2016}]{Wisotzki2016}
{Wisotzki} L.,  et~al., 2016, \mn@doi [\aap] {10.1051/0004-6361/201527384},
  \href {https://ui.adsabs.harvard.edu/abs/2016A&A...587A..98W} {587, A98}

\bibitem[\protect\citeauthoryear{{Zahedy}, {Chen}, {Rauch}, {Wilson}  \&
  {Zabludoff}}{{Zahedy} et~al.}{2016}]{Zahedy2016}
{Zahedy} F.~S.,  {Chen} H.-W.,  {Rauch} M.,  {Wilson} M.~L.,   {Zabludoff} A.,
  2016, \mn@doi [\mnras] {10.1093/mnras/stw484}, \href
  {https://ui.adsabs.harvard.edu/abs/2016MNRAS.458.2423Z} {458, 2423}

\bibitem[\protect\citeauthoryear{{Zahedy} et~al.,}{{Zahedy}
  et~al.}{2021}]{Zahedy2021}
{Zahedy} F.~S.,  et~al., 2021, \mn@doi [\mnras] {10.1093/mnras/stab1661}, \href
  {https://ui.adsabs.harvard.edu/abs/2021MNRAS.506..877Z} {506, 877}

\bibitem[\protect\citeauthoryear{{da {\^A}ngela} et~al.,}{{da {\^A}ngela}
  et~al.}{2008}]{daAngela2008}
{da {\^A}ngela} J.,  et~al., 2008, \mn@doi [\mnras]
  {10.1111/j.1365-2966.2007.12552.x}, \href
  {https://ui.adsabs.harvard.edu/abs/2008MNRAS.383..565D} {383, 565}

\bibitem[\protect\citeauthoryear{{den Brok} et~al.,}{{den Brok}
  et~al.}{2020}]{denBrok2020}
{den Brok} J.~S.,  et~al., 2020, \mn@doi [\mnras] {10.1093/mnras/staa1269},
  \href {https://ui.adsabs.harvard.edu/abs/2020MNRAS.495.1874D} {495, 1874}

\bibitem[\protect\citeauthoryear{{van de Voort}, {Schaye}, {Booth}  \& {Dalla
  Vecchia}}{{van de Voort} et~al.}{2011}]{vanDeVoort2011}
{van de Voort} F.,  {Schaye} J.,  {Booth} C.~M.,   {Dalla Vecchia} C.,  2011,
  \mn@doi [\mnras] {10.1111/j.1365-2966.2011.18896.x}, \href
  {https://ui.adsabs.harvard.edu/abs/2011MNRAS.415.2782V} {415, 2782}

\makeatother
\end{thebibliography}



\appendix
\section{Effect of Star-Formation density threshold on \texteta} \label{app:effectSFden}
\begin{figure}
	\centering
	\includegraphics[width=\columnwidth]{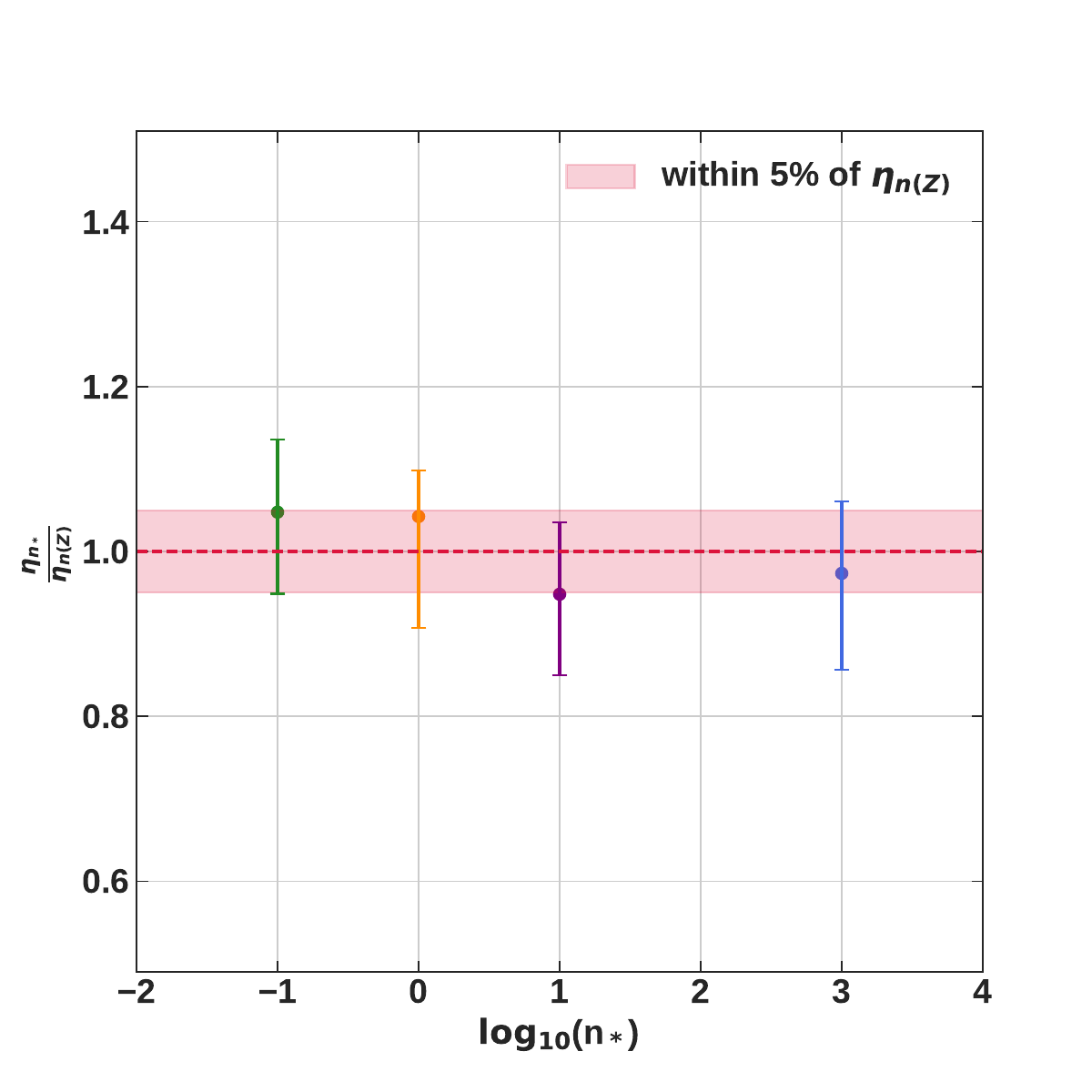}
    \caption{The median ratios between $\eta^{140-200}_{40-100}$ calculated for Ly$\alpha$ nebulae with a fixed star-formation density threshold and $\eta^{140-200}_{40-100}$ using a metallicity dependent star-formation density threshold. The dotted line marks the ratio 1.0 and the red shaded area marks the region between 0.95 and 1.05. The points mark the median ratios for n$_*$ = 0.1 cm$^{-3}$ (green),  n$_*$ = 1 cm$^{-3}$ (orange),  n$_*$ = 10 cm$^{-3}$ (purple) \&  n$_*$ = 10$^3$ cm$^{-3}$ (blue), with the error bars marking the extent of the 25th and 75th percentiles.}
    \label{fig:etaRatioSFRden}
\end{figure}

As explained in Section \ref{subsec:CosmSims}, the gas in the multi-phase ISM is not resolved in cosmological simulations. The EAGLE and ENGINE simulations deal with this by allowing the gas above a metallicity dependent density threshold (see Equation \ref{eq:SFthr}) to be star-forming and deriving its properties from an effective equation of state and not from its hydrodynamics. To check whether our empirical relation depends on the star-formation density threshold chosen, we re-generate the mock cubes for a random subset of halos at $z\sim3.5$ in the mass bin $10^{12} \mathrm{M}_{\odot}$ - $10^{12.4} \mathrm{M}_{\odot}$ using various star-formation density thresholds. We find that the median $\eta^{140-200}_{40-100}$ of the Ly$\alpha$ nebulae generated using the star-formation density thresholds 0.1, 1, 10 and 10$^3$ cm$^{-3}$ are all within 5 \% of the median $\eta^{140-200}_{40-100}$ of the Ly$\alpha$ nebulae generated using the metallicity dependent star-formation density threshold, as demonstrated in Figure \ref{fig:etaRatioSFRden}. Our results are thus independent of this particular choice, which would be however very relevant for the SB values as will be discussed in detail in Paper II.

\section{Effect of observational noise on the intrinsic velocity dispersion} \label{app:noiseLevel}
\begin{figure}
	\centering
	\includegraphics[width=\columnwidth]{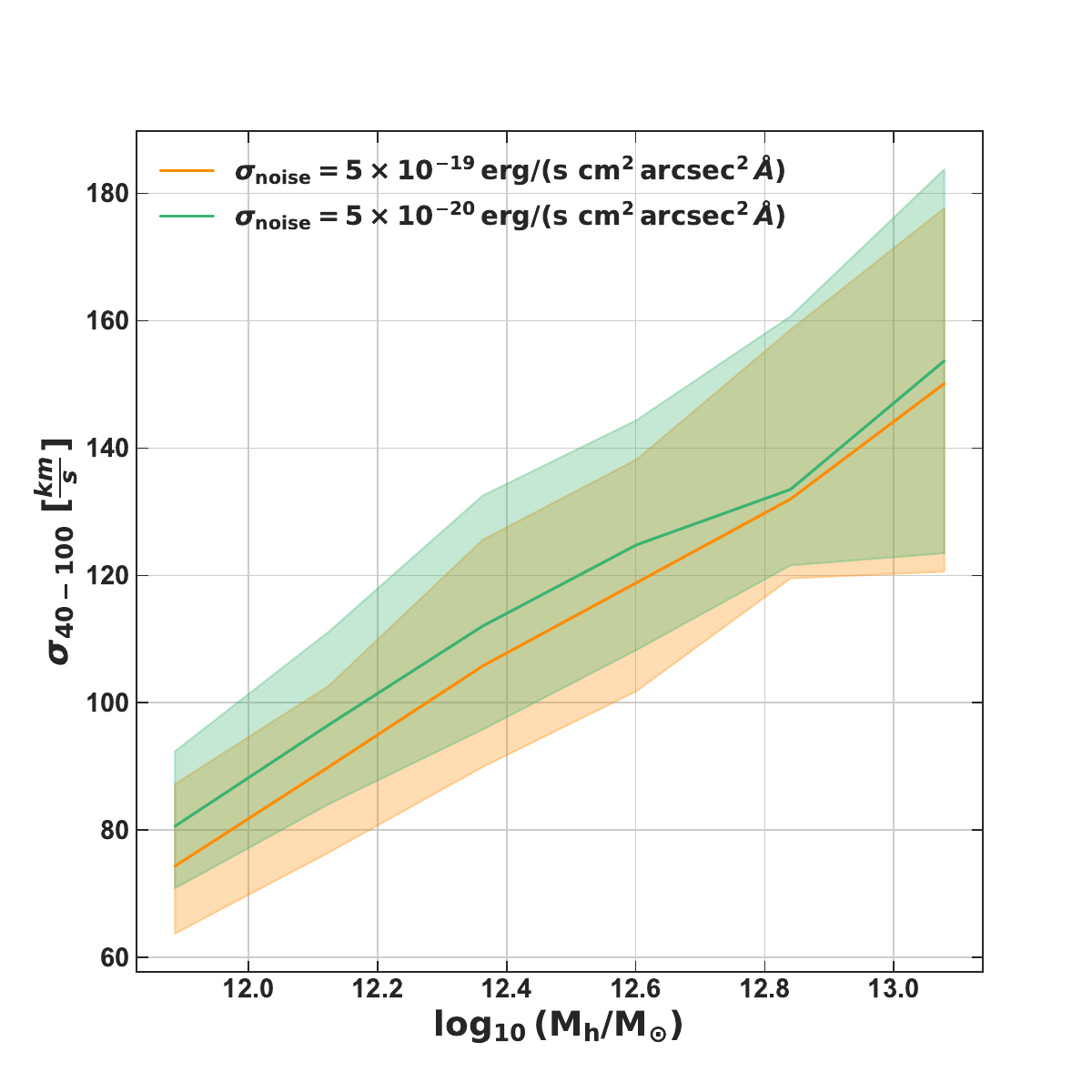}
    \caption{The median intrinsic velocity dispersion in the inner annulus, $\sigma_{40-100}$, as a function of halo mass for all three simulations and both redshifts  ($z\sim 3.5$ and $z\sim 3$) included in this analysis. We calculate $\sigma_{40-100}$ for Ly$\alpha$ nebulae extracted from mock cubes with $\sigma_{\mathrm{noise}} = 5\times 10^{-19}\,\, \mathrm{erg}/(\mathrm{s} \,\, \mathrm{cm}^2 \, \mathrm{arcsec}^2 \, \si{\angstrom})$ (orange line) and $\sigma_{\mathrm{noise}} = 5\times 10^{-20}\,\, \mathrm{erg}/(\mathrm{s} \,\, \mathrm{cm}^2 \, \mathrm{arcsec}^2 \, \si{\angstrom})$ (green line). The shaded regions indicate the 25th and 75th percentiles. Despite $\sigma_{\mathrm{noise}}$ varying by an order of magnitude the two median $\sigma_{40-100}$ are consistent with each other.}
    \label{fig:effectNoise}
\end{figure}

We quantify the effect of varying the amount of noise added to the mock cubes, as described in Section \ref{subsec:mockCubes}, by comparing the median intrinsic velocity dispersion in the inner annulus, $\sigma_{40-100}$, of Ly$\alpha$ nebulae extracted from mock cubes generated using Gaussian noise with a standard deviation of $\sigma_{\mathrm{noise}} = 5\times 10^{-19}\,\, \mathrm{erg}/(\mathrm{s} \,\, \mathrm{cm}^2 \, \mathrm{arcsec}^2 \, \si{\angstrom})$ and $\sigma_{\mathrm{noise}} = 5\times 10^{-20}\,\, \mathrm{erg}/(\mathrm{s} \,\, \mathrm{cm}^2 \, \mathrm{arcsec}^2 \, \si{\angstrom})$. The inner annulus extends from $\sim$40 ckpc to $\sim$100 ckpc. In Figure \ref{fig:effectNoise} we plot the median $\sigma_{40-100}$ of all three simulations and both redshifts ($z\sim 3.5$ \& $z\sim 3$) included in this analysis as a function of halo mass. The orange solid line denotes the median $\sigma_{40-100}$ calculated from mock cubes with $\sigma_{\mathrm{noise}} = 5\times 10^{-19}\,\, \mathrm{erg}/(\mathrm{s} \,\, \mathrm{cm}^2 \, \mathrm{arcsec}^2 \, \si{\angstrom})$ and the  green solid line corresponds to the $\sigma_{\mathrm{noise}}$ used in this analysis: $5\times 10^{-20}\,\, \mathrm{erg}/(\mathrm{s} \,\, \mathrm{cm}^2 \, \mathrm{arcsec}^2 \, \si{\angstrom})$. The shaded area denotes the 25th and 75th percentiles. Despite a slight decrease in $\sigma_{40-100}$ with the higher  $\sigma_{\mathrm{noise}}$, the behaviour of $\sigma_{40-100}$ as a function of halo mass is independent of the noise level chosen and the two median $\sigma_{40-100}$ are consistent with each other, indicating that the behaviour of the intrinsic velocity dispersion is not dominated by the Gaussian noise added to the mock cubes.

\section{Effect of SB normalisation on the intrinsic velocity dispersion} \label{app:HeIIvelDisp}
As discussed in the Introduction and Section \ref{subsec:mockCubes}, the actual value of the Ly$\alpha$ SB depends on several factors, including the sub-grid clumpiness of the medium which is in turn dependent on the simulation's spatial resolution and physics included in the model. On the other hand, some of the observations used here have noise values which would make part of the CGM  undetectable with respect to the mock observations presented in this work. Similarly, the HeII-H$\alpha$ emission is a factor of a few fainter than Ly$\alpha$ emission. How do these SB variations affects the velocity dispersion maps used in this work? 

\begin{figure}
	\centering
	\includegraphics[width=\columnwidth]{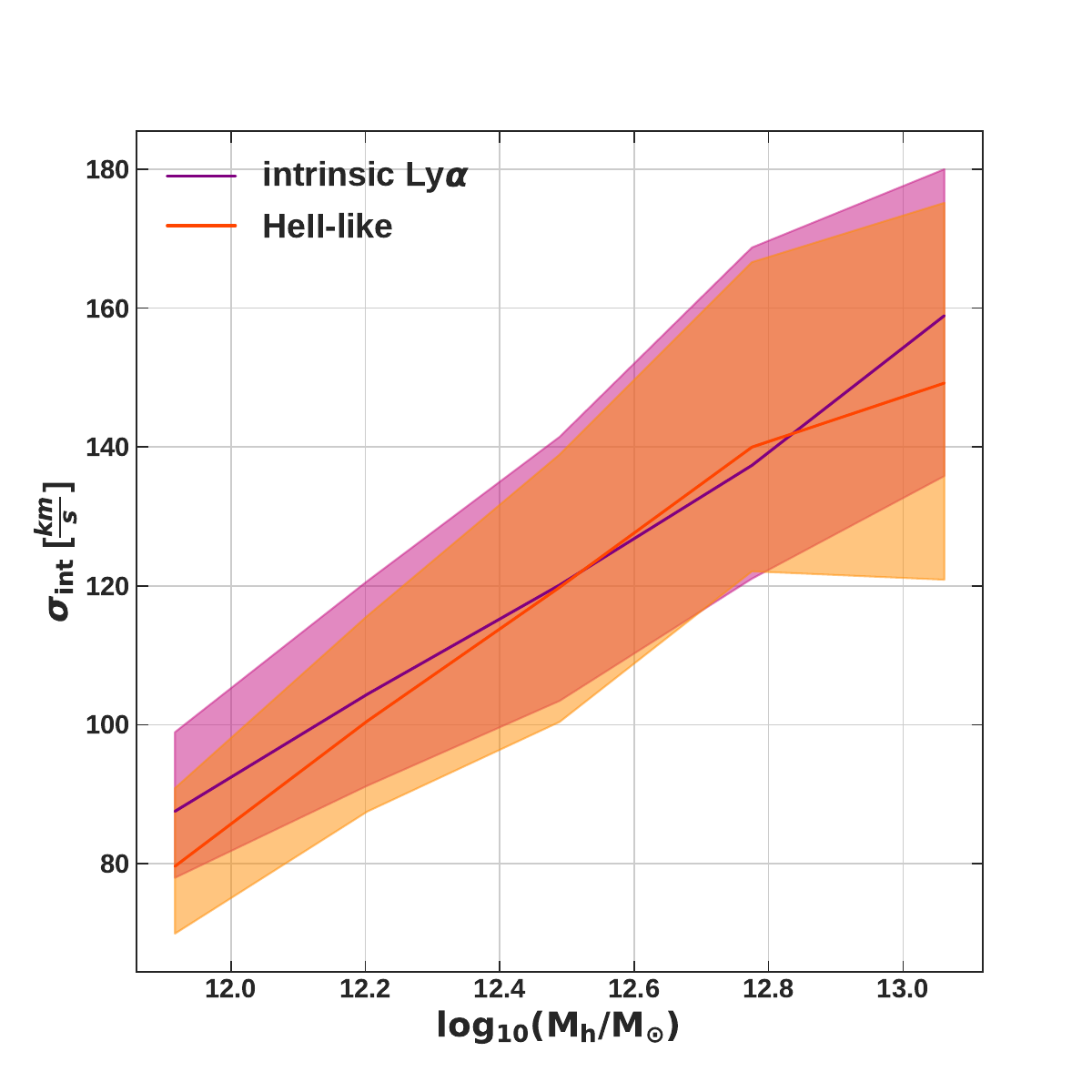}
    \caption{The median intrinsic velocity dispersion in the inner annulus, $\sigma_{40-100}$, with (orange line) and without (purple line) a re-normalisation of the SB values in the mock observations by a factor of 0.2. The shaded areas indicate the extent of the 25th and 75th percentiles. The re-normalisation leads to fainter, ''HeII-like'' nebulae. The good agreement of the two median $\sigma_{40-100}$ as a function of halo mass demonstrates that the intrinsic velocity dispersion is not qualitatively affected by any normalisation of the surface brightness, provided the nebula is still detected with two spectral layers.}
    \label{fig:HeIIvelDisp}
\end{figure}

In order to address this question, we re-generate a subset of the mock observations with a lower SB normalisation. In particular, we produce ``HeII-like'' nebulae in the mock observations by re-normalising the calculated emissivity values by a factor of 0.2 before adding Gaussian noise and applying Gaussian smoothing as detailed in Section \ref{subsec:mockCubes}. We then compare the median intrinsic velocity dispersion in the inner annulus, $\sigma_{40-100}$, of the simulated Ly$\alpha$ nebulae with the $\sigma_{40-100}$ of the ``HeII-like'' nebulae. This comparison is shown in Figure \ref{fig:HeIIvelDisp}, where the median $\sigma_{40-100}$ of the Ly$\alpha$ nebulae as a function of halo mass is plotted with a solid purple line and the median $\sigma_{40-100}$ of the ``HeII-like'' nebulae as a function of halo mass is plotted with a solid orange line. The respective 25th and 75th percentiles are indicated by the shaded areas. The close agreement of the two medians indicates that the behaviour of the intrinsic velocity dispersion is not affected by a re-normalisation of the SB values, provided the nebula is still detected with two spectral layers. 

\section{Effect of AGN-feedback on the Gas Kinematics} \label{app:AGNfeedbackEffect}
\begin{figure}
	\centering
	\includegraphics[width=\columnwidth]{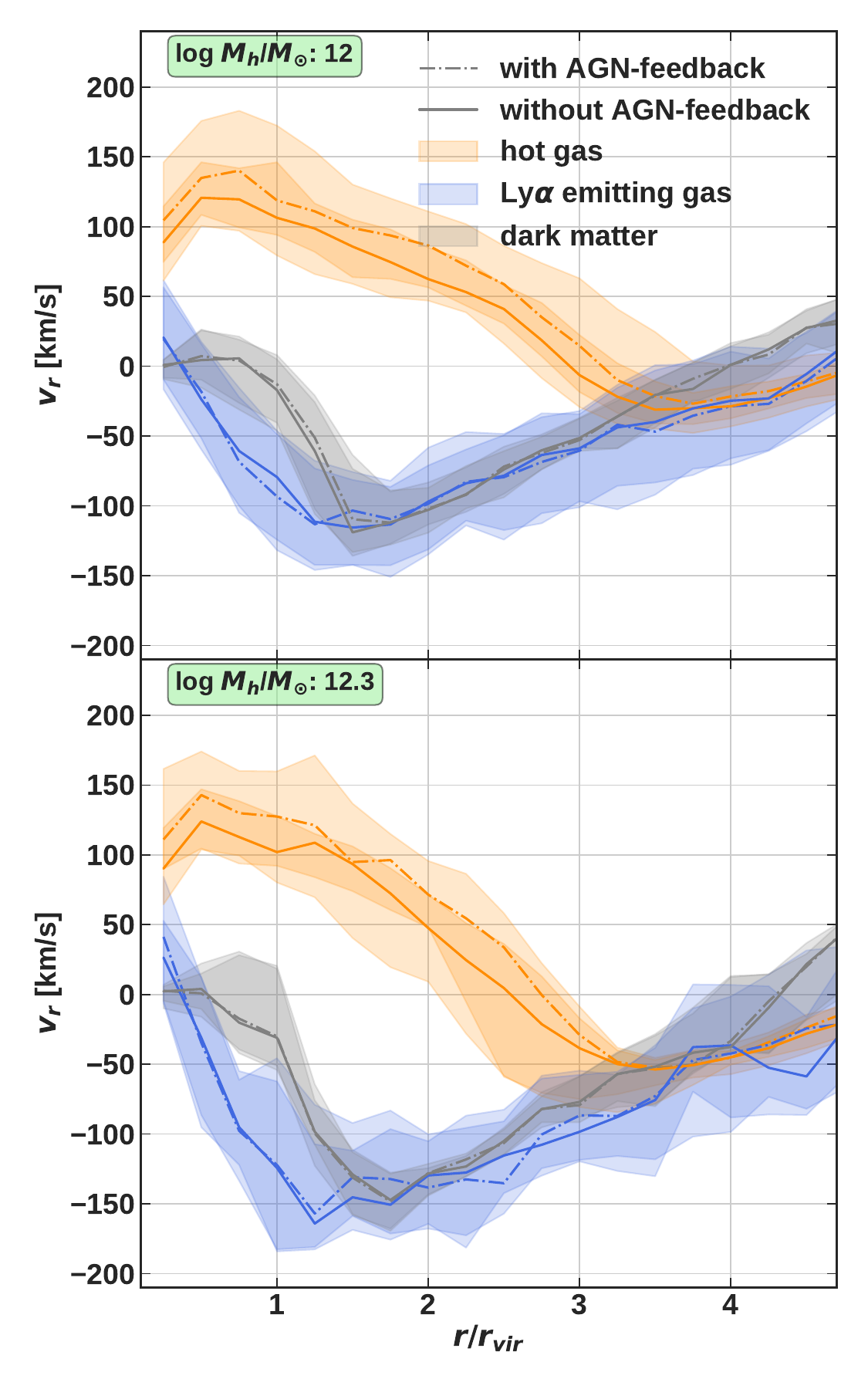}
    \caption{Spherically averaged radial velocity profiles of the dark matter (grey), Ly$\alpha$ emitting gas (blue) and hot ($>10^5K$) gas (orange) with (dash-dotted line) and without (solid line) AGN-feedback at redshift $z\sim 3.5$.}
    \label{fig:radVel_AGNvsNoAGNz3p528}
\end{figure}

\begin{figure}
	\centering
	\includegraphics[width=\columnwidth]{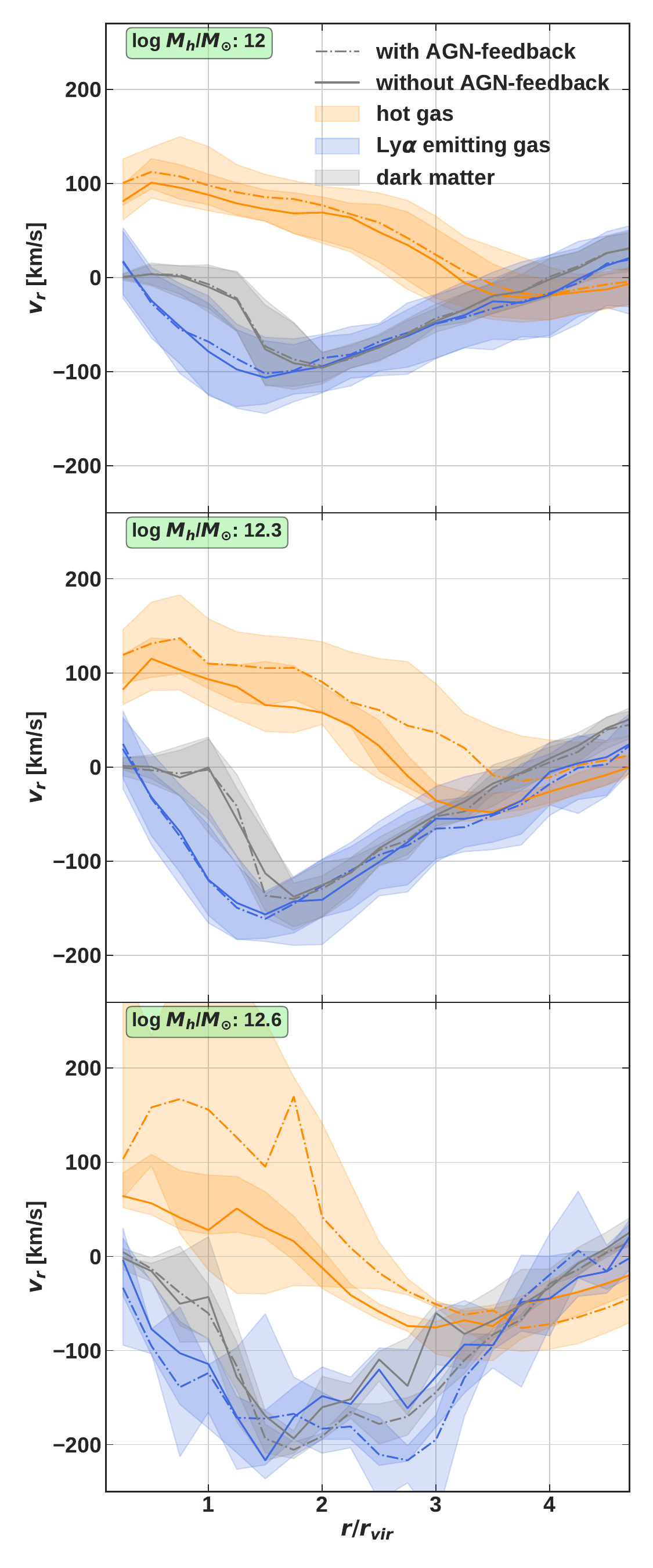}
    \caption{Spherically averaged radial velocity profiles of the dark matter (grey), Ly$\alpha$ emitting gas (blue) and hot ($>10^5K$) gas (orange) with (dash-dotted line) and without (solid line) AGN-feedback at redshift $z\sim 3$.}
    \label{fig:radVel_AGNvsNoAGNz3p017}
\end{figure}

To quantify the effect of the AGN-feedback on the kinematics of the gas in the ENGINE simulations, we directly compare the radial velocity profiles of the gas and dark matter in the NoAGN and RECAL simulations at redshifts $z \sim 3.5$ (Figure \ref{fig:radVel_AGNvsNoAGNz3p528}) and $z \sim 3$ (Figure \ref{fig:radVel_AGNvsNoAGNz3p017}). Analogously to Figure \ref{fig:radVel}, the median radial velocity profiles for different halo mass bins of the hot gas ($> 10^5K$) are shown in orange, those of the Ly$\alpha$ emitting gas in blue and those of the dark matter in grey. The 25th and 75th percentiles are indicated by the shaded regions. The profiles with the dash-dotted lines are extracted from the RECAL simulation (where AGN-feedback is implemented) and those with solid lines from the NoAGN simulation. Unsurprisingly, the radial profiles of the dark matter are completely unaffected by the change in baryonic physics implementation. 
Although stellar feedback is main cause for the hot outflows at these redshifts and at these halo masses, the AGN-feedback leads to higher outflow velocities for the hot gas and this effect increases with increasing halo mass. 
In contrast, the radial velocity profiles of the Ly$\alpha$ emitting gas are not significantly altered by the inclusion of AGN-feedback, implying that the AGN-feedback, as implemented in the ENGINE (and 
EAGLE) simulations has little to no effect on the kinematics of the Ly$\alpha$ emitting gas. As we are only looking at the ENGINE simulations here (NoAGN and RECAL), the highest halo mass bins are only $10^{12.3}$ and $10^{12.6}$ $M_{\odot}$ at redshifts $z\sim3.5$ and $z\sim3$ due to the smaller box sizes of the NoAGN and RECAL simulations compared to the EAGLE Ref simulation box.

\section{Number of detected LAEs compared to \texteta} \label{app:checkClustering}
\begin{figure}
	\centering
	\includegraphics[width=\columnwidth]{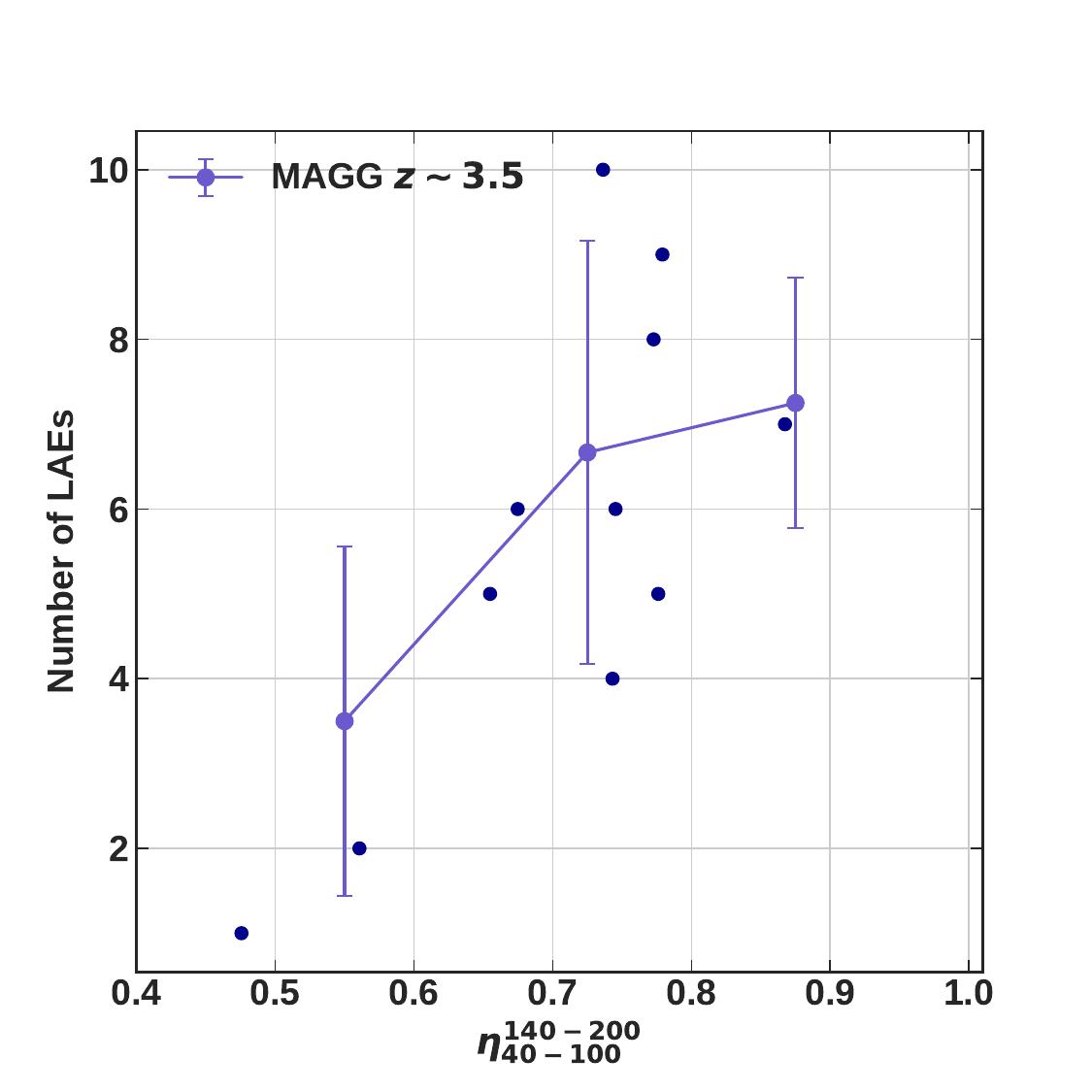}
    \caption{The number of LAEs detected in the MUSE field of view for each Ly$\alpha$ nebulae in the MAGG $z\sim3.5$ sample as a function of the nebulae's measured $\eta^{140 - 200}_{40-100}$. The dark blue points represent the individual nebulae and solid line with errorbars shows the median number of LAEs if the Ly$\alpha$ nebulae are binned in $\eta^{140 - 200}_{40-100}$.}
    \label{fig:checkClustering}
\end{figure}

Here we compare the $\eta^{140 - 200}_{40-100}$'s with the number of detected LAEs in each field of the MAGG $z\sim3.5$ sample, where $\eta^{140 - 200}_{40-100} < 1$. We exclude Ly$\alpha$ nebulae with  $\eta^{140 - 200}_{40-100} > 1$ as this implies that the velocity dispersion of that nebulae increases with the distance to the centre of the halo, indicating superposition effects and not high halo masses, as already mentioned in Sections \ref{subsec:velDispAnnuli} \& \ref{subsec:haloMassEstObs}. We do not include the MAGG $z\sim4.1$ sample in this analysis, as the number of LAEs detected decreases strongly with redshift, see \citet{Fossati2021} for a discussion on the causes for this. As can be seen in Figure \ref{fig:checkClustering}, the number of detected LAEs does increase slightly with the measured $\eta^{140 - 200}_{40-100}$ value of the individual Ly$\alpha$ nebulae. 


\bsp	
\label{lastpage}
\end{document}